 \documentclass[11pt]{article}
\usepackage[square]{natbib}
\usepackage{graphicx}% Include figure files
\usepackage{dcolumn}% Align table columns on decimal point
\usepackage{bm}% bold math
\usepackage{amsmath}
\usepackage{amsxtra}
\usepackage{amstext}
\usepackage{amssymb}
\usepackage{latexsym}
\begin{document}
\title{
\vskip-2.5truecm
\rightline{\small{\tt ULB-TH/08-02}}
\vskip2.5truecm
Nonlinear Transport Processes in Tokamak Plasmas 
\vskip0truecm
Part I: The Collisional Regimes}
\author{ Giorgio SONNINO and Philippe PEETERS}
\date{}
\maketitle
\begin{center}
{\it 
EURATOM - Belgian State Fusion Association

Free University of Brussels (U.L.B.)

Blvd du Triomphe, Campus de la Plaine, C.P. 231, Building  NO}

Brussels, B-1050, Belgium\\
E-mail: {\tt gsonnino@ulb.ac.be}\\
E-mail: {\tt ppeeters@ulb.ac.be}\\
\end{center}
\begin{abstract}
An application of the thermodynamic field theory (TFT) to transport processes in L-mode tokamak plasmas is presented. The nonlinear corrections to the linear ("Onsager") transport coefficients in the collisional regimes are derived. A quite encouraging result is the appearance of an asymmetry between the Pfirsch-Schl{\"{u}}ter (P-S) ion and electron transport coefficients: the latter presents a nonlinear correction, which is absent for the ions, and makes the radial electron coefficients much larger than the former. Explicit calculations and comparisons between the neoclassical results and the TFT predictions for JET plasmas are also reported. We found that the nonlinear electron P-S transport coefficients exceed the values provided by neoclassical theory by a factor, which may be of the order $10^2$. The nonlinear classical coefficients exceed the neoclassical ones by a factor, which may be of order $2$. The expressions of the ion transport coefficients, determined by the neoclassical theory in these two regimes, remain unaltered. 

\noindent The low-collisional regimes i.e., the plateau and the banana regimes, are analyzed in the second part of this work. 
\end{abstract}
\section{Introduction}

The thermodynamic field theory (TFT) was proposed in 1999, to describe the behaviour of thermodynamic systems beyond the linear ("Onsager") region \cite{sonnino}. Attempts to derive a generally covariant thermodynamic field theory (GTFT) can be found in refs \cite{sonnino5}. The TFT extends the theory previously formulated by Prigogine in 1954, which was applied only to thermodynamic systems close to equilibrium. The characteristic feature of this theory is its purely macroscopic nature. We do not mean a formulation based on the macroscopic evolution equations, but rather a purely thermodynamic formulation starting solely from the entropy production and from the transport equations, i.e., the flux-force relations. The latter provide the possibility of defining an abstract space, (the {\it thermodynamic space}) whose metric is given by the transport matrix. The law of evolution is not the dynamical law of motion of particle, or the set of two-fluid macroscopic equations of plasma dynamics. The evolution in the thermodynamic configurations is rather determined by postulating three purely geometrical principles: the {\it Shortest Path Principle}, the {\it Closeness of the Thermodynamic Field Strength}, and the {\it Principle of Least Action}. From theses principles, a set of field equations, constraints, and boundary conditions are derived. These equations, referred to as the {\it thermodynamic field equations}, determine the nonlinear corrections to the linear ("Onsager") transport coefficients. 

\noindent The validity of this theory in the weak-field approximation has been successfully tested in many examples of $\alpha-\alpha$ and $\beta-\beta$ processes, such as the thermoelectric effect and the unimolecular triangular chemical reaction \cite{sonnino}\footnote{Here, we adopt the terminology of De Groot and Mazur \cite{degroot}, i.e., when the velocity distribution function is an even (odd) function of the velocities of the particles, a processes is said to be an $\alpha$-processes ($\beta$-processes). It is possible to show that this definition implies that $\alpha$ processes only involve the symmetric part of the Onsager tensor, whereas $\beta$ processes only involve the skew-symmetric one.\label{processes}}. The thermodynamic field equations, in the weak-field approximation, have also been applied to several $\alpha-\beta$ processes. For example, the Field-K{\"{o}}r{\"{o}}s-Noyes model, in which the thermodynamic forces and flows are related by an asymmetric tensor, was analyzed in ref. \cite{peeters}. Even on this case, the numerical solutions of the model are in agreement with the theoretical predictions of the TFT. More recently, the Hall effect \cite{sonnino7} has been analyzed in the nonlinear region. In each of these papers, it was shown that the TFT successfully describes the known physics in the nonlinear region and also predicts new interesting effects, such as the nonlinear Hall effect. The theoretical predictions of the nonlinear Hall effect have been confirmed experimentally \cite{hall}.

\noindent The TFT has also been used to study transport processes in magnetically confined plasmas. Preliminary results can be found in refs \cite{sonnino8} and  \cite{sonnino8a}. The study of the behaviour of a plasma in the presence of an inhomogeneous and curved magnetic field is one of the main objects of the neoclassical theory. One of the most important results of the neoclassical theory is that the global geometry of the magnetic field has a very strong influence on the transport processes. The neoclassical theory is able to derive the transport coefficients for ions and electrons when the plasma is magnetically confined in tokamak reactors. In spite of its elegant and coherent formulation, the theoretical predictions of the neoclassical theory are in strong disagreement with experience. The experimental ion heat flux measured in tokamak plasmas is roughly in agreement with the neoclassical theory. However, the electron particle flux and the electron heat flux are about $10^2\div10^3$ greater than the values computed by the neoclassical theory. This difference between the experimental and the neoclassical flux is referred to as the {\it anomalous flux}. For many physicists, the origin of this discrepancy is mainly attributed to turbulent phenomena existing in tokamak plasmas. Fluctuations in plasmas can become unstable and therefore amplified. According to this interpretation, fluctuations will successively interact in a nonlinear way leading the plasma to a state, which is far away from equilibrium. In this condition, the transport properties are supposed to change significantly.  

\noindent In this one and the successive paper, we shall analyze the influence of the nonlinear contributions, evaluated by the thermodynamic field theory, on the transport processes in tokamak plasmas. As mentioned above, the neoclassical theory (i.e., the linear theory) fails with a factor $10^2\div10^3$ and thus magnetically confined plasmas represent an ideal case for testing the validity of our thermodynamic approach. For simplicity, in our calculations we deal with {\it fully ionized plasmas} defined as a collection of electrons and positively charged ions \cite{balescu1}. In the present and in the subsequent work \cite{sonnino9}, we analyzed in great detail transport processes in tokamak plasmas in the collisional and in low-collisional regimes. In this first part, we limit ourselves to a confined plasma in collisional regimes, i.e., in the classical and Pfirsch-Schl{\"u}ter regimes, providing the complete set of the ion and electron nonlinear transport equations and analyzing the solution for JET-plasmas. We report a new set of nonlinear transport equations for plasmas in collisional regimes different from those established in refs \cite{sonnino8} and \cite{sonnino8a}. This is due to the fact that here we adopt a definition of the Pfirsch-Schl{\"u}ter thermodynamic forces different from that reported in refs \cite{sonnino8} and  \cite{sonnino8a}. The work is organized as follows: in sections (\ref{entropy}) and (\ref{linear}) we derive the expressions of the entropy productions for ions and electrons in magnetically confined plasmas and the transport equations in the linear region. After a brief summary of the general theory of the thermodynamic field (TFT), we adapt the formalism in order to obtain the thermodynamic field equations for a fully ionized plasma. The solution of these equations determine the nonlinear corrections to the linear ("Onsager") transport coefficients. The analysis of the classical and Pfirsch-Schl{\"u}ter transport is conducted in sections (\ref{ps}) and (\ref{NLps}): in section (\ref{ps}) we derive the gauge-invariant form of the solutions, which are explicitly analyzed in section (\ref{NLps}). As general result we find that, in the Pfirsch-Schl{\"{u}}ter regime, the thermodynamic fluxes are linked to the thermodynamic forces by an {\it amplification factor} ${\mathcal F}_{ps}$ times the Onsager matrix. In order to interpret the mathematical results found in section (\ref{NLps}) in terms of collisional mechanisms, a kinetic model can be used originally introduced in ref. \cite{sonnino10}. Through this model, it was possible to show that, in the collisional regimes, the transport coefficients are approximatively given by the linear (Onsager) transport coefficients times a function, which is proportional to the inverse of the electron collision time $\tau_e$. At the end of section (\ref{NLps}) we also find specific calculations for JET plasmas. The main conclusion of our analysis is that the electron nonlinear Pfirsch-Schl{\"{u}}ter transport coefficients exceed the values provided by neoclassical theory by a factor ${\mathcal F}_{ps}$, which may be of order $10^2$. The values of the ion transport coefficients remain, however, unaltered. These results are in line with experimental observations. The definition of gauge invariance of the field equations for our problem and the proof that our choice of boundary conditions respects the principle of covariance is reported in appendix (\ref{gauge}). Details on the derivation of the appropriate boundary conditions for our problem can be found in appendix (\ref{Solutions}). The explicit solution of the field equations, submitted to the appropriate boundary conditions, is given in appendix (\ref{2-solution})

\section{Entropy Production of Magnetically Confined Plasmas}\label{entropy}

We consider a plasma consisting of electrons and a single species of ions, in the presence of an external axisymmetric confining magnetic field. As it is known, neglecting the drift and modified drift mechanism, in magnetically confined plasmas, we have three distinct mechanisms of transport: banana, plateau and the Pfirsch-Schl{\"u}ter transport regimes (see, for example, ref. \cite{balescu2}). In ref. \cite{balescu2} we can find the evaluation of the entropy production $\sigma^{\alpha}$ of species $\alpha$ ($\alpha =e$ for electrons and $\alpha=i$ for ions). Let us first introduce the {\it dimensionless (density of) entropy productions of species} $\alpha$, $\Sigma^{\alpha}$:
\begin{equation}\label{entropy1}
\Sigma^{\alpha}\equiv \frac{\tau_{\alpha}}{n_{\alpha}}\sigma^{\alpha}
\end{equation}
\noindent where $\tau_{\alpha}$ and $n_{\alpha}$ indicate the {\it relaxation time} and the {\it number density of particles} of species $\alpha$, respectively. The expression for $\Sigma^{\alpha}$ is \cite{balescu2}:
\begin{equation}\label{entropy2}
\Sigma^{\alpha}=-q_r^{(1)}\tau_{\alpha}G_r^{\alpha (1)}-\sum_{n=1}^{N}q_r^{\alpha (2n+1)}\tau_{\alpha}G_r^{\alpha (2n+1)}
\end{equation}
\noindent The upper limit is $\infty$, but in practice the sum is truncated at $n=N$. The $G_r^{\alpha (2n+1)}$ and the $q_r^{\alpha (2n+1)}$ indicate the {\it dimensionless collisional terms} (often also referred to as the {\it dimensionless generalized frictions}) and the {\it dimensionless irreducible Hermitian moments}, respectively. $q_r^{(1)}$ is not an Hermitian moment (i.e., a moment defined by the deviation from the local Maxwellian) but indicates the {\it dimensionless electric current density}. Index $r$ denotes the components of this vector. Expression (\ref{entropy2}) is completely general: it is based only on the assumption that the deviation from the local equilibrium is small and that tensor Hermitian moments provide negligible contributions. The specificity of the neoclassical theory appears when we express the generalized frictions $G_r^{\alpha (2n+1)}$ in terms of the dimensionless Hermitian moments $q_r^{\alpha (2n+1)}$ by making use of the approximated moment equations. These equations are obtained assuming the validity of the {\it drift approximation}. Indicating with $\varepsilon$ the drift parameter i.e., the ratio of the Larmor radius to the length of the gradient of the magnetic field, all terms of order $\varepsilon^2$ or higher appearing in the moment equations can be neglected. This means that the time-derivative $\tau_{\alpha}\partial_t q_r^{\alpha (n)}$ (of order $\varepsilon^2$) as well as the nonlinear terms (of order $\varepsilon^2$ at most) give negligible contributions. Up to the order $\varepsilon^2$, the final moment equations read \cite{balescu2}
\begin{eqnarray}\label{entropy3}
&&\Omega_{\alpha}\tau_{\alpha}\epsilon_{rmn}q_m^{\alpha (1)}b_n+\tau_{\alpha}G_r^{\alpha (1)}+g_r^{\alpha (1)}+{\bar g}_r^{\alpha (1)}=0+O(\varepsilon^2)\nonumber\\
&&\Omega_{\alpha}\tau_{\alpha}\epsilon_{rmn}q_m^{\alpha (3)}b_n+\tau_{\alpha}G_r^{\alpha (3)}+g_r^{\alpha (3)}+{\bar g}_r^{\alpha (3)}=0+O(\varepsilon^2)\\
&&\Omega_{\alpha}\tau_{\alpha}\epsilon_{rmn}q_m^{\alpha (5)}b_n+\tau_{\alpha}G_r^{\alpha (5)}+{\bar g}_r^{\alpha (5)}=0+O(\varepsilon^2)\nonumber
\end{eqnarray}
\noindent The moment equations link the generalized frictions $G_r^{\alpha (n)}$ with the {\it purely thermodynamic forces} $g_r^{(1)}$, $g_r^{\alpha(3)}$ and the {\it generalized stresses} ${\bar g}_r^{\alpha (2n+1)}$. $b_n$ is a unit vector along the magnetic field {\bf B} i.e., $b_n\equiv B_n/B$, $\epsilon_{rmn}$ is the completely antisymmetric Levi-Civita symbol and $\Omega_{\alpha}$ is the electron ($\alpha=e$) or the ion ($\alpha=i$) Larmor frequency. From Eq.(\ref{entropy3}) we obtain
\begin{eqnarray}\label{entropy4}
&&\Sigma^e=q_r^{(1)}g_r^{(1)}+q_r^{e(3)}g_r^{e(3)}-q_r^{(1)}{\bar g}_r^{e(1)}+\sum_{n=1}^Nq_r^{e(2n+1)}{\bar g}_r^{e(2n+1)}\nonumber\\
&&\Sigma^i=q_r^{i(3)}g_r^{i(3)}+\sum_{n=1}^Nq_r^{i(2n+1)}{\bar g}_r^{i(2n+1)}
\end{eqnarray}
\noindent The expression for the entropy production is therefore an infinite bilinear form, truncated at $n=N$. We draw attention to the fact that expression (\ref{entropy4}) is valid up to the leading order in $\mu=m_e/m_i$ and, therefore, the ion-electron entropy production, giving contribution of relative order $\mu=m_e/m_i$, has been neglected. This also explains why the expression for the electronic entropy production differs from the ionic one. In literature, expression (\ref{entropy4}) is referred to as the {\it quasi-thermodynamic form of the entropy production} \cite{balescu2}. If we work in the twenty-one moment (21 M) approximation (i.e., $N=2$), Eqs (\ref{entropy4}) can be cast into the form
\begin{eqnarray}\label{entropy5}
&&\Sigma^e=q_r^{(1)}(g_r^{(1)}-{\bar g}_r^{e(1)})+q_r^{e(3)}(g_r^{e(3)}+{\bar g}_r^{e(3)})+q_r^{e(5)}{\bar g}_r^{e(5)}\nonumber\\
&&\Sigma^i=q_r^{i(3)}(g_r^{i(3)}+{\bar g}_r^{i(3)})+q_r^{i(5)}{\bar g}_r^{i(5)}
\end{eqnarray}
\noindent For easy reference, we list here explicitly the relations between the dimensionless and the corresponding dimensional Hermitian moments:
\begin{eqnarray}\label{entropy6}
&&q_r^{\alpha (1)}=\Bigl(\frac{m_{\alpha}}{T_{\alpha}}\Bigr)^{1/2}\frac{1}{n_{\alpha}}\Gamma_r^{\alpha}\nonumber\\
&&q_r^{(1)}=-q_r^{e(1)}+\Bigl(\frac{m_e}{T_e}\Bigr)^{1/2} u_r=\frac{1}{en_e}\Bigl(\frac{m_e}{T_e}\Bigr)^{1/2}j_r\nonumber\\
&&q_r^{\alpha (3)}=\sqrt{\frac{2}{5}}\Bigl(\frac{m_{\alpha}}{T_{\alpha}}\Bigr)^{1/2}\frac{1}{T_{\alpha}n_{\alpha}}Q_r^{\alpha}\\
&&q_r^{\alpha (5)}=\frac{1}{n_{\alpha}}\Bigl(\frac{m_{\alpha}}{T_{\alpha}}\Bigr)^{1/2}L_r^{\alpha}\nonumber
\end{eqnarray}
\noindent where $m_{\alpha}$ and $T_{\alpha}$ are respectively the mass and the temperature of species $\alpha$ and ${\bf u}$ denotes the {\it centre-of-mass velocity} of the plasma. Moreover, $j_r$, $\Gamma_r^{\alpha}$ and $Q_r^{\alpha}$ indicate the {\it electric current}, the {\it particle fluxes} and the {\it heat fluxes}, respectively. $L_r^{\alpha}$ is the dimensional fifth-order Hermitian moment corresponding to $q_r^{\alpha (5)}$. For completeness, we also report the relation between the {\it pressure tensor} $\pi_{rs}^{\alpha}$ and the {\it second-order tensor Hermitian moment} $q_{rs}^{\alpha (2)}$
\begin{equation}\label{entropy7}
q_{rs}^{\alpha (2)}=\frac{1}{\sqrt{2}n_{\alpha}T_{\alpha}}\pi_{rs}^{\alpha}
\end{equation}
\noindent The dimensionless generalized forces $g_{r}^{\alpha (2n+1)}$ and the dimensionless generalized source terms are defined as
\begin{eqnarray}\label{entropy8}
&&\!\!\!\!\!\!\!\!\!\!\!\!\!g_r^{(1)}=\tau_e\Bigl(\frac{m_e}{T_e}\Bigr)^{1/2}\Bigl(\frac{e}{m_e}E_r-\Omega_e\epsilon_{rmn}u_mb_n+\frac{1}{m_en_e}\nabla_r(n_eT_e)\Bigr)\nonumber\\
&&\!\!\!\!\!\!\!\!\!\!\!\!\!g_r^{\alpha (1)}=\tau_{\alpha}\Bigl(\frac{m_{\alpha}}{T_{\alpha}}\Bigr)^{1/2}\Bigl(\frac{e_{\alpha}}{m_{\alpha}}E_r-\frac{1}{m_{\alpha}n_{\alpha}}\nabla_r(n_{\alpha} T_{\alpha})\Bigr)\nonumber\\
&&\!\!\!\!\!\!\!\!\!\!\!\!\!g_r^{\alpha (3)}=-\sqrt{\frac{5}{2}}\tau_{\alpha}\Bigl(\frac{T_{\alpha}}{m_{\alpha}}\Bigr)^{1/2}\frac{1}{T_{\alpha}}\nabla_rT_{\alpha}\nonumber\\
&&\!\!\!\!\!\!\!\!\!\!\!\!\!{\bar g}_r^{\alpha (1)}=-\sqrt{2}\tau_{\alpha}\Bigl(\frac{m_{\alpha}}{T_{\alpha}}\Bigr)^{1/2}\frac{1}{m_{\alpha}n_{\alpha}}\nabla_s(n_{\alpha}T_{\alpha}q_{rs}^{\alpha (2)})\\
&&\!\!\!\!\!\!\!\!\!\!\!\!\!{\bar g}_r^{\alpha (3)}=-\sqrt{\frac{2}{5}}\tau_{\alpha}\Bigl(\frac{T_{\alpha}}{m_{\alpha}}\Bigr)^{1/2}\!
\Bigl[\sqrt{7}\frac{1}{n_{\alpha}T^2_{\alpha}}\nabla_s(n_{\alpha}T^2_{\alpha}q_{rs}^{\alpha (4)})\!+\!\sqrt{2}T_{\alpha}^{-7/2}\nabla_s(T_{\alpha}^{7/2}q_{rs}^{\alpha (2)})\Bigr]\nonumber\\
&&\!\!\!\!\!\!\!\!\!\!\!\!\!{\bar g}_r^{\alpha (5)}=-\sqrt{\frac{2}{5}}\tau_{\alpha}\Bigl(\frac{T_{\alpha}}{m_{\alpha}}\Bigr)^{1/2}\Bigl[
\frac{3}{n_{\alpha}T^3_{\alpha}}\nabla_s(n_{\alpha}T^3_{\alpha}q_{rs}^{\alpha (6)})\nonumber\\
&&\qquad\qquad\qquad\qquad\qquad+2T_{\alpha}^{-11/2}\nabla_s(T_{\alpha}^{11/2}q_{rs}^{\alpha (4)})+
\sqrt{14}q_{rs}^{\alpha (2)}T_{\alpha}^{-1}\nabla_sT_{\alpha}\Bigr]\nonumber
\end{eqnarray}
\noindent Indicating with $e$ the {\it absolute value of the charge of the electron} and with $Z$ the {\it charge number} of the ions, we have $e_{\alpha}=-e$ for electrons and $e_{\alpha}=+Ze$ for ions. {\bf E} is the electric field of the plasma. In the {\it local dynamical triad} it is possible to show the validity of the following relations \cite{balescu2}:
\begin{eqnarray}\label{entropy9}
&&g_{\parallel}^{\alpha (n)}=O(\varepsilon);\quad {\bar g}_{\parallel}^{\alpha (n)}=O(\varepsilon)
;\quad g_{\rho}^{\alpha (n)}=O(\varepsilon);\quad {\bar g}_{\wedge}^{\alpha (n)}=O(\varepsilon)\nonumber\\
&&{\tilde L}_{\wedge}=O(\varepsilon);\quad {\tilde L}_{\perp}=O(\varepsilon^2)
\end{eqnarray}
\noindent where ${\tilde L}$ is any dimensionless transport coefficient. Taking into account relation (\ref{entropy9}), expressions (\ref{entropy5}) reduce to
\begin{eqnarray}\label{entropy10}
&&\!\!\!\!\!\!\!\!\!\!\Sigma^e=q_{\parallel}^{(1)}(g_{\parallel}^{(1)}-{\bar g}_{\parallel}^{e(1)})+
q_{\parallel}^{e(3)}(g_{\parallel}^{e(3)}+{\bar g}_{\parallel}^{e(3)})+q_{\parallel}^{e(5)}{\bar g}_{\parallel}^{e(5)}-{\hat q}_{\rho cl}^{e(1)}g_{\rho}^{(1)}+
{\hat q}_{\rho cl}^{e(3)}g_{\rho}^{e(3)}\nonumber\\
&&\!\!\!\!\!\!\!\!\!\!\Sigma^i=q_{\parallel}^{i(3)}(g_{\parallel}^{i(3)}+{\bar g}_{\parallel}^{i(3)})+q_{\parallel}^{i(5)}{\bar g}_{\parallel}^{i(5)}+{\hat q}_{\rho cl}^{i(3)}g_{\rho}^{i(3)}
\end{eqnarray}
\noindent Eqs (\ref{entropy10}) are valid up to order $\varepsilon^2$. ${\hat q}_{\rho cl}^{e(1)}$, ${\hat q}_{\rho cl}^{e(3)}$ and ${\hat q}_{\rho cl}^{i(3)}$ are the radial Hermitian moments {\it truncated} at the order $\epsilon^2$. Notice that $\Sigma^e_{cl}\equiv-{\hat q}_{\rho cl}^{e(1)}g_{\rho}^{(1)}+
{\hat q}_{\rho cl}^{e(3)}g_{\rho}^{e(3)}$ and $\Sigma^i_{cl}\equiv {\hat q}_{\rho cl}^{i(3)}g_{\rho}^{i(3)}$ are the classical contributions to the entropy productions. $g_{\parallel}^{(1)}$ can further be decomposed as
\begin{eqnarray}\label{entropy11}
&&g_{\parallel}^{(1)} =- g_{\parallel}^{e(1)P}-g_{\parallel}^{e(1)\Phi}-g_{\parallel}^{e(1)A}=-g_{\parallel}^{e(1)}\nonumber\\
&&g_{\parallel}^{e(1)P}\equiv -\Bigl(\frac{m_e}{T_e}\Bigr)^{1/2}\tau_e\frac{1}{m_en_e}\nabla_{\parallel}P_e\nonumber\\
&&g_{\parallel}^{e(1)\Phi}\equiv -\Bigl(\frac{m_e}{T_e}\Bigr)^{1/2}\tau_e\frac{e}{m_e}\nabla_{\parallel}\Phi\\
&&g_{\parallel}^{e(1)A}\equiv \Bigl(\frac{m_e}{T_e}\Bigr)^{1/2}\tau_e\frac{e}{m_e}E_{\parallel}^{(A)}\nonumber
\end{eqnarray}
\noindent where $P_e$ is the {\it electron pressure}, $-\nabla_r \Phi$ represents the self-consistent electric field built inside the plasma and $E_r^{(A)}$ represents the electric field induced by purely external means (such as transformer coils) in the confined plasma. The self-consistent electric field is, usually, much smaller than the external electric field and for this it can be neglected. 

It is useful to collect here the hypotheses adopted to obtain expressions (\ref{entropy10}):
\begin{description}
\item[1)] {\it The state of the plasma is not too far from the reference local equilibrium state};
\item[2)] {\it The drift approximation is applicable};
\item[3)] {\it The ratio $\mu=m_e/m_i\ll 1$ and therefore quantities of order $\mu$ have been neglected (but not quantities of order less that $\mu$)};
\item[4)] {\it The tensor Hermitian moments have been neglected};
\item[5)] {\it The expressions have been evaluated in the 21 M approximation}.
\end{description}
\noindent Assumption 1) deserves an additional comment. The expression for the entropy production has been obtained making use of the so-called {\it local equilibrium principle}. However, the validity of this expression goes beyond the validity of the hypothesis of local equilibrium i.e., Eq. (\ref{entropy10}) {\it remains correct also when the local equilibrium principle is invalid}. Indeed, it is possible to show that the expression of the entropy production, written as sum of products of thermodynamic forces and their conjugate fluxes, is valid {\it throughout the whole range of thermodynamics} \cite{prigogine}-\cite{prigogine1}. More generally, we can state that the limit of validity for the expression of the entropy production, written in a bilinear form, establishes the limit of validity of the {\it thermodynamic description of a physical system} \cite{prigogine1}. In ref.\cite{jou} we can find many physical examples where the local equilibrium principle is violated and the expression for the entropy production can still be brought in a bilinear form. Then, the only precaution that we have to take is to truncate the expressions at the first order of the drift parameter $\varepsilon$.

From the physical point of view, we have taken into account one assumption and the {\it plasmadynamical balance equations}:

\begin{description}
\item[a)] {\it Collisions are the only source of irreversibility or dissipation};
\item[b)] {\it The plasmadynamical equations, expressing the conservation of mass, energy and momentum of plasmas, have been taken into account};
\item[c)] {\it The plasma is in mechanical equilibrium i.e., $\frac{d{\bf u}}{dt}=0$}.
\end{description}
\noindent Assumption a) is valid whenever the plasma is quiescent. Additional sources of dissipation must be taken into account if the plasma becomes unstable and turbulent \cite{balescu3}. The hydrodynamic equations have been used to obtain the Hermitian moment equations. As already mentioned, the truncation of the hydrodynamic equations is based on the smallness of the drift parameter $\varepsilon$ and {\it not} on the hydrodynamic parameter $\lambda_H$, defined as the ratio between the shortest mean free path of the particles and the largest length of the hydrodynamic gradients. The momentum conservation implies the following relation
\begin{equation}\label{entropy12}
P_e\Bigl(\frac{m_e}{T_e}\Bigr)^{3/2}G_r^{e(1)}+P_i\Bigl(\frac{m_i}{T_i}\Bigr)^{3/2}G_r^{i(1)}=0
\end{equation}
\noindent where $P_i$ denotes the {\it ion pressure}. In terms of dimensionless generalized thermodynamic forces, Eq.(\ref{entropy12}) can be cast in the form \cite{balescu2}:
\begin{equation}\label{entropy13}
g_{\parallel}^{e(1)}+{\hat a}g_{\parallel}^{i(1)}+{\bar g}_{\parallel}^{e(1)}+{\hat a}{\bar g}_{\parallel}^{i(1)}=0
\end{equation}
\noindent where ${\hat a}\equiv \frac{\tau_e}{\tau_i}\frac{1}{Z}\Bigl(\frac{T_i}{\mu T_e}\Bigr)^{1/2}$. 

The Hermitian moments $q_{\parallel}^{\alpha (n)}$ are linked to the {\it poloidal fluxes} $\omega_n^{\alpha}$ through the following relations \cite{balescu2}:
\begin{eqnarray}\label{entropy14}
&&q_{\parallel}^{(1)} =- K_e\frac{\beta_0}{B}g_{\rho}^{(1)P}+\frac{B}{\beta_0}\!\omega_1
\qquad\qquad\quad\omega_1\equiv a\omega_1^i-\omega_1^e\nonumber\\
&&q_{\parallel}^{\alpha (3)} =K_{\alpha}\frac{\beta_0}{B}g_{\rho}^{\alpha (3)}+\frac{B}{\beta_0}\omega_3^{\alpha}\qquad\qquad\quad K_{\alpha}=\frac{B_{\xi}}{B_{\theta}}\Bigl(\frac{1}{\Omega_{\alpha 0}\tau_\alpha}\Bigr)\\
&&q_{\parallel}^{\alpha (5)} =\frac{B}{\beta_0}\omega_5^{\alpha}\nonumber
\end{eqnarray}
\noindent where $a=(\mu T_i/T_e)^{1/2}$ and $\Omega_{\alpha 0}$ is the {\it Larmor frequency associated with the average magnetic field}. $\beta_0\equiv <B^2>^{1/2}$ where $<\dots>$ denotes the {\it magnetic-surface averaging operation}. $B_{\xi}$ and $B_{\theta}$ are the components of the magnetic field in the {\it local geometric triad} (often called the Hinton-Hazeltine coordinates \cite{hinton}) i.e., ${\bf B}=B_{\theta}{\bf e_{\theta}}+B_{\xi}{\bf e_{\xi}}$.  ${\bf e}_{\theta}$ and ${\bf e}_{\xi}$ are the unit vectors in the {\it local geometrical triad}. Quantity $g_{\rho}^{(1)P}$ is defined as 
\begin{equation}\label{entropy15}
g_{\rho}^{(1)P}=-\tau_e\Bigl(\frac{T_e}{m_e}\Bigr)^{1/2}\Bigl(1+\frac{P_i}{P_e}\Bigr)\frac{\nabla_{\rho}P}{P}
\end{equation}
\noindent where $P$ indicates the {\it total pressure} i.e., $P=P_e+P_i$. The poloidal fluxes are surface quantities, independent of the poloidal angle $\theta$. In general, the quantity $K_{\alpha}$ is not a surface quantity but it becomes one for some configurations of the magnetic field as, for example, in the standard model. In this paper we decide to make the assumption that also quantity $K_{\alpha}$ is a surface quantity.

The parallel fluxes $q_{\parallel}^{\alpha (n)}$ and  $q_{\parallel}^{(1)}$ can be further decomposed in three contributions
\begin{eqnarray}\label{entropy16}
&&q_{\parallel}^{\alpha (n)}=q_{\parallel cl}^{\alpha (n)}+q_{\parallel ps}^{\alpha (n)}+q_{\parallel bp}^{\alpha (n)}\nonumber\\
&&q_{\parallel}^{(1)}=q_{\parallel cl}^{ (1)}+q_{\parallel ps}^{(n)}+q_{\parallel bp}^{(1)}\
\end{eqnarray}
\noindent The first contribution, $q_{\parallel cl}^{\alpha (n)}$, represents the parallel flux evaluated by the classical transport theory. The second contribution, $q_{\parallel ps}^{\alpha (n)}$, is the parallel Pfirsch-Schl{\"{u}}ter fluxes, which we define as
\begin{equation}\label{entropy17}
<Bq_{\parallel ps}^{\alpha (n)}>=0\qquad\quad\forall\ \alpha ,\ n
\end{equation}
\noindent The differences between the total parallel fluxes and the classical plus the Pfirsch-Schl{\"{u}}ter fluxes will be referred to as the banana/plateau flux $q_{\parallel bp}^{\alpha (n)}$. We can easily calculate the value $\omega_n^{\alpha\ast}$ of the poloidal fluxes such that Eq.(\ref{entropy17}) is verified. Indeed, from Eqs (\ref{entropy14}) and (\ref{entropy16}) we find
\begin{eqnarray}\label{entropy18}
&&\!\!\!\!\!\!\!\!\!\!\!\!\!\!\!\!\!\!\!\!<Bq_{\parallel}^{(1)}> =<B(- K_e\frac{\beta_0}{B}g_{\rho}^{(1)P}+\frac{B}{\beta_0}\omega_1^{\ast})>=0\ \ \Longrightarrow\quad \omega_1^{\ast}=K_eg_{\rho}^{(1)P}\nonumber\\
&&\!\!\!\!\!\!\!\!\!\!\!\!\!\!\!\!\!\!\!\!<Bq_{\parallel}^{\alpha (3)}>=<B (K_{\alpha}\frac{\beta_0}{B}g_{\rho}^{\alpha (3)}+\frac{B}{\beta_0}\omega_3^{\alpha\ast})>=0\ \ \Longrightarrow\quad \omega_3^{\alpha\ast}=-K_{\alpha}g_{\rho}^{\alpha(3)}\\
&&\!\!\!\!\!\!\!\!\!\!\!\!\!\!\!\!\!\!\!\!<B(q_{\parallel}^{\alpha (5)})>=<B( \frac{B}{\beta_0}\omega_5^{\alpha\ast})>=0\qquad\qquad\qquad\!\Longrightarrow\quad\omega_5^{\alpha\ast}=0\nonumber
\end{eqnarray}
\noindent From Eq.(\ref{entropy14}), we have
\begin{eqnarray}\label{entropy19}
&&q_{\parallel}^{(1)} =\Bigl(\frac{B}{\beta_0}-\frac{\beta_0}{B}\Bigr)K_eg_{\rho}^{(1)P}+q_{\parallel bpcl}^{(1)} \nonumber\\
&&q_{\parallel}^{\alpha (3)} =-\Bigl(\frac{B}{\beta_0}-\frac{\beta_0}{B}\Bigr)K_{\alpha}g_{\rho}^{\alpha (3)}+q_{\parallel bpcl}^{\alpha (3)}\\
&&q_{\parallel}^{\alpha (5)} =q_{\parallel bpcl}^{\alpha (5)} \nonumber
\end{eqnarray}
\noindent where
\begin{eqnarray}\label{entropy20}
&&q_{\parallel bpcl}^{\alpha (n)}=q_{\parallel bp}^{\alpha (n)}+q_{\parallel cl}^{\alpha (n)}\nonumber\\
&&q_{\parallel bpcl}^{(1)}=q_{\parallel bp}^{(1)}+q_{\parallel cl}^{(1)}
\end{eqnarray}
\noindent and 
\begin{eqnarray}\label{entropy21}
&&q_{\parallel ps}^{(1)} =\Bigl(\frac{B}{\beta_0}-\frac{\beta_0}{B}\Bigr)K_eg_{\rho}^{(1)P}\nonumber\\
&&q_{\parallel ps}^{\alpha (3)} =-\Bigl(\frac{B}{\beta_0}-\frac{\beta_0}{B}\Bigr)K_{\alpha}g_{\rho}^{\alpha (3)}\\
&&q_{\parallel ps}^{\alpha (5)} =0\nonumber
\end{eqnarray}
\noindent As we can see, using Eq.(\ref{entropy17}), we have easily obtained the expression of the parallel fluxes in the Pfirsch-Schl{\"{u}}ter regime. In literature, the contribution to electric current $\beta_0<Bq_{\parallel}^{(1)} >$, produced by the pressure and the thermal gradients, but {\it not} by the parallel electric field, is referred to as the {\it bootstrap current} \cite{bikerton}, \cite{kadomtsev} and \cite{wesson}. The presence of the bootstrap current leads to the attractive idea of operating a tokamak in a steady state, with ${\bf E^{(A)}}=0$. From Eq.({\ref{entropy17}) we immediately found that there is no contribution to the bootstrap current in the Pfirsch-Schl{\"{u}}ter transport. On the other hand, the classical transport contributes to the parallel electric flux only through the presence of the parallel electric field. Therefore, the bootstrap current exists only in the banana regime. Taking into account Eqs (\ref{entropy16}) and (\ref{entropy20}), the expression of the entropy production, Eq.(\ref{entropy10}), can be brought into the following form
\begin{eqnarray}\label{entropy22}
&&\!\!\!\!\!\!\!\!\!\!\Sigma^e=q_{\parallel ps}^{(1)}(g_{\parallel}^{(1)}-{\bar g}_{\parallel}^{e(1)})+
q_{\parallel ps}^{e(3)}(g_{\parallel}^{e(3)}+{\bar g}_{\parallel}^{e(3)})+q_{\parallel bpcl}^{(1)}(g_{\parallel}^{(1)}-{\bar g}_{\parallel}^{e(1)})
\nonumber\\
&&\ \ 
+q_{\parallel bpcl}^{e(3)}(g_{\parallel}^{e(3)}+{\bar g}_{\parallel}^{e(3)})
+q_{\parallel bpcl}^{e(5)}{\bar g}_{\parallel}^{e(5)}+{\hat q}_{\rho cl}^{e(1)}g_{\rho}^{(1)P}+
{\hat q}_{\rho cl}^{e(3)}g_{\rho}^{e(3)}\nonumber\\
&&\!\!\!\!\!\!\!\!\!\!\Sigma^i=q_{\parallel ps}^{i(3)}(g_{\parallel}^{i(3)}+{\bar g}_{\parallel}^{i(3)})+q_{\parallel bpcl}^{i(3)}(g_{\parallel}^{i(3)}+{\bar g}_{\parallel}^{i(3)})+q_{\parallel bpcl}^{i(5)}{\bar g}_{\parallel}^{i(5)}+{\hat q}_{\rho cl}^{i(3)}g_{\rho}^{i(3)}
\end{eqnarray}
\noindent where we have used the identity $g_{\rho}^{(1)}=-g_{\rho}^{(1)P}$ \cite{balescu2}.

The quantities of major interest are the radial fluxes averaged over a magnetic surface $<q_{\rho}^{\alpha (n)}>$. Indeed, these are the quantities measuring the leakage of matter and energy through the confinement region. It is possible to show that the average radial fluxes can be decomposed in six contributions \cite{balescu2} and \cite{hirshman}
\begin{eqnarray}\label{entropy23}
&&\!\!\!\!\!\!\!\!\!\!\!\!\!\!\!\!\!\!\!\!<q_{\rho}^{\alpha (n)}>=<q_{\rho}^{\alpha (n)}>_{cl}+<q_{\rho}^{\alpha (n)}>_{ps}+<q_{\rho}^{\alpha (n)}>_{b}+<q_{\rho}^{\alpha (n)}>_{p}\nonumber\\
&&\qquad\qquad\qquad\qquad\qquad+\delta_{n,1}<q_{\rho}^{\alpha (n)}>_{dr}+\delta_{n,1}<q_{\rho}^{\alpha (n)}>_{mdr}
\end{eqnarray}
\noindent In general the {\it electric drift fluxes} $<q_{\rho}^{\alpha (n)}>_{dr}$ and the {\it modified electric drift fluxes} $\delta_{n,1}<q_{\rho}^{\alpha (n)}>_{mdr}$ are very small contributions compared to the other fluxes. Therefore, they will be neglected in the forthcoming sections. The total average radial fluxes are simply the sum of the classical, the Pfirsch-Schl{\"{u}}ter, the banana and the plateau averaged radial  fluxes:
\begin{equation}\label{entropy24}
<q_{\rho}^{\alpha (n)}>\simeq<q_{\rho}^{\alpha (n)}>_{cl}+<q_{\rho}^{\alpha (n)}>_{ps}+<q_{\rho}^{\alpha (n)}>_{b}+<q_{\rho}^{\alpha (n)}>_{p}
\end{equation}
\noindent In section (\ref{nlps}) we shall provide the nonlinear corrections to the averaged radial fluxes in the nonlinear classical and Pfirsch-Schl{\"{u}}ter regimes.

Finally, it is important to note that our formalism starts from the concept of the entropy production. Thus when the plasma is far from equilibrium, {\it the entropy production is no longer a potential and its extremal property does not determine the non-equilibrium steady states of the plasma}. A correct theory cannot be based on the {minimum entropy production theorem} for finding steady states when the plasma is far from equilibrium. In fact, a plasma far from equilibrium must satisfy the {\it Universal Criterion of Evolution} established in 1954 by Glansdorff and Prigogine \cite{prigogine2}. This criterion however is not derived from a variational principle and by itself is not able to provide the corrections to the Onsager theory. The nonlinear corrections will be obtained by solving the {\it field equations} \cite{sonnino} combined with nonequilibrium statistical mechanics, establishing in this way the link between micro- and macro-levels. In section (\ref{tft}) we briefly summarize the Thermodynamic Field Theory and we write the field equations for a fully ionized plasma beyond the Onsager region.

\section{Linear Analysis}\label{linear}

In the linear region, the relations between the dimensionless Hermitian moment and the dimensionless thermodynamic forces can be brought into the following form:
\begin{eqnarray}\label{linear1}
&& q_{\parallel}^{(1)}={\tilde{\sigma}}_{\parallel}(g_{\parallel}^{(1)}-{\bar g}^{e(1)}_{\parallel})+{\tilde{\alpha}}_{\parallel}(g_{\parallel}^{e(3)}+{\bar g}^{e(3)}_{\parallel})\nonumber\\
&&q_{\parallel}^{e(3)}={\tilde{\alpha}}_{\parallel}(g_{\parallel}^{(1)}-{\bar g}^{e(1)}_{\parallel})+{\tilde{\kappa}}^e_{\parallel}(g_{\parallel}^{e(3)}+{\bar g}^{e(3)}_{\parallel})
\end{eqnarray}
\noindent for the parallel electron fluxes, and 
\begin{equation}\label{linear2}
q_{\parallel}^{i(3)}={\tilde{\kappa}}^i_{\parallel}(g_{\parallel}^{i(3)}+{\bar g}^{i(3)}_{\parallel})
\end{equation}
\noindent for the parallel ion fluxes. Coefficients ${\tilde{\sigma}}_{\parallel}$, ${\tilde{\alpha}}_{\parallel}$, ${\tilde{\kappa}}^{\iota}_{\parallel}$ are the dimensionless parallel component of the {\it electronic conductivity}, the {\it thermoelectric coefficient} and the {\it electric} ($\iota =e$) or {\it ion} ($\iota =i$) {\it thermal conductivity}, respectively. The relations between the dimensional and the dimensionless transport coefficients are
\begin{eqnarray}\label{linear6}
&&\sigma_{n}=\frac{e^2n_e}{m_e}\tau_e{\tilde\sigma}_{n}\nonumber\\
&&\alpha_{n}=\sqrt{\frac{5}{2}}\frac{en_e}{m_e}\tau_e{\tilde\alpha}_{n}\\
&&\kappa^{\alpha}_{n}=\frac{5}{2}\frac{n_{\alpha}T_{\alpha}}{m_{\alpha}}\tau_{\alpha}{\tilde\kappa}^{\alpha}_{n}\nonumber
\end{eqnarray}
\noindent The parallel transport coefficients define a definite positive matrix \cite{balescu2}
\begin{eqnarray}\label{linear7}
&&\!\!\!\!\!\!\!\!\!\!\!\!\!\!\!\!\!\!\!\!{\tilde\sigma}_{\parallel},\ {\tilde\kappa}^e_{\parallel},\ {\tilde\kappa}^i_{\parallel} >0\nonumber\\
&&\!\!\!\!\!\!\!\!\!\!\!\!\!\!\!\!\!\!\!\!{\tilde\sigma}_{\parallel}{\tilde\kappa}^e_{\parallel}-{\tilde\alpha}^2_{\parallel}\ >0
\end{eqnarray}
\noindent Eq.(\ref{linear1}) must be supplemented by the {\it solubility conditions} expressing the mechanical equilibrium of the plasma:
\begin{equation}\label{linear3}
 g_{\parallel}^{e(1)}+{\hat a} g_{\parallel}^{i(1)}+{\bar g}^{e(1)}_{\parallel}+{\hat a}{\bar g}^{i(1)}_{\parallel}=0
\end{equation}
\noindent Up to the order $\epsilon^2$, the classical electron and ion radial Hermitian moments are linked to the thermodynamic forces by the following relations
\begin{eqnarray}\label{linear5}
&&{\hat q}_{\rho cl}^{e(1)}={\tilde{\sigma}}_{\perp}g_{\rho}^{(1)P}-{\tilde{\alpha}}_{\perp}g_{\rho}^{e(3)}+O(\epsilon^2)\nonumber\\
&&{\hat q}_{\rho cl}^{e(3)}=-{\tilde{\alpha}}_{\perp}g_{\rho}^{(1)P}+{\tilde{\kappa}}^e_{\perp}g_{\rho}^{e(3)}+O(\epsilon^2)\\
&&{\hat q}_{\rho cl}^{i(3)}={\tilde{\kappa}}^i_{\perp}g_{\rho}^{i(3)}+O(\epsilon^2)\nonumber
\end{eqnarray}
\noindent where ${\tilde{\sigma}}_{\perp}$, ${\tilde{\alpha}}_{\perp}$ and ${\tilde{\kappa}}^{\alpha}_{\perp}$ are the dimensionless perpendicular component of the transport coefficients. These coefficients satisfy the relations
\begin{eqnarray}\label{linear7}
&&{\tilde\sigma}_{\perp},\ {\tilde\kappa}^e_{\perp},\ {\tilde\kappa}^i_{\perp}\ >0\nonumber\\
&&{\tilde\sigma}_{\perp}{\tilde\kappa}^e_{\perp}-{\tilde\alpha}^2_{\perp}\ >0
\end{eqnarray}
\noindent The transport equations are derived by taking into account the definition of the fluxes in the different regimes. In the linear regime, the classical transport equations are immediately provided by Eq.(\ref{linear5}):
\begin{eqnarray}\label{linear8}
&&<{\hat q}_{\rho cl}^{e(1)}>=<{\tilde{\sigma}}_{\perp}>g_{\rho}^{(1)P}-<{\tilde{\alpha}}_{\perp}>g_{\rho}^{e(3)}\nonumber\\
&&<{\hat q}_{\rho cl}^{e(3)}>=-<{\tilde{\alpha}}_{\perp}>g_{\rho}^{(1)P}+<{\tilde{\kappa}}^e_{\perp}>g_{\rho}^{e(3)}\\
&&<{\hat q}_{\rho cl}^{i(3)}>=<{\tilde{\kappa}}^i_{\perp}>g_{\rho}^{i(3)}\nonumber
\end{eqnarray}
\noindent Notice that the classical dimensionless radial fluxes can also be obtained by the following definition (see, for example refs \cite{balescu2} and \cite{nishikawa})
\begin{equation}\label{linear9}
<{\hat q}_{\rho cl}^{\alpha (n)}>=-<\frac{1}{\Omega_{\alpha}}G_{\wedge}^{\alpha (n)}>
\end{equation}
\noindent The dimensionless radial Pfirsch-Schl{\"{u}}ter fluxes are defined as \cite{pfirsch}:
\begin{equation}\label{linear10}
<{q}_{\rho}^{\alpha (n)}>_{ps}=-K_{\alpha}\tau_{\alpha}<\frac{\beta_0}{B}\Bigl(1-\frac{B^2}{\beta_0^2}\Bigr)G_{\parallel}^{\alpha (n)}>
\end{equation}
\noindent From Eqs (\ref{entropy3}), (\ref{linear1}) and (\ref{linear10}), we easily find
\begin{eqnarray}\label{linear11}
&&<{q}_{\rho}^{e(1)}>_{ps}=K_e^2(g-1)(c_{11}^eg_{\rho}^{(1)P}-c_{13}^eg_{\rho}^{e(3)})\nonumber\\
&&<{q}_{\rho}^{e(3)}>_{ps}=K_e^2(g-1)(-c_{13}^eg_{\rho}^{(1)P}+c_{33}^eg_{\rho}^{e(3)})\\
&&<{q}_{\rho}^{i(3)}>_{ps}=K_i^2(g-1)c_{33}^ig_{\rho}^{i(3)}\nonumber
\end{eqnarray}
\noindent where factor $g$ is defined as
\begin{equation}\label{linear11a}
g\equiv\beta_0^2<\frac{1}{B^2}>
\end{equation}
\noindent and $c_{11}$, $c_{13}$, $c_{33}$ are the electron collision matrix elements
\begin{equation}\label{linear11b}
c_{11}^e=\frac{{\tilde{\kappa}}^e_{\parallel}}{{\tilde\sigma}_{\parallel}{\tilde{\kappa}}^e_{\parallel}-{\tilde\alpha}_{\parallel}^2}\quad c_{13}^e=-
\frac{{\tilde{\alpha}}_{\parallel}}{{\tilde\sigma}_{\parallel}{\tilde{\kappa}}^e_{\parallel}-{\tilde\alpha}^2_{\parallel}}\quad c_{33}^e=
\frac{{\tilde{\sigma}}_{\parallel}}{{\tilde\sigma}_{\parallel}{\tilde{\kappa}}^e_{\parallel}-{\tilde\alpha}^2_{\parallel}}
\end{equation}
\noindent From now on, we shall adapt a more compact notation. We shall label with $X_{A}^{\mu}$ ($\mu=1,2$) the electron thermodynamic forces and with $Y_{A}$ the ion thermodynamic forces. Symbol $A$ distinguishes the different regimes $A=(cl,\ ps)$. The conjugate electron and ion thermodynamic fluxes will be denoted with symbols $J_{\mu A}^{(e)}$ and $J_A^{(i)}$, respectively. We have
\begin{eqnarray}\label{linear19}
&&\!\!\!\!\!\!\!\!\!\!\!\!\!\!\!\!\!\!\!\!\!\!\!\!\!\!\!\!X_{cl}^{\mu}=
\begin{pmatrix}g_{\rho}^{(1)P}\\ g_{\rho}^{e(3)}
\end{pmatrix}\qquad\qquad\qquad\qquad\quad\qquad Y_{cl}=g_{\rho}^{i(3)}\\
&&\!\!\!\!\!\!\!\!\!\!\!\!\!\!\!\!\!\!\!\!\!\!\!\!\!\!\!\!X_{ps}^{\mu}=\begin{pmatrix}g_{\parallel}^{(1)}-{\bar g}^{e(1)}_{\parallel}\\ g_{\parallel}^{e(3)}+{\bar g}^{e(3)}_{\parallel}
\end{pmatrix}\qquad\qquad\qquad\qquad
Y_{ps}=g_{\parallel}^{i(3)}+{\bar g}^{i(3)}_{\parallel}\\
&&\!\!\!\!\!\!\!\!\!\!\!\!\!\!\!\!\!\!\!\!\!\!\!\!\!\!\!\!J_{\mu cl}^{(e)}=
\begin{pmatrix}{\hat q}_{\rho cl}^{e(1)}\\{\hat q}_{\rho cl}^{e(3)}
\end{pmatrix}\qquad\qquad\qquad\qquad\qquad\quad\  \ \!\!\!J_{cl}^{(i)}={\hat q}_{\rho cl}^{i(3)}\\
&&\!\!\!\!\!\!\!\!\!\!\!\!\!\!\!\!\!\!\!\!\!\!\!\!\!\!\!\!J_{\mu ps}^{(e)}=
\Bigl(\frac{B}{\beta_0}-\frac{\beta_0}{B}\Bigr)K_e
\begin{pmatrix}g_{\rho}^{(1)P}\\ -g_{\rho}^{e(3)}
\end{pmatrix}
\qquad\quad\  \ \ \!J_{ps}^{(i)}=-\Bigl(\frac{B}{\beta_0}-\frac{\beta_0}{B}\Bigr)K_i g_{\rho}^{i(3)}
\end{eqnarray}
\noindent and the Onsager matrixes read
\begin{eqnarray}\label{linear21}
&&\!\!\!\!\!\!\!\!\!\!\!\!\!\!\!\!\!\!\!\!\!\!\!\!\!\!\!\!\!\!\!\!\!\!\!\!\!\!\!\!\!\!\!\!\!\!\!\!\!\!\!\!\!\!\!\!\!\!\!\!\!\!\!\!\!\!\!
L_{(e)cl \mu\nu}=
\begin{pmatrix}
{\tilde\sigma}_{\perp} & -{\tilde\alpha}_{\perp}\\
-{\tilde\alpha}_{\perp} & {\tilde\kappa}^e_{\perp}
\end{pmatrix}\qquad\qquad\quad\quad L_{(i) cl}={\tilde\kappa}^i_{\perp}\\
&&\!\!\!\!\!\!\!\!\!\!\!\!\!\!\!\!\!\!\!\!\!\!\!\!\!\!\!\!\!\!\!\!\!\!\!\!\!\!\!\!\!\!\!\!\!\!\!\!\!\!\!\!\!\!\!\!\!\!\!\!\!\!\!\!\!\!\!
L_{(e)ps \mu\nu}=
\begin{pmatrix}
{\tilde\sigma}_{\parallel}& {\tilde\alpha}_{\parallel}\\
{\tilde\alpha}_{\parallel} & {\tilde\kappa}^e_{\parallel}
\end{pmatrix}\qquad\qquad\qquad\quad\quad \! L_{(i) ps}={\tilde\kappa}^i_{\parallel}
\end{eqnarray}
\noindent Our objective is to evaluate the nonlinear terms that should be added to the expressions (\ref{linear1}), (\ref{linear2}) and (\ref{linear5}) when the plasma is far from equilibrium. These nonlinear corrections are provided by the solutions of the thermodynamic field equations. With these solutions, through the expressions (\ref{linear9}) and  (\ref{linear10}), we shall provide the nonlinear transport equations in the different regimes. The definitions (\ref{linear9}) and (\ref{linear10}) for radial fluxes in the different regimes assumes {\it ambipolar diffusion} i.e., the electric current circulates only on the magnetic surfaces: $<\Gamma_{\rho}^e>=Z<\Gamma_{\rho}^i>$. From the neoclassical theory we find that not only the total average fluxes, but also the separate classical, banana and Pfirsch-Schl{\"{u}}ter fluxes are ambipolar. The nonlinear corrections are determined by combining the solutions of the thermodynamic field equations with the expressions (\ref{linear9}) and (\ref{linear10}) and, therefore, the classical and Pfirsch-Schl{\"{u}}ter fluxes (as well as the banana flux) remain ambipolar also in the nonlinear regime. 

\section{The Thermodynamic Field Equations for Magnetically Confined Plasmas}\label{tft}

it is known that the validity of the Onsager reciprocity relations (\ref{linear1}), (\ref{linear2}) and (\ref{linear5}) is limited to the so called {\it linear region}. As already mentioned in the introduction, a covariant thermodynamic field theory (TFT) has been proposed in 1999 in order to evaluate how the relations flux-forces {\it deform} when the thermodynamic system is far from the linear ("Onsager") region. Let us briefly summarize the main aspects of this theory when the transport coefficients are fully symmetric. 

\noindent Clearly, one cannot develop a transport theory without a knowledge of microscopic dynamical laws. Transport theory is only but an aspect of non-equilibrium statistical mechanics, which provides the links between micro and macro-levels. This link appears indirectly in the "unperturbed" metric, i.e., the linear classical and neoclassical transport coefficients used as an input in the equations: these coefficients have to be calculated in the usual way by the methods of kinetic theory. In the TFT description, a thermodynamic configuration corresponds to a point in the thermodynamic space $Ts$, defined as an $n$-dimensional manifold covered by the $n$ independent thermodynamic forces $X^\mu$. The elements of the metric tensor $g_{\mu\nu}$ are the symmetric components of the transport coefficients $\tau_{\mu\nu}$. The conjugate flows, $J_{\mu}$, are dual to the thermodynamic forces through the relation $J_{\mu}=\tau_{\mu\nu}X^{\nu}$. In this equation, as in the remainder of this paper, the Einstein summation convention on the repeated indexes is adopted. The positive definiteness of the matrix $g_{\mu\nu}$ ensures the validity of the second principle of thermodynamics \cite{sonnino}. The theory is constructed with the following requirements (\cite{sonnino}, \cite{sonninor}):

\begin{itemize}
\item {\it The theorems valid when a generic thermodynamic system is out from equilibrium should be satisfied};
\item {\it The thermodynamic field equations should be invariant under the thermodynamic coordinate transformations (TCT)}.
\end{itemize}
\noindent With the elements of the transport coefficients we construct two objects: {\it operators}, which may act on thermodynamic tensorial objects and {\it thermodynamic tensorial objects}, which under coordinate (forces) transformations, obey well specified transformation rules.
\vskip 0.2truecm
\noindent{\bf Operators }
\vskip 0.2truecm
We introduce two operators, the {\it entropy production operator} $\sigma$ and the {\it dissipative quantity operator} $P(X)$  considering the transport coefficients as elements of $n$ x $n$ matrixes $\tau$ and $g$, which multiply the thermodynamic forces $X$ expressed as $n$ x $1$ column matrices:
\begin{eqnarray}\label{pts2a}
&&\sigma (X):\rightarrow\sigma(X)\equiv X gX^T\nonumber\\
&& P(X):\rightarrow P(X)\equiv X\tau {\dot X}^T
\end{eqnarray}
\noindent [The dot symbol stands for derivative with respect to parameter $\eta$, defined in Eq.~(\ref{pts8a})]. Thermodynamic states $X_{s}$ such that
\begin{eqnarray}\label{pts2b}
P(X_{s})=0
\end{eqnarray}
\noindent are referred to as {\it steady-states}. These are physical quantities and should remain invariant under thermodynamic coordinate transformations. Eqs~(\ref{pts2a}) {\it should not} be interpreted as the metric tensor $g_{\mu\nu}$, which acts on the coordinates. The metric tensor {\it acts only on} elements of the tangent space (like $dX^\mu$) or on the thermodynamic tensorial objects.
\vskip 0.2truecm
\noindent{\bf Transformation Rules of Entropy Production, Forces, Flows and Thermodynamic Tensorial Objects}
\vskip 0.2truecm
According to Prigogine's statement \cite{prigogine}, {\it thermodynamic systems are thermodynamically equivalent if, under transformation of fluxes and forces, the bilinear form of the entropy production $\sigma$ remains unaltered}. In mathematical terms, this implies:
 \begin{equation}\label{pts3}
 \sigma=J_\mu X^\mu=J'_\mu X'^\mu
 \end{equation}
 \noindent This condition requires that the transformed thermodynamic forces and flows satisfy the relation
\begin{equation}\label{pts8}
X'^\mu=\frac{\partial X'^{\mu}}{\partial X^\nu} X^\nu\qquad\quad J'_\mu=\frac{\partial X^{\nu}}{\partial X'^\mu}J_\nu
\end{equation}
\noindent These transformations are referred to as {\it Thermodynamic Coordinate Transformations} (TCT).  

\noindent By direct inspection, it is easy to verify that the general solutions of equations (\ref{pts8}) are 
\begin{equation}\label{pts14a}
X'^\mu=X^1F^\mu\Bigl(\frac{X^2}{X^1},\ \frac{X^3}{X^2},\ \cdots\ \frac{X^n}{X^{n-1}}\Bigr)
\end{equation}
\noindent where $F^\mu$ is an {\it arbitrary function} of variables $X^j/X^{j-1}$ with ($j=2,\dots, n$). We immediately note that $dX^\mu$ and $\partial/\partial X^{\mu}$ transform like a contra-variant and a covariant vector, respectively. Vectors $dX^\mu$ define then the {\it tangent space} to $Ts$. It also follows that the operator $P(X)$, i.e. the dissipation quantity, and in particular the definition of steady-states, are invariant under TCT. Moreover, it is easy to check that the metric tensor $g_{\mu\nu}$ transforms like a {\it thermodynamic tensor of second rank}. In particular, parameter $\eta$, defined as 
\begin{equation}\label{pts8a}
d\eta^2=g_{\mu\nu}dX^\mu dX^\nu
\end{equation}
\noindent is a scalar under TCT.

\noindent Operator $\mathcal{O}$
\begin{equation}\label{pts14d}
\mathcal{O}\equiv X^\mu\frac{\partial}{\partial X^\mu}=X'^\mu\frac{\partial}{\partial X'^\mu}
\end{equation}
\noindent is also invariant under TCT. This operator plays an important role in the formalism. 
\vskip 0.2truecm
\noindent{\bf The Principle of Least Action}
\vskip 0.2truecm
We introduce now the following postulate:

\noindent {\it There exists a thermodynamic action $I$, scalar under $TCT$, which is stationary with respect to arbitrary variations in $g_{\mu\nu}$}. 

\noindent We want to construct an action, scalar under $TCT$, from the metric tensor and its first and second derivatives, that has linear second derivatives. It is easily checked that the only action with these requirements is (\cite{sonnino} and \cite{sonninor})
\begin{equation}\label{pts14g}
I=\int\Bigl( R_{\mu\nu}g^{\mu\nu}+\lambda_\sigma X^\mu X^\nu\frac{\partial g_{\lambda\kappa}}{\partial X^\mu}\frac{\partial g^{\lambda\kappa}}{\partial X^\nu}\Bigr)dV=
\int\Bigl(R+\lambda_\sigma\mathcal{O}(g_{\lambda\kappa})\mathcal{O}(g^{\lambda\kappa})\Bigr)dV
\end{equation}
\noindent where
\begin{eqnarray}\label{pts15}
&& \lambda_\sigma={\rm constant\ having\ dimension\ the\ inverse\ of\ the\ entropy\ production}\nonumber\\
&&R_{\mu\nu}\equiv\Gamma^\lambda_{\mu\lambda,\nu}-\Gamma^\lambda_{\mu\nu,\lambda}+
\Gamma_{\mu\lambda}^\eta\Gamma_{\eta\nu}^{\lambda}-\Gamma_{\mu\nu}^\eta\Gamma_{\eta\lambda}^\lambda\qquad {\rm (The\ Ricci\ Tensor)}\nonumber\\
&&\Gamma_{\mu\nu}^\lambda = {\rm affine\ connection}\nonumber\\
&&dV={\rm infinitesimal\ volume\ element\ of}\ Ts
\end{eqnarray}
\noindent To avoid misunderstanding, while it is correct to mention that this postulate affirms the possibility of deriving the thermodynamic field equations by a variational principle it does not state that the solution of the thermodynamic field equations can also be derived by a variational principle. In particular the well-known {\it Universal Criterion of Evolution} established by Glansdorff-Prigogine {\it cannot} be derived by a variational principle.
\pagebreak
\vskip 0.2truecm
\noindent{\bf The Thermodynamic Field Equations}
\vskip 0.2truecm
We immagine $g_{\mu\nu}$ to be subject to an infinitesimal variation $g_{\mu\nu}\rightarrow g_{\mu\nu}+\delta g_{\mu\nu}$ where $\delta g_{\mu\nu}$ is arbitrary, except that it is required to vanish as $\mid X^\mu\mid\rightarrow\infty$. By imposing that the action (\ref{pts14g}) is stationary with respect to arbitrary variations in $g_{\mu\nu}$, we find
\begin{equation}\label{pts16a}
R^{\mu\nu}+\lambda_\sigma S^{\mu\nu}=0
\end{equation}
\noindent where
\begin{eqnarray}\label{pts16b}
&&S^{\mu\nu}=-2\mathcal{O}^2(g^{\mu\nu})+2g_{\lambda\kappa}\mathcal{O}(g^{\mu\kappa})\mathcal{O}(g^{\nu\lambda})+\frac{1}{2}g^{\mu\nu}\mathcal{O}(g_{\lambda\kappa})\mathcal{O}(g^{\lambda\kappa})\nonumber\\
&&\qquad\quad-[\mathcal{O}(\ln g)+4]\mathcal{O}(g^{\mu\nu})
\end{eqnarray}
\noindent Close to the Onsager region, having metric $L_{\mu\nu}$, we can write
\begin{eqnarray}\label{pts17}
&&g_{\mu\nu}=L_{\mu\nu}+h_{\mu\nu}+O(\epsilon^2)\nonumber\\
&&\lambda_\sigma=O(\epsilon)\qquad {\rm with}\quad \epsilon=Max\Big\{\frac{Eigenvalues[g_{\mu\nu}-L_{\mu\nu}]}{Eigenvalues [L_{\mu\nu}]}\Big\}\ll 1
\end{eqnarray}
\noindent and therefore, $h_{\mu\nu}$ are small variations with respect to Onsager's coefficients satisfying the equations \cite{sonnino}
\begin{equation}\label{pts18}
L^{\lambda\kappa}\frac{\partial^2 h_{\mu\nu}}{\partial X^{\lambda}\partial X^{\kappa}}+
L^{\lambda\kappa}\frac{\partial^2 h_{\lambda\kappa}}{\partial X^{\mu}\partial X^{\nu}}-
L^{\lambda\kappa}\frac{\partial^2 h_{\lambda\nu}}{\partial X^{\kappa}\partial X^{\mu}}-
L^{\lambda\kappa}\frac{\partial^2 h_{\lambda\mu}}{\partial X^{\kappa}\partial X^{\nu}}=0
\end{equation}
\noindent Eqs~(\ref{pts18}) should be solved with the appropriate gauge-choice and boundary conditions.

\noindent Looking at the expression of the entropy production for electrons and ions, Eq.(\ref{entropy10}), we immediately realize that the thermodynamic space of a fully ionized plasma possesses $11$-dimensions. Notice that in this case, the coefficients of the reciprocity relations are fully symmetric and, therefore, the metric tensor coefficients $g_{\mu\nu}$ coincide with the transport coefficients of the plasma. Luckily, the thermodynamic space results to be the direct sum of six independent subspaces: three subspaces for electrons and three subspaces for ions. The electron subspaces have dimensions $2$ for the classical and Pfirsch-Schl{\"{u}}ter transport and $3$ for the plateau/banana regimes. The ionic subspaces have dimensions $1$ for the classical and the Pfirsch-Schl{\"{u}}ter transport and $2$ for the banana regime. Below, we report the schematic decomposition of the thermodynamic space:
\begin{eqnarray}\label{plasma1}
&& Ts^{11}=Ts^{7}_{(e)}\oplus Ts^{4}_{(i)}\qquad\qquad\qquad\quad\mathrm{where}\nonumber\\
&& Ts_{(e)}^{7}=Ts^{2}_{cl(e)}\oplus Ts^{2}_{ps(e)}\oplus Ts^{3}_{b/p(e)}\nonumber\\
&&Ts_{(i)}^{4}=Ts^{1}_{cl(i)}\oplus Ts^{1}_{ps(i)}\oplus Ts^{2}_{b/p(i)}
\end{eqnarray}
\noindent In sections (\ref{entropy}) and (\ref{linear}) we have found the expressions for the entropy of a fully ionized plasma and the linear relations between the thermodynamic forces and the conjugate fluxes. 

\noindent Considering Eq.(\ref{entropy10}), the electron and ion weak-field equations read
\begin{eqnarray}\label{plasma9}
&&\!\!\!\!\!\!\!\!\!\!\!L_{(e)A}^{\lambda\kappa}\frac{\partial^2 h_{\mu\nu}^{(e)A}}{\partial X_A^{\lambda}\partial X_A^{\kappa}}\!+\!
L_{(e)A}^{\lambda\kappa}\frac{\partial^2 h_{\lambda\kappa}^{(e)A}}{\partial X_A^{\mu}\partial X_A^{\nu}}\!-\!
L_{(e)A}^{\lambda\kappa}\frac{\partial^2 h_{\lambda\nu}^{(e)A}}{\partial X_A^{\kappa}\partial X_A^{\mu}}\!-\!
L_{(e)A}^{\lambda\kappa}\frac{\partial^2 h_{\lambda\mu}^{(e)A}}{\partial X_A^{\kappa}\partial X_A^{\nu}}\!=0\nonumber\\
&&\!\!\!\!\!\!\!\!\!\!\!L_{(i)A}^{\lambda\kappa}\frac{\partial^2 h_{\mu\nu}^{(i)A}}{\partial Y_A^{\lambda}\partial Y_A^{\kappa}}+
L_{(i)A}^{\lambda\kappa}\frac{\partial^2 h_{\lambda\kappa}^{(i)A}}{\partial Y_A^{\mu}\partial Y_A^{\nu}}-
L_{(i)A}^{\lambda\kappa}\frac{\partial^2 h_{\lambda\nu}^{(i)A}}{\partial Y_A^{\kappa}\partial Y_A^{\mu}}-
L_{(i)A}^{\lambda\kappa}\frac{\partial^2 h_{\lambda\mu}^{(i)A}}{\partial Y_A^{\kappa}\partial Y_A^{\nu}}=0\nonumber\\
&&
\end{eqnarray}
\noindent where $A=(cl,\ ps,\ p,\ b)$, corresponds to classical ($cl$), Pfirsch-Schl{\"{u}}ter ($ps$), plateau ($p$) and banana ($b$) regimes. To solve Eqs (\ref{plasma9}), we have to chose the particular system where the calculations [i.e., the {\it gauge choice}, see appendix (\ref{gauge})] will be performed and then to solve the resulting P.D.E. with the appropriate boundary conditions. These tasks will be accomplished in the next sections. Notice that, it follows from the second principle of thermodynamics that these P.D.E. are of {\it elliptic type}.

\section{The Electron and Ion Gauge-invariant Solutions for the Collisional Regimes}\label{ps}

To solve Eqs~(\ref{plasma9}) it is convenient to work in the harmonic coordinates. We label $X_A^{\mu}$ ($\mu = 1,2$ and $A=cl,ps$) the electron thermodynamic forces and with $h_{\mu\nu}^{(e)cl}(X_{cl})$ and $h_{\mu\nu}^{(e)ps}(X_{ps})$ the small field perturbations $h_{\mu\nu}$  when we refer to the classical and Pfirsch-Schl{\"{u}}ter electron regimes, respectively. In harmonic coordinates these fields satisfy the following equations
\begin{eqnarray}\label{ps1}
L_{(e)cl}^{\mu\lambda}\frac{{\partial}h_{\lambda\nu}^{(e)cl}}{\partial X_{cl}^{\mu}}&=&
\frac{1}{2}L_{(e)cl}^{\mu\lambda}\frac{{\partial}h_{\mu\lambda}^{(e)cl}}{\partial X_{cl}^{\nu}}\nonumber\\
L_{(e)ps}^{\mu\lambda}\frac{{\partial}h_{\lambda\nu}^{(e)ps}}{\partial X_{ps}^{\mu}}&=&
\frac{1}{2}L_{(e)ps}^{\mu\lambda}\frac{{\partial}h_{\mu\lambda}^{(e)ps}}{\partial X_{ps}^{\nu}}
\end{eqnarray}
\noindent where  $L_{(e)cl}^{\mu\lambda}$ and $L_{(e)ps}^{\mu\lambda}$ indicate the corresponding electron Onsagers's matrices in the classical and Pfirsch-Schl{\"{u}}ter regimes, respectively. In the next section is shown that in the weak field approximation, the ion classical and  Pfirsch-Schl{\"{u}}ter regimes remain unperturbed (i.e., $h_{\mu\nu}^{(i)A}(Y_{A})=0$). If $h_{\mu\nu}^{(e)A}(X_{A})$ are cast into the form $h_{\mu\nu}^{(e)A}(X_{A})=c^{(e)A}_{\mu\nu}f^{(e)A}(X_{A})$ with $c^{(e)A}_{\mu\nu}$ constant and $f^{(e)A}(X_A)$ functions depending on the electron thermodynamic forces, Eqs (\ref{ps1}) give
\begin{eqnarray}\label{ps2}
&&\biggr (L_{(e)cl}^{\mu\lambda}c^{(e)cl}_{\lambda\nu}-\frac{1}{2}L_{(e)cl}^{\lambda\kappa}c^{(e)cl}_{\lambda\kappa}{\delta}_{\nu}^{\mu}\biggr )\frac{{\partial}f(X_{cl})}{\partial X_{cl}^{\mu}}=0\nonumber\\
&&\biggr (L_{(e)ps}^{\mu\lambda}c^{(e)ps}_{\lambda\nu}-\frac{1}{2}L_{(e)ps}^{\lambda\kappa}c^{(e)ps}_{\lambda\kappa}{\delta}_{\nu}^{\mu}\biggr )\frac{{\partial}f(X_{ps})}{\partial X_{ps}^{\mu}}=0
\end{eqnarray}
\noindent which are verified for {\it any} value of $\frac{{\partial}f(X_A)}{\partial X_{A}^{\mu}}$ only if 
\begin{eqnarray}\label{ps3}
c^{(e)cl}_{\mu\nu}&=&\chi_{cl} L_{(e)cl\mu\nu}\nonumber\\
c^{(e)ps}_{\mu\nu}&=&\chi_{ps} L_{(e)ps\mu\nu}
\end{eqnarray}
\noindent whit $\chi_{cl}$ and $\chi_{ps}$ being parameters independent of the thermodynamic forces.

\section{The Nonlinear  Classical and Pfirsch - Schl{\"{u}}ter Transport Regimes}\label{NLps}
In the nonlinear region, the relations between the dimensionless Hermitian moments and the dimensionless thermodynamic forces can be cast into the form:
\begin{eqnarray}\label{nonlinearcl}
&&\!\!\!\!\!\!\!\!\!\!\!\!\!\!\!\!\!\!\!\!{\hat q}_{\rho cl}^{e(1)}=(1+\chi_{cl}f^{(e)cl}(X_{cl}))\Bigl({\tilde{\sigma}}_{\perp}g_{\rho}^{(1)P}-{\tilde{\alpha}}_{\perp}g_{\rho}^{e(3)}\Bigr)\nonumber\\
&&\!\!\!\!\!\!\!\!\!\!\!\!\!\!\!\!\!\!\!\!{\hat q}_{\rho cl}^{e(3)}=(1+\chi_{cl}f^{(e)cl}(X_{cl}))\Bigl(-{\tilde{\alpha}}_{\perp}g_{\rho}^{(1)P}+{\tilde{\kappa}}^e_{\perp}g_{\rho}^{e(3)}
\Bigr)\\
&&\!\!\!\!\!\!\!\!\!\!\!\!\!\!\!\!\!\!\!\!{\hat q}_{\rho cl}^{i(3)}=({\tilde{\kappa}}^i_{\perp}+h_{33}^{(i)cl})g_{\rho}^{i(3)}\nonumber
\end{eqnarray}
\noindent for the classical regime, and
\begin{eqnarray}\label{nonlinearps}
&&\!\!\!\!\!\!\!\!\!\!\!\!\!\!\!\!\!\!\!\!
q_{\parallel}^{(1)}=(1+\chi_{ps}f^{(e)ps}(X_{ps}))\Bigl({\tilde{\sigma}}_{\parallel}(g_{\parallel}^{(1)}-{\bar g}^{e(1)}_{\parallel})+{\tilde{\alpha}}_{\parallel}(g_{\parallel}^{e(3)}+{\bar g}^{e(3)}_{\parallel})\Bigr)\nonumber\\
&&\!\!\!\!\!\!\!\!\!\!\!\!\!\!\!\!\!\!\!\!
q_{\parallel}^{e(3)}=(1+\chi_{ps}f^{(e)ps}(X_{ps}))\Bigl({\tilde{\alpha}}_{\parallel}(g_{\parallel}^{(1)}-{\bar g}^{e(1)}_{\parallel})+{\tilde{\kappa}}^e_{\parallel}(g_{\parallel}^{e(3)}+{\bar g}^{e(3)}_{\parallel})\Bigr)\\
&&\!\!\!\!\!\!\!\!\!\!\!\!\!\!\!\!\!\!\!\!
q_{\parallel}^{i(3)}=({\tilde{\kappa}}^i_{\parallel}+h_{33}^{(i)ps})(g_{\parallel}^{i(3)}+{\bar g}^{i(3)}_{\parallel})\nonumber 
\end{eqnarray}
\noindent for the Pfirsch - Schl{\"{u}}ter regime. In these equations, we have taken into account that the perturbations $h_{\mu\nu}^{(e)A}$ are of the form 
\begin{equation}\label{nonlinear}
h_{\mu\nu}^{(e)A}(X_A)=\chi_{A}L_{(e)A\mu\nu}f^{(e)A}(X_A)\qquad({\rm with}\quad A=cl,ps)
\end{equation}
\noindent Let us now determine $h_{\mu\nu}^{(\alpha)A}$ ($\alpha =i,e$) by solving the field equations. In harmonic coordinates, the ion field equations for the classical and Pfirsch-Schl{\"{u}}ter regimes take the form
\begin{eqnarray}\label{nonlinearps1}
&&L_{(i)cl}^{-1}\frac{\partial^2h_{33}^{(i)cl}}{\partial Y_{cl}^2}=0\nonumber\\
&&L_{(i)ps}^{-1}\frac{\partial^2h_{33}^{(i)ps}}{\partial Y_{ps}^2}=0
\end{eqnarray}
\noindent The solutions of Eqs (\ref{nonlinearps1}) are
\begin{eqnarray}\label{nonlinearps2}
&&h_{33}^{(i)cl}=a_1Y_{cl}+b_1\nonumber\\
&&h_{33}^{(i)ps}=a_2Y_{ps}+b_2
\end{eqnarray}
\noindent where $a_n$ and $b_n$ are constant coefficients. The nonlinear solutions must coincide with the Onsager solutions when the system approaches equilibrium. This condition imposes that $b_1=b_2=0$. Moreover, solutions (\ref{nonlinearps2}) must verify the harmonic gauge conditions:
\begin{eqnarray}\label{nonlinearps3}
&&L_{(i)cl}^{-1}\frac{\partial h_{33}^{(i)cl}}{\partial Y_{cl}}=\frac{1}{2}L_{(i)cl}^{-1}\frac{\partial h_{33}^{(i)cl}}{\partial Y_{cl}}\nonumber\\
&&L_{(i)ps}^{-1}\frac{\partial h_{33}^{(i)ps}}{\partial Y_{ps}}=\frac{1}{2}L_{(i)ps}^{-1}\frac{\partial h_{33}^{(i)ps}}{\partial Y_{ps}}
\end{eqnarray}
\noindent Inserting solutions (\ref{nonlinearps2}) into Eqs (\ref{nonlinearps3}), we find $a_1=a_2=0$. Therefore, {\it the Thermodynamic Field Theory does not correct the expressions provided by the neoclassical transport theory in the classical and Pfirsch-Schl{\"{u}}ter transport}. Notice that we could have obtained the same results simply by observing that in the classical and Pfirsch-Schl{\"{u}}ter regimes the ion field equations [i.e., the second equation in Eqs~(\ref{plasma9})] vanish identically.

The electric field equations are P.D.E. of elliptic type,
\begin{eqnarray}\label{nonlinearps4}
&&L_{(e)cl}^{\mu\nu}\frac{\partial h_{\mu\nu}^{(e)cl}}{\partial X_{cl}^{\mu}\partial X_{cl}^{\nu}}=0\nonumber\\
&&L_{(e)ps}^{\mu\nu}\frac{\partial h_{\mu\nu}^{(e)ps}}{\partial X_{ps}^{\mu}\partial X_{ps}^{\nu}}=0
\end{eqnarray}
\noindent These equations can be reduced to Laplacian equations
\begin{eqnarray}\label{nonlinearps5}
&&{\Delta}^{2} h_{\mu\nu}^{(e)cl} =0\nonumber\\
&&{\Delta}^{2} h_{\mu\nu}^{(e)ps} =0
\end{eqnarray}
\noindent by performing the following coordinate transformations
\begin{eqnarray}\label{nonlinearps5a}
&&\begin{pmatrix} 
X_{cl}^{1} \\ X_{cl}^{2}
\end{pmatrix}\longrightarrow\nonumber
 \begin{pmatrix}
X1_{cl} \\ X2_{cl}
\end{pmatrix}\\
&&\begin{pmatrix} X_{ps}^{1} \\ X_{ps}^{2}
\end{pmatrix}\longrightarrow 
\begin{pmatrix} 
X1_{ps} \\ X2_{ps}
\end{pmatrix}
\end{eqnarray}
\noindent The "scaled" variables $X1,\!2_{cl}$ and $X\!1,\!2_{ps}$ are defined as follows
\begin{eqnarray}\label{nonlinearps6}
&&\begin{pmatrix} \frac{X1_{cl}}{\lambda_{+ cl}} \\ \frac{X2_{cl}}{\lambda_{- cl}}
\end{pmatrix}={\bf U}^T_{cl}\ 
\begin{pmatrix} X_{cl}^{1} \\ X_{cl}^{2}
\end{pmatrix}\nonumber\\
&&\begin{pmatrix} \frac{X1_{ps}}{\lambda_{+ ps}} \\ \frac{X2_{ps}}{\lambda_{- ps}}
\end{pmatrix}= {\bf U}^T_{ps}\ \begin{pmatrix} X_{ps}^{1} \\ X_{ps}^{2}
\end{pmatrix}
\end{eqnarray}
\noindent where $\lambda_{\pm cl}$ and ${\bf U}_{cl}$ are, respectively, the eigenvalues and the matrix formed by the eigenvectors ${\bf u}_{+cl}, {\bf u}_{-cl}$ (put in column) of the contravariant Onsager matrix $L_{(e)cl}^{\mu\nu}$ i.e.,
\begin{eqnarray}\label{nonlinearps7}
&&\!\!\!\!\!\!\!\!\!\!\!\!L_{(e)cl}^{\mu\nu}=\frac{1}{{\tilde\kappa}^e_{\perp}{\tilde\sigma}_{\perp}-{\tilde\alpha}^2_{\perp}}
\begin{pmatrix}{\tilde\kappa}^e_{\perp} & {\tilde\alpha}_{\perp}\\ {\tilde\alpha}_{\perp} & {\tilde\sigma}_{\perp}
\end{pmatrix}\nonumber\\
&&\!\!\!\!\!\!\!\!\!\!\!\!\lambda_{\pm  cl}=\frac{{\tilde\kappa}^e_{\perp}+{\tilde\sigma}_{\perp}\pm\sqrt{\Delta_{cl}}}{2({\tilde\kappa^e}_{\perp}{\tilde\sigma}_{\perp}-{\tilde\alpha}^2_{\perp})}\quad\quad\ \! {\rm with}\quad\Delta_{cl}\equiv ({\tilde\kappa}^e_{\perp}-{\tilde\sigma}_{\perp})^2+4{\tilde\alpha}^2_{\perp}\\
&&\!\!\!\!\!\!\!\!\!\!\!\!{\bf U}_{cl}=({\bf u}_{+cl}, {\bf u}_{-cl})\qquad\qquad\quad\mathrm{where}\nonumber\\
&&\!\!\!\!\!\!\!\!\!\!\!\!{\bf u}_{\pm cl}=
\frac{\sqrt{2}{\tilde\alpha}_{\perp}}{\sqrt{({\tilde\kappa}^e_{\perp}-{\tilde\sigma}_{\perp})^2+4{\tilde\alpha}^2_{\perp}\pm ({\tilde\kappa}^e_{\perp}-{\tilde\sigma}_{\perp})\sqrt{\Delta_{cl}}}}
\begin{pmatrix}\frac{{\tilde\kappa}^e_{\perp}-{\tilde\sigma}_{\perp}\pm\sqrt{\Delta_{cl}}}{2{\tilde\alpha}_{\perp}}  \\  1 
\end{pmatrix}\nonumber
\end{eqnarray}
\noindent The corresponding relations for the Pfirsch-Schl{\"{u}}ter regime are
\begin{eqnarray}\label{nonlinearps8}
&&\!\!\!\!\!\!\!\!\!\!\!\!L_{(e)ps}^{\mu\nu}=\frac{1}{{\tilde\sigma}_{\parallel}{\tilde\kappa^e}_{\parallel}-{{\tilde\alpha}_{\parallel}}^2}
\begin{pmatrix}{\tilde\kappa^e}_{\parallel}& -{\tilde\alpha}_{\parallel}\\ -{\tilde\alpha}_{\parallel}&{\tilde\sigma}_{\parallel}
\end{pmatrix}\nonumber\\
&&\!\!\!\!\!\!\!\!\!\!\!\!\lambda_{\pm ps}=\frac{{\tilde\sigma}_{\parallel}+{\tilde\kappa^e}_{\parallel}\pm\sqrt{\Delta_{ps}}}{2({\tilde\sigma}_{\parallel}{\tilde\kappa^e}_{\parallel}-{{\tilde\alpha}_{\parallel}}^2)}\quad\quad\ \! {\rm with}\quad\Delta_{ps}\equiv ({\tilde\sigma}_{\parallel}-{\tilde\kappa^e}_{\parallel})^2+4{{\tilde\alpha}_{\parallel}}^2\\
&&\!\!\!\!\!\!\!\!\!\!\!\!{\bf U}_{ps}=({\bf u}_{+ps}, {\bf u}_{-ps})\qquad\qquad\quad\! \!\mathrm{where}\nonumber\\
&&\!\!\!\!\!\!\!\!\!\!\!\!{\bf u}_{\pm ps}=\frac{\sqrt{2}{\tilde\alpha}_{\parallel}}{\sqrt{({\tilde\sigma}_{\parallel}-{\tilde\kappa^e}_{\parallel})^2+4{{\tilde\alpha}_{\parallel}}^2\pm ({\tilde\sigma}_{\parallel}-{\tilde\kappa^e}_{\parallel})\sqrt{\Delta_{ps}}}}
\begin{pmatrix}\frac{{\tilde\sigma}_{\parallel}-{\tilde\kappa^e}_{\parallel}\pm \sqrt{\Delta_{ps}}}{2{\tilde\alpha}_{\parallel}} \\  1
\end{pmatrix}\nonumber
\end{eqnarray}
\noindent The solutions of Eq.(\ref{nonlinearps5}) are submitted to the following conditions:
\begin{description}
\item[a)] The solutions vanish at the origin of the axes ; 
\item[b)] The solutions are constant, with value $\chi_{cl} L_{(e)cl\mu\nu}$ in the classical regime, and $\chi_{ps} L_{(e)ps\mu\nu}$ in the Pfirsch-Schl{\"{u}}ter regime, on the circumferences with radii equal to $R_{0 cl}$ and $R_{0 ps}$, respectively;
\item[c)] The solutions are invariant with respect to the permutation of the axes $X1_{cl}$ and $X2_{cl}$ for the first equation (\ref{nonlinearps5}) and of the axes $X1_{ps}$ and $X2_{ps}$ for the second equation (\ref{nonlinearps5});
\item[d)] Inside the circumferences, the solutions are of class $\mathit{C^{2}}$ (possibly, except in a set of measure zero where they should be at least of class $\mathit{C^{0}}$)
\item[e)] The profiles for the losses (matter and energy) are, at least, functions of class $\mathit{C^{0}}$.
\end{description}
\noindent The first condition is due to the fact that our solution must coincide with the Onsager matrix when the system approaches equilibrium. Point b) provides the boundary condition of Eqs~(\ref{nonlinearps5}) at thelimit of a bounded region. The existence of a bounded region, where we have to solve Eqs~(\ref{nonlinearps5}), is warranted by the fact that the amplitudes of the thermodynamic forces $X_A^\mu$ are, in reality, bounded (see figs~\ref{g1P}$\div$\ref{gi3} for a JET plasma). Therefore, $\exists M_A\in{\mathcal R^+}\ \mid\ M_A=max\{X_A^\mu,\ \mu=1,2\}<\infty$. Since the thermodynamic forces are bounded and independent, symmetry implies that they span a (finite) ellipse in the thermodynamic space. On the other hand, it is always possible to perform a coordinate transformation (\ref{nonlinearps5a}) such that the thermodynamic field equations take the symmetric form of Eqs~(\ref{nonlinearps5}) and the ellipse transforms onto a circle of radius $R_0$. After the coordinate transformation (\ref{nonlinearps5a}), the new thermodynamic forces become perfectly equivalent while at the boundary, one direction cannot be privileged with respect to another. This condition requires that the solution is constant at the boundary. The third condition ensures that the two axes are treated in an equivalent manner. The next to last condition requires that the only solutions, acceptable from the physical point of view, are functions continuous in {\it {all points}} inside the circle (i.e., excluding the boundary) and, which are solutions of Eqs~(\ref{nonlinearps5}) (possibly, except in a set of measure zero). Finally, condition e) requires that the losses profiles are continuous functions. 

\noindent The first task is to construct a well-posed Dirichlet problem from the above-mentioned conditions. This has been accomplished in appendix (\ref{Solutions}). In appendix (\ref{gauge}) it is shown that, up to the first order, i.e., in the limit of validity of the weak-field approximation, the boundary conditions derived by conditions a)-e), respect the principle of covariance. Successively, we have to find the solutions of the P.D.E. (\ref{nonlinearps5}) submitted to the Dirichlet problem. These are obtained in appendix (\ref{2-solution}) as
\begin{eqnarray}\label{nonlinearps9}
&&h_{\mu\nu}^{(e)cl}(X1_{cl}, X2_{cl}) =\chi_{cl} L_{(e)cl\mu\nu}f^{(e)cl}(X1_{cl}, X2_{cl})\\
&&h_{\mu\nu}^{(e)ps}(X1_{ps}, X2_{ps}) =- L_{(e)ps\mu\nu}f^{(e)ps}(X1_{ps}, X2_{ps})
\end{eqnarray}
\noindent with
\begin{eqnarray}\label{nonlinearps9a}
&&\!\!\!\!\!\!\!\!\!\!\!\!\!\!\!\!\!\!\!\! f^{(e)cl}(X1_{cl}, X2_{cl}) = \frac {2}{\pi}\arctan \biggl [\frac{4R_{0 cl}^2  \mid X1_{cl} X2_{cl} \mid}{R_{0 cl}^4-(X1_{cl}^{2}+X2_{cl}^{2})^2}\biggr ]\nonumber\\
&&\!\!\!\!\!\!\!\!\!\!\!\!\!\!\!\!\!\!\!\! f^{(e)ps}(X1_{ps}, X2_{ps}) = \frac {2}{\pi}\arctan \biggl [\frac{4R_{0 ps}^2  \mid X1_{ps} X2_{ps} \mid}{R_{0 ps}^4-(X1_{ps}^{2}+X2_{ps}^{2})^2}\biggr ]
\end{eqnarray}
\noindent Notice that, $\chi_{cl}$ is still undetermined at this stage while the general condition e) imposes $\chi_{ps}=-1$. Parameter $\chi_{cl}$ is easily determined by imposing that, for large values of the thermodynamic forces, the expressions for the diffusion coefficient evaluated by the TFT and the kinetic theory must coincide. Through a simple kinetic model, it is possible to show that $\chi_{cl}=1$ \cite{sonnino10}.

\noindent {\it In the classical regime, the nonlinear transport equations} are immediately obtained from Eqs~(\ref{nonlinearcl}). They read
\begin{eqnarray}\label{nonlinearps10}
&&\!\!\!\!\!\!\!\!\!\!<{\hat q}_{\rho cl}^{e(1)}>=\Bigl((<\!{\tilde{\sigma}}_{\perp}\!>\!+\frac{2}{\pi}<\!{\tilde{\sigma}}_{\perp}
\!\arctan\biggl [\frac{4R_{0 cl}^2  \mid X1_{cl} X2_{cl} \mid}{R_{0 cl}^4-(X1_{cl}^{2}+X2_{cl}^{2})^2}\biggr ]
\!>\Bigr)g_{\rho}^{(1)P}\nonumber\\
&&\qquad-\Bigl(<\!{\tilde{\alpha}}_{\perp}>\!+\frac{2}{\pi}<\!{\tilde{\alpha}}_{\perp}
\!\arctan\biggl [\frac{4R_{0 cl}^2  \mid X1_{cl} X2_{cl} \mid}{R_{0 cl}^4-(X1_{cl}^{2}+X2_{cl}^{2})^2}\biggr ]\!>\Bigr)g_{\rho}^{e(3)}\nonumber\\
&&\!\!\!\!\!\!\!\!\!\!<{\hat q}_{\rho cl}^{e(3)}>=
-\Bigl(<\!{\tilde{\alpha}}_{\perp}>\!+\frac{2}{\pi}<\!{\tilde{\alpha}}_{\perp}
\!\arctan\biggl [\frac{4R_{0 cl}^2  \mid X1_{cl} X2_{cl} \mid}{R_{0 cl}^4-(X1_{cl}^{2}+X2_{cl}^{2})^2}\biggr ]\!>\Bigr)g_{\rho}^{(1)P}\nonumber\\
&&\qquad+\Bigl(<\!{\tilde{\kappa}}^e_{\perp}>\!+
\frac{2}{\pi}<{\tilde{\kappa}}^e_{\perp}
\!\arctan\biggl [\frac{4R_{0 cl}^2  \mid X1_{cl} X2_{cl} \mid}{R_{0 cl}^4-(X1_{cl}^{2}+X2_{cl}^{2})^2}\biggr ]\!>\Bigr)
g_{\rho}^{e(3)}\nonumber\\
&&\!\!\!\!\!\!\!\!\!\!<{\hat q}_{\rho cl}^{i(3)}>={\tilde{\kappa}}^i_{\perp}g_{\rho}^{i(3)}
\end{eqnarray}
\noindent where $X1,\!2_{cl}$ are given by Eqs (\ref{nonlinearps6}) and (\ref{nonlinearps7}). The nonlinear Pfirsch-Schl{\"{u}}ter transport equations are obtained recalling Eqs~(\ref{entropy3}) and (\ref{linear10}). They suggest that
\begin{eqnarray}\label{nonlinearps10}
&& <{q}_{\rho}^{e(1)}>_{ps}=-K_e<\frac{\beta_0}{B}\Bigl(1-\frac{B^2}{\beta_0^2}\Bigr)(g_{\parallel}^{(1)}-{\bar g}_{\parallel}^{e(1)})>\nonumber\\
&&<{q}_{\rho}^{e(3)}>_{ps}=K_e<\frac{\beta_0}{B}\Bigl(1-\frac{B^2}{\beta_0^2}\Bigr)(g_{\parallel}^{e(3)}+{\bar g}_{\parallel}^{e(3)})>\\
&&<{q}_{\rho}^{i(3)}>_{ps}=K_i<\frac{\beta_0}{B}\Bigl(1-\frac{B^2}{\beta_0^2}\Bigr)(g_{\parallel}^{i(3)}+{\bar g}_{\parallel}^{i(3)})>\nonumber
\end{eqnarray}
\noindent which, combined with Eqs~(\ref{nonlinearps}), give
\begin{eqnarray}\label{nonlinearps11}
&&<{q}_{\rho}^{e(1)}>_{ps}=K_e^2({\tilde g}_{ps}-1)\Bigl(c^e_{11}g_{\rho}^{(1)P}-c^e_{13} g_{\rho}^{e(3)}\Bigr)\nonumber\\
&&<{q}_{\rho}^{e(3)}>_{ps}=K_e^2({\tilde g}_{ps}-1)\Bigl(-c^e_{13}g_{\rho}^{(1)P}+c^e_{33} g_{\rho}^{e(3)}\Bigr)\\
&&<{q}_{\rho}^{i(3)}>_{ps}=K_i^2(g-1)c_{33}^ig_{\rho}^{i(3)}\nonumber
\end{eqnarray}
\noindent where $g$ and the collision matrix elements $c_{\mu\nu}$ are given by Eq.~(\ref{linear11a}) and Eq.~(\ref{linear11b}) respectively and ${\tilde g}_{ps}$ is defined as
\begin{equation}\label{nonlinearps11a}
{\tilde g}_{ps}\equiv 1+<\frac{\Bigl(1-\frac{B^2}{\beta_0^2}\Bigr)\Bigl(\frac{\beta_0^2}{B^2}-1\Bigr)}{1-\frac{2}{\pi}\arctan\biggl [\frac{4R_{0 ps}^2  \mid X1_{ps} X2_{ps} \mid}{R_{0 ps}^4-(X1_{ps}^{2}+X2_{ps}^{2})^2}\biggr]}>
\end{equation}
\noindent with $X1_{ps}$ and $X2_{ps}$ given by Eqs (\ref{nonlinearps6}) and (\ref{nonlinearps8}).

\noindent It is useful to re-write Eq.~(\ref{nonlinearps11}) as follows
\begin{eqnarray}\label{nonlinearps11c}
&&<{q}_{\rho}^{e(1)}>_{ps}=\mathcal{F}_{{ps}}K_e^2(g-1)\Bigl(c^e_{11}g_{\rho}^{(1)P}-c^e_{13} g_{\rho}^{e(3)}\Bigr)\nonumber\\
&&<{q}_{\rho}^{e(3)}>_{ps}=\mathcal{F}_{{ps}}K_e^2(g-1)\Bigl(-c^e_{13}g_{\rho}^{(1)P}+c^e_{33} g_{\rho}^{e(3)}\Bigr)\\
&&<{q}_{\rho}^{i(3)}>_{ps}=K_i^2(g-1)c_{33}^ig_{\rho}^{i(3)}\nonumber
\end{eqnarray}
\noindent where we have introduced the {\it amplification factor} $\mathcal{F}_{ps}$ defined as
\begin{equation}\label{nonlinearps11d}
\mathcal{F}_{{ps}}\equiv\frac{{\tilde g}_{ps}-1}{g-1}
\end{equation}
\noindent Note that the amplification factor $\mathcal{F}_{{ps}}$ is equal to $1$ when the perturbation function $f^{(e)ps}(X_{ps})$ is set to zero and it may assume very large values as $X1^2_{ps}+X2^2_{ps}\rightarrow R^2_{0 ps}$. 

 \noindent An estimation of the radii $R_{0 cl}$ and $R_{0 ps}$ can be provided. We define with $L_{H}$ the {\it minimum length of the gradient} of the macroscopic quantities, namely density, the velocity, the pressure or the temperature:
\begin{equation}\label{nonlinearps12}
\frac{1}{L_H}={\rm Max}\frac{\mid\nabla A(X)\mid}{\mid A(X)\mid}
\end{equation}
\noindent Without going into more accurate estimates, we note that these gradients pertain to macroscopic devices. Thus, the length of the gradients is very large compared to molecular dimensions. From Eqs (\ref{entropy8}) and (\ref{entropy15}) we find after a little algebra:
\begin{eqnarray}\label{nonlinearps15}
&&\!\!\!\!\!\!\!\!\!\!\!\!\!\!\!\!\!\!\!\!\!\!\!\!\!\!\!\!\!\!\!R_{0 cl}={\rm Max}\Bigl\{(X1_{cl}+X2_{cl})^{1/2}\Bigr\}\simeq\\
&&\!\!\!\!\!\!\!\!\!\!\!\!\!\!\!\!\!\frac{1}{L_H}{\rm Max}\Bigl\{
\Bigl(\frac{T_e}{m_e}\Bigr)^{1/2}\tau_e\Bigl[a\Bigl(1+\frac{P_i}{P_e}\Bigr)^2+\sqrt{10}\ b\Bigl(1+\frac{P_i}{P_e}\Bigr)+\frac{5}{2}c\Bigr]^{1/2}\Big\}\nonumber
\end{eqnarray}
\noindent \noindent where
\begin{eqnarray}\label{nonlinearps15a}
&&a=\alpha_{+}\beta^2_{+}+\alpha_{-}\beta^2_{-}\qquad  b=\alpha_{+}\beta_{+}+\alpha_{-}\beta_{-}\qquad c= \alpha_{+}+\alpha_{-}\nonumber\\
&&\frac{1}{L_H}={\rm Max}\Bigl\{\frac{\mid\nabla_{\rho} P\mid}{P},\ \frac{\mid\nabla_{\rho} T_e\mid}{T_e}\Bigr\}
\end{eqnarray}
\noindent with
\begin{eqnarray}\label{nonlinearps16}
&&\alpha_\pm=\frac{{\tilde\alpha}_{\perp}^2[{{\tilde\kappa}^{e}_{\perp}}\!{}^2+{\tilde\sigma}^2_{\perp}+2{\tilde\alpha}^2_{\perp}\pm({\tilde\kappa}^e_{\perp}+{\tilde\sigma}_{\perp})\sqrt{\Delta_{cl}}]}
{({\tilde\kappa}^e_{\perp}{\tilde\sigma}_{\perp}-{\tilde\alpha}^2_{\perp})^2[({\tilde\kappa}^e_{\perp}-{\tilde\sigma}_{\perp})^2+4{\tilde\alpha}^2_{\perp}\pm ({\tilde\kappa}^e_{\perp}-{\tilde\sigma}_{\perp})\sqrt{\Delta_{cl}}]}
\nonumber\\
&&\beta_\pm=\frac{{\tilde\kappa}^e_{\perp}-{\tilde\sigma}_{\perp}\pm\sqrt{\Delta_{cl}}}{2{\tilde\alpha}_{\perp}}
\qquad\qquad \Delta_{cl}= ({\tilde\kappa}^e_{\perp}-{\tilde\sigma}_{\perp})^2+4{\tilde\alpha}^2_{\perp}
\end{eqnarray}
\noindent Therefore we find 
\begin{equation}\label{nonlinearps16a}
R_{0 cl}{\big\arrowvert}_{TFT}=R_{0 cl}{\big\arrowvert}_{Neocl.}
\end{equation}
\noindent i.e., the value of $R_{0 cl}$ evaluated by the TFT coincides with that one estimated by the neoclassical theory. Similarly, from Eqs (\ref{entropy8}), (\ref{entropy15}) and (\ref{entropy21}) we obtain, for $Z=1$,
\begin{eqnarray}\label{nonlinearps16b}
&&\!\!\!\!\!\!\!\!\!\!\!\!R_{0 ps}={\rm Max}\Bigl\{(X1_{ps}+X2_{ps})^{1/2}\Bigr\}\simeq\\
&&\!\!\!\!\!\!\!\!\!\!\!\!\frac{1}{L_H}{\rm Max}\Bigl\{\mathcal{F}_{{ps}}\mid K_e\mid\tau_e
\frac{\beta_0}{B}{\Bigg\arrowvert}1-\frac{B^2}{\beta_0^2}{\Bigg\arrowvert}
\Bigl(\frac{T_e}{m_e}\Bigr)^{1/2}
\Bigl[0.932\Bigl(1+\frac{P_i}{P_e}\Bigr)^2+2.331\Bigr]^{1/2}\Big\}\nonumber
\end{eqnarray}
\noindent The relation between the TFT and the neoclassical theory estimations is then
\begin{equation}\label{nonlinearps16c}
R_{0 ps}{\big\arrowvert}_{TFT}=R_{0 ps}{\big\arrowvert}_{Neocl.}{\rm Max}\bigl\{\mathcal{F}_{{ps}}\big\}
\end{equation}
\noindent The exact values of the parameters $R_{0 cl}$ and $R_{0 ps}$ can be obtained numerically if the magnetic configuration and the profiles of the thermodynamic forces are specified. 

\noindent Fig.(\ref{2Dsolution}) depicts, in polar coordinates, the function $f^{(e)A}(X_A)$ (with $A=cl, ps$). We note that, in the weak-field approximation, {\it the TFT does not correct the pure effects} i.e., it does not correct the pure Fick's law and the pure Fourier's law. {\it The TFT provides nonlinear corrections only for the cross-effects}.
%%%%%%%%%%%%%%%%%%%%%%%%%%%%%%%%%%%%%%%%%%%%%%%%%%%%%%%%%%%%%%%%%%%%%%%%%%%%%
\begin{figure*}[htb] 
\hspace{3cm}\includegraphics[width=8cm,height=6.5cm]{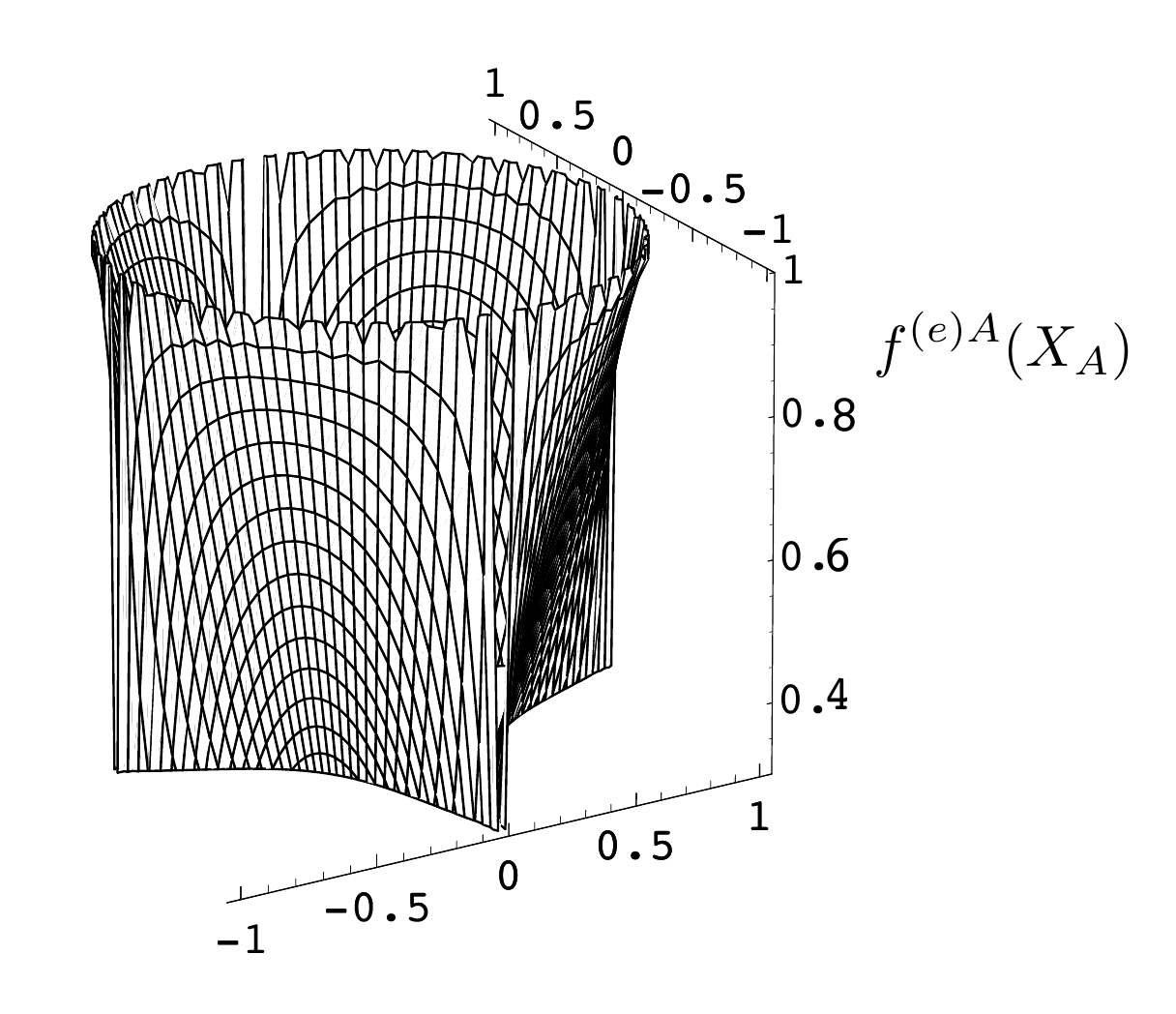}
\caption{ \label{2Dsolution} $f^{(e)A}(X_A)$ (with $A=cl, ps$) in polar coordinates. This perturbation function satisfies the condition $0\leq f^{(e)A}(X_A)\leq1$.}
\end{figure*}
%%%%%%%%%%%%%%%%%%%%%%%%%%%%%%%%%%%%%%%%%%%%%%%%%%%%%%%%%%%%%%%%%%%%%%%%%%%%%%%
%%%%%%%%%%%%%%%%%%%%%%%%%%%%%%%%%%%%%%%%%%%%%%%%%%%
\noindent At the end of this section we shall see that, for a JET plasma in L-mode, the fluxes of electron energy (heat flow) is further enhanced in the nonlinear classical and Pfirsch-Schl{\"{u}}ter (P-S) transports. The main conclusion of the present analysis is thus:
\begin{description}
\item[1)] In the nonlinear regime, the transport coefficients are given by the neoclassical values supplemented by extra terms evaluated by the TFT. These corrective terms are given by an {\it amplification function} times the Onsager matrix. This function corresponds to the arc-tangent function of the thermodynamic forces for the classical regime and to the amplification factor ${\mathcal F}_{ps}$ for the Pfirsch-Schl{\"{u}}ter transport. We have already mentioned that, through a kinetic model introduced in ref. \cite{sonnino10}, it is possible to show that, in the collisional transport regimes, the electric diffusion coefficient and the electrical thermal coefficient are approximatively given by the linear (Onsager) transport coefficients times a universal function expressed by the inverse of the electron collision time $\tau_e$. Since, the collision time decreases as the intensity of the thermodynamic forces increases, the model predicts that the values of the electrical thermal and diffusion coefficients {\it increase} as the intensity of the thermodynamic forces increases;
\item[2)]  In the classical and Pfirsch-Schl{\"{u}}ter regimes, the TFT does not correct the expressions of the ion heat fluxes evaluated by the neoclassical theory;
\item[3)] The electron nonlinear classical transport coefficients may exceed the linear ones by a factor, which may be of order $\chi_{cl}+1=2$. The electron nonlinear Pfirsch-Schl{\"{u}}ter transport coefficients exceed the values evaluated by the neoclassical theory by the factor ${\mathcal F}_{ps}$. Concrete calculations, for a JET plasma in L-mode, show that the amplification factor may assume values of order $10^2$ (see below). These nonlinear corrections refer to the cross-effects; the pure Fick's and Fourier's laws remain, however, unaltered. 
\end{description}
\noindent This dissymmetry between the P-S ion and electron transport coefficients is encouraging since it is in qualitative agreement with experiment. Fig.~(\ref{nlps}) schematically illustrates the physical interpretation of the {\it nonlinear Pfirsch-Schl{\"{u}}ter effect} \cite{sonnino10}. 
%%%%%%%%%%%%%%%%%%%%%%%%%%%%%%%%%%%%%%%%%%%%%%%%%%%%%%%%%%%%%%%%%%%%%%%%%%%%%
\begin{figure*}[htb] 
\hspace{2cm}\includegraphics[width=9cm]{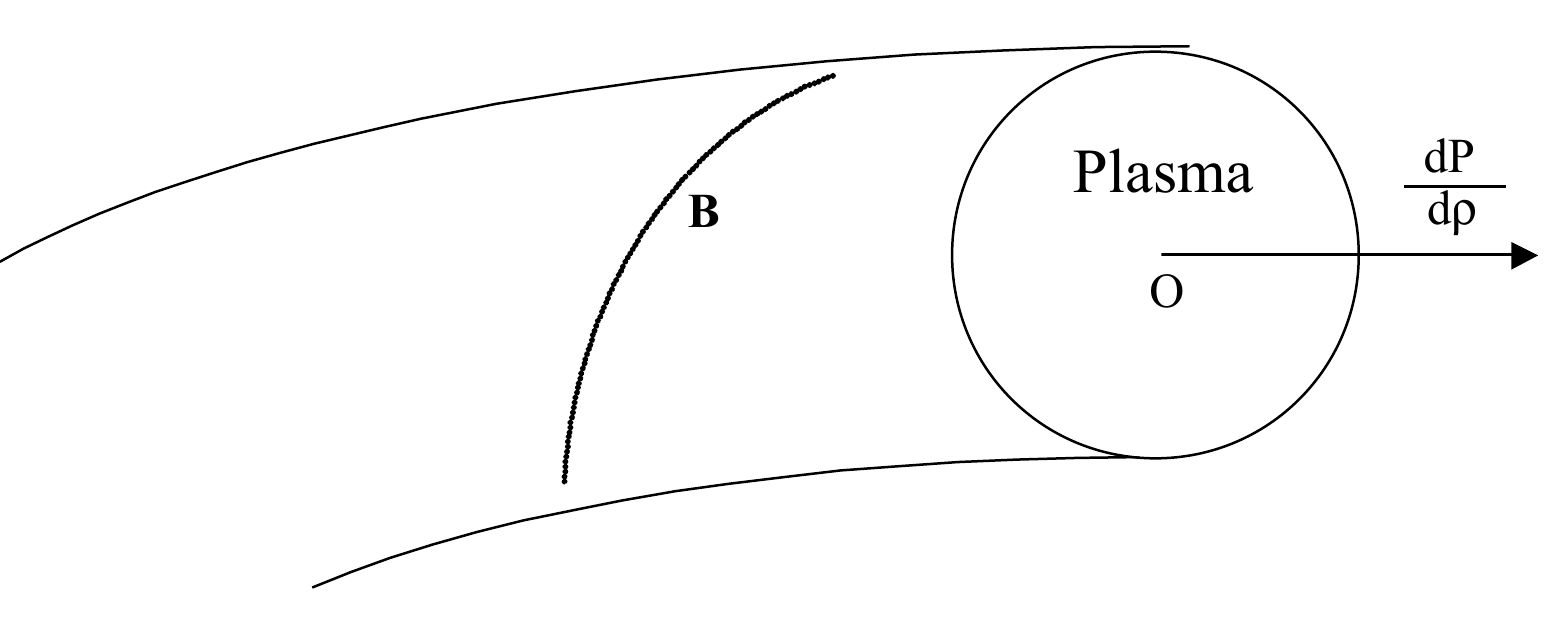}
\caption{ \label{nlps} The nonlinear Pfirsch-Schl{\"{u}}ter effect. As the thermodynamic forces increase their magnitudes along the major radius, the resulting Pfirsch-Schl{\"{u}}ter current must also increase. The intensity of this current is maximized when the thermodynamic forces reach their maxima values (i.e., $X1^2_{ps}+X2^2_{ps}\rightarrow R^2_{0 ps}$).}
\end{figure*}
%%%%%%%%%%%%%%%%%%%%%%%%%%%%%%%%%%%%%%%%%%%%%%%%%%%%%%%%%%%%%%%%%%%%%%%%%%%%%%%
\noindent As the thermodynamic forces increase their magnitudes, after some collisions, the plasma pressure, through toroidal geometry, tends to be larger at larger major radii. The resulting increased pressure must be balanced by the internal magnetic force. As a consequence, the currents, which produce this force, must increase their magnitudes producing an increment of charge accumulation in the upper and lower parts of the plasma. Charge accumulation is prevented by an increment of the intensity of the Pfirsch-Schl{\"{u}}ter current. Consistent with intuitive expectations, this effect is maximized when the thermodynamic forces reach their maxima values ($X1^2_{ps}+X2^2_{ps}\rightarrow R^2_{0 ps}$). In this limiting scenario, a large amount of energy loss (heat flux) and matter (particles flux) is possible.

\noindent We conclude this section reporting the comparison between the neoclassical results and the TFT predictions for the electron mass flux and electron heat flow for L-mode JET plasmas. The numerical estimations of the electron fluxes, given by Eqs (\ref{nonlinearps10}) and (\ref{nonlinearps11}), can be evaluated by specifying the magnetic field configuration and the expressions for the thermodynamic forces. In the local triad $({\bf e}_r,{\bf e}_\theta,{\bf e}_\zeta)$, the magnetic field, in the standard hight aspect ratio, low $\beta$, circular tokamak equilibrium model, here referred to as the {\it standard model}, reads
\begin{equation}\label{nonlinearps17}
{\bf B}=B_0G(r)\ {\bf e}_\theta+\frac{B_0}{1+(r/R)\cos\theta}\ {\bf e}_\zeta\quad\quad{\rm with}\quad G(r)\equiv\frac{r}{Rq(r)}
\end{equation}
\noindent where $B_0$ is a constant having the dimension of a magnetic field, $q(r)$ denotes the {\it safety factor} and $R$ is the major radius. For JET, $B_0=34500\ Gauss$ and the major and the minor radii of the reactor are, respectively, $R=296\ cm$ and $a=125\ cm$. The experimental profile of the  $q(r)$, for a JET plasma in L-mode, can be found in fig.~\ref{q}.
%%%%%%%%%%%%%%%%%%%%%%%%%%
\begin{figure*}[htb] 
\hspace{2cm}\includegraphics[width=7.5cm,height=5cm]{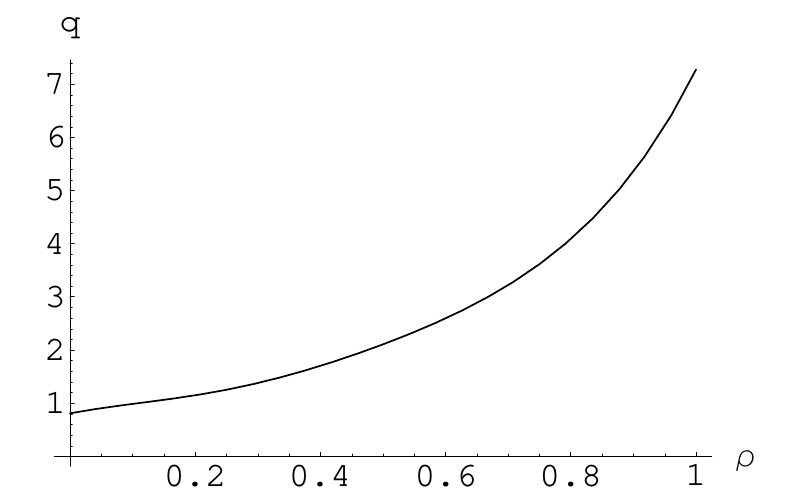}
\caption{ \label{q} The experimental safety factor versus the re-normalized minor radius $\rho$ for L-mode JET plasmas \cite{sozzi}.}
\end{figure*}
%%%%%%%%%%%%%%%%%%%%%%%%%%
\noindent The specification of thermodynamic forces $g_{\rho}^{(1)P}$ and $g_{\rho}^{\alpha (3)}$ requires the knowledge of the electron and ion temperature and density profiles. Figs~\ref{temperature} and \ref{Density} show typical experimental profiles of the electron temperature and the electron density of a L-mode JET plasma\footnote{At this stage, we can assume that the ion temperature profile coincides with the electron one. Notice that, for $Z=1$, the electroneutrality condition imposes that $n^i\simeq n^e$.\label{temp1}}\cite{sozzi}.
%%%%%%%%%%%%%%%%%%%%%%%%%%%%%%%%%%%%%%%%%%%%%%%%%%%
\begin{figure*}[htb]
\hspace{2.5cm}\includegraphics[width=7cm]{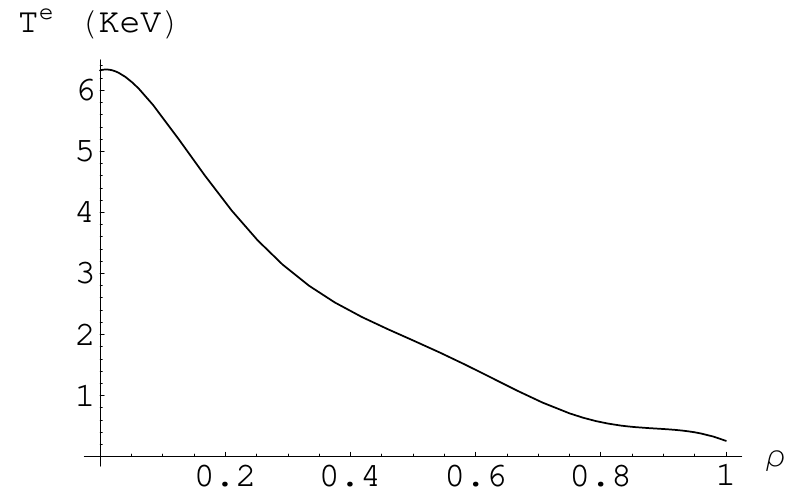}
\caption{ \label{temperature} Typical experimental electron temperature profile versus the re-normalized minor radius $\rho$ for L-mode JET plasmas \cite{sozzi}.}
\end{figure*}
%%%%%%%%%%%%%%%%%%%%%%%%%%%%%%%%%%%%%%%%%%%%%%%%%%%
%%%%%%%%%%%%%%%%%%%%%%%%%%%%%%%%%%%%%%%%%%%%%%%%%%%%%%
\begin{figure*}[htb] 
\hspace{2cm}\includegraphics[width=7.5cm,height=5cm]{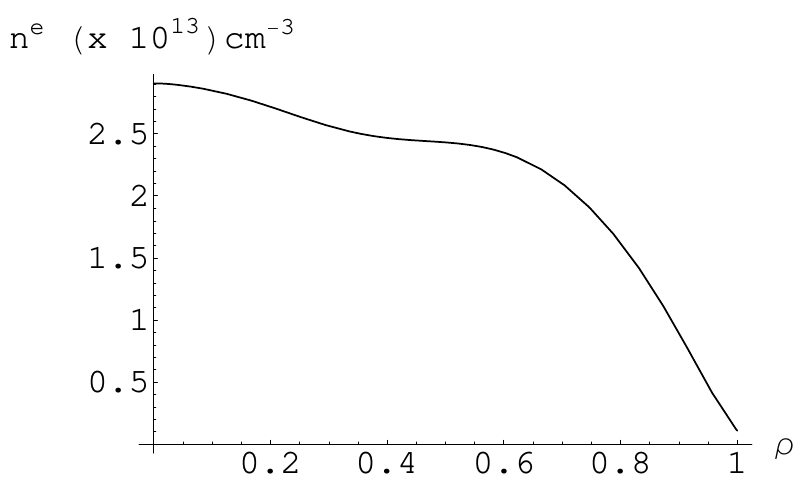}
\caption{ \label{Density} Typical experimental electron density profile versus the re-normalize minor radius $\rho$ for L-mode JET plasmas \cite{sozzi}.}
\end{figure*}
%%%%%%%%%%%%%%%%%%%%%%%%%%%%%%%%%%%%%%%%%%%%%%%%%%%%%%%%%%%%%%%%%%%%%%%%%%%%%%%
\noindent In the standard configuration, the {\it averaging formula} of a physical quantity $A(r,\theta)$ simply reads \cite{balescu2}
\begin{equation}\label{nonlinearps18}
<A(r,\theta)>=\frac{1}{2\pi}\int_0^{2\pi}d\theta \Bigl(1+\frac{r}{R}\cos\theta\Bigr)A(r,\theta)
\end{equation}
\noindent Figures~\ref{g1P}, \ref{ge3} and \ref{gi3} report the profiles of the thermodynamic forces for a L-mode JET plasma against the re-normalized minor radius $\rho=\frac{r}{a}$. In these calculations $Z=1$. The theoretical predictions of the Thermodynamic Field Theory (TFT) and the neoclassical results, for a JET plasma in the two collisional transport regimes, are compared in figs~\ref{Q1_CL}$\div$\ref{Q3_PF}. In particular, Figs~\ref{Q1_CL} and \ref{Q1_PF} report the (radial) electron fluxes evaluated by the TFT and the neoclassical theory, for plasmas in the classical and Pfirsch-Schl{\"{u}}ter regimes, respectively. The profiles of the (radial) electron heat fluxes, determined by these two theories in the classical and Pfirsch-Schl{\"{u}}ter regimes, are illustrated in figs~\ref{Q3_CL} and \ref{Q3_PF}, respectively. It is seen that, in the core of the plasma, the nonlinear classical transport coefficients exceed the linear ones by a factor, which may be of order $2$. The electron nonlinear Pfirsch-Schl{\"{u}}ter transport coefficients exceed the linear ones by a factor, which may be of order $10^2$. Although having similar shapes, the curves strongly differ in magnitude (see Figs~\ref{comp_e_flux} and \ref{comp_heat_flux}).
%%%%%%%%%%%%%%%%%%%%%%%%%%
\begin{figure*}[htb] 
\hspace{2cm}\includegraphics[width=7.5cm,height=5cm]{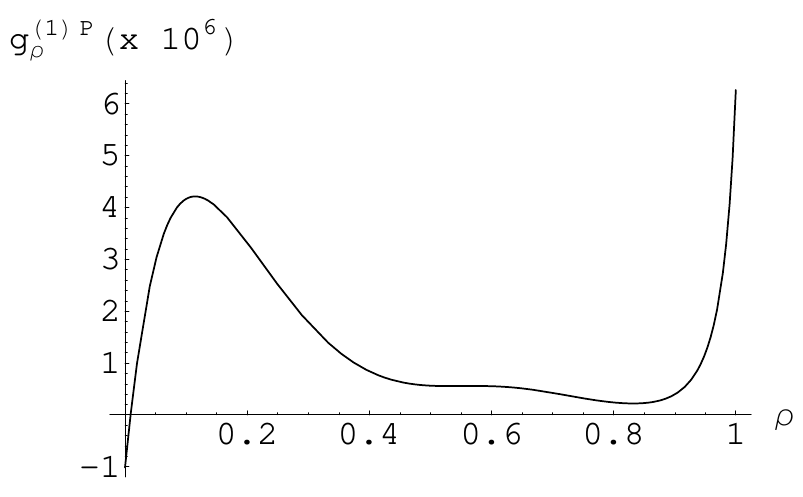}
\caption{ \label{g1P} The dimensionless thermodynamic force $g^{(1)P}_{\rho}$ versus the re-normalized minor radius $\rho$ for a L-mode JET plasma. This force is proportional to the gradient of the total pressure.}
\end{figure*}
%%%%%%%%%%%%%%%%%%%%%%%%%%%%%%%%%%%%%%%%%%%%%%%%%%%%%%%%%%%%%%%%%%%%%%%%%%%%%%%
%%%%%%%%%%%%%%%%%%%%%%%%%%
\begin{figure*}[htb] 
\hspace{2cm}\includegraphics[width=7.5cm,height=5cm]{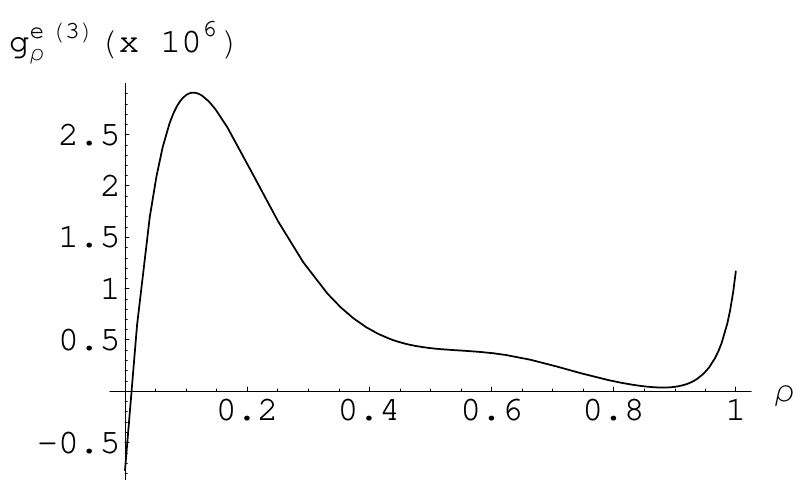}
\caption{ \label{ge3} The dimensionless thermodynamic force $g^{e(3)}_{\rho}$ versus the re-normalized minor radius $\rho$ for a L-mode JET plasma. This force is proportional to the electron temperature gradient.}
\end{figure*}
%%%%%%%%%%%%%%%%%%%%%%%%%%%%%%%%%%%%%%%%%%%%%%%%%%%%%%%%%%%%%%%%%%%%%%%%%%%%%%%
%%%%%%%%%%%%%%%%%%%%%%%%%%
\begin{figure*}[htb] 
\hspace{2cm}\includegraphics[width=7.5cm,height=5cm]{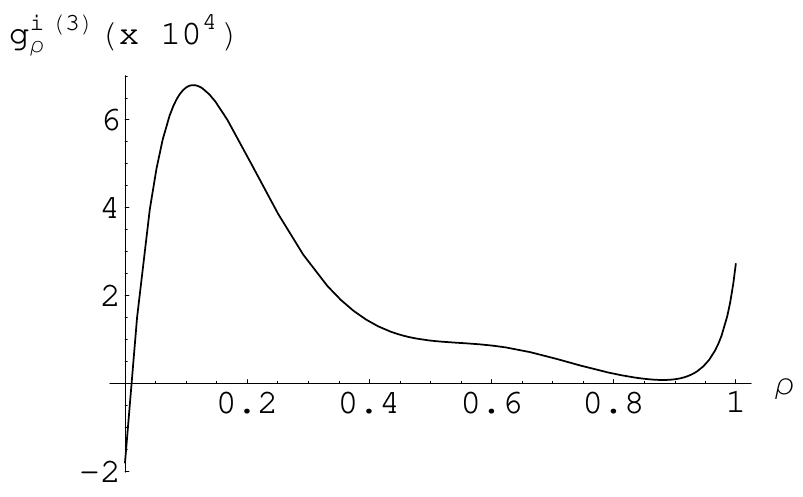}
\caption{ \label{gi3} The dimensionless thermodynamic force $g^{i(3)}_{\rho}$ versus the re-normalized minor radius $\rho$ for a L-mode JET plasma. This force is proportional to the ion temperature gradient.}
\end{figure*}
%%%%%%%%%%%%%%%%%%%%%%%%%%%%%%%%%%%%%%%%%%%%%%%%%%%
%%%%%%%%%%%%%%%%%%%%%%%%%%%%
%%%%%%%%%%%%%%%%%%%%%%%%%%
\begin{figure*}[htb] 
\hspace{2cm}\includegraphics[width=7.0cm,height=6cm]{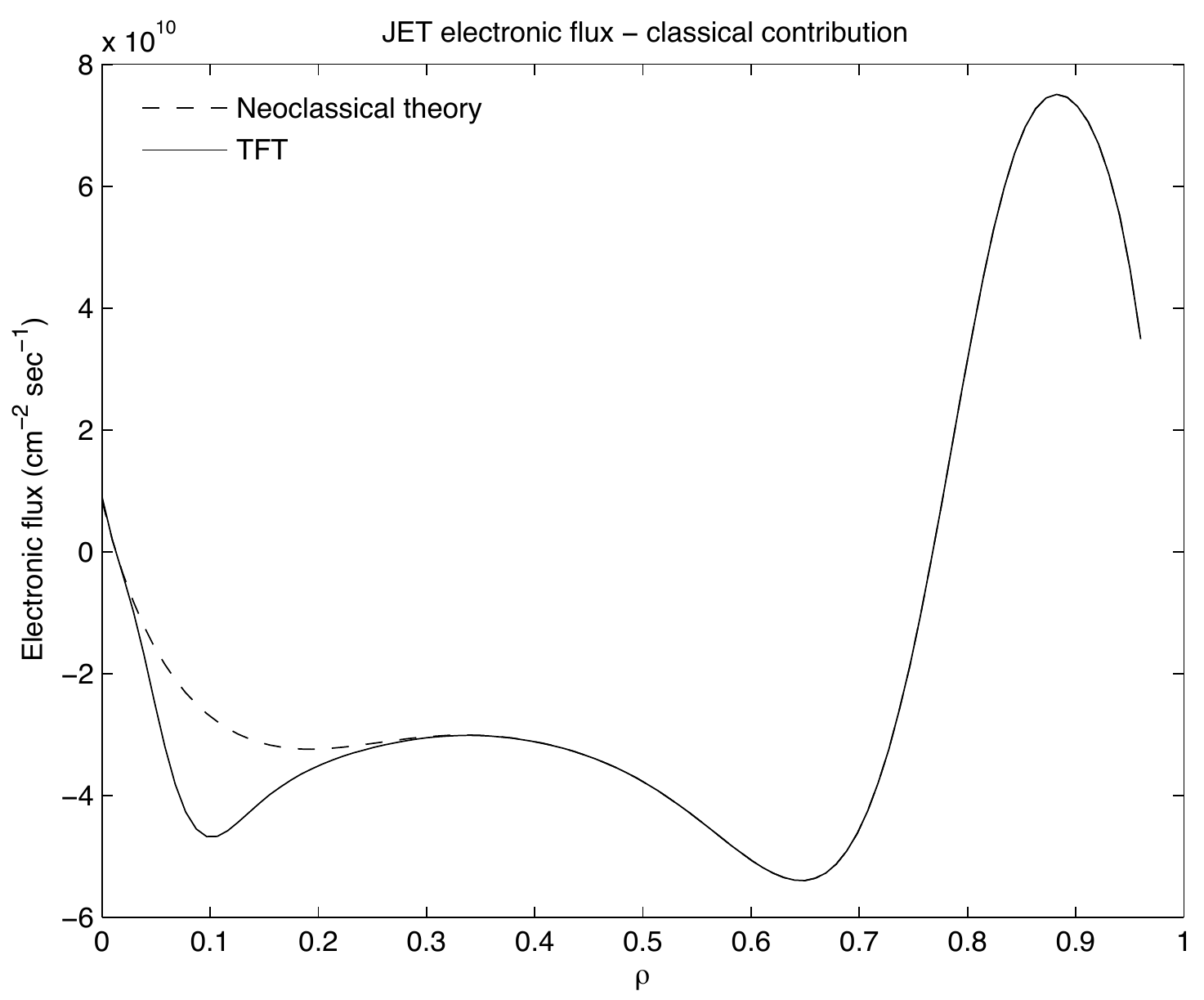}
\caption{ \label{Q1_CL}   L-mode JET plasmas, with $Z=1$, in the classical regime. Predictions of the radial electron fluxes profiles according to the Thermodynamic Field Theory (TFT) and the neoclassical theory.}
\end{figure*}
%%%%%%%%%%%%%%%%%%%%%%%%%%%%%%%%%%%%%%%%%%%%%%%%%%%

%%%%%%%%%%%%%%%%%%%%%%%%%%
\begin{figure*}[htb] 
\hspace{2cm}\includegraphics[width=7.0cm]{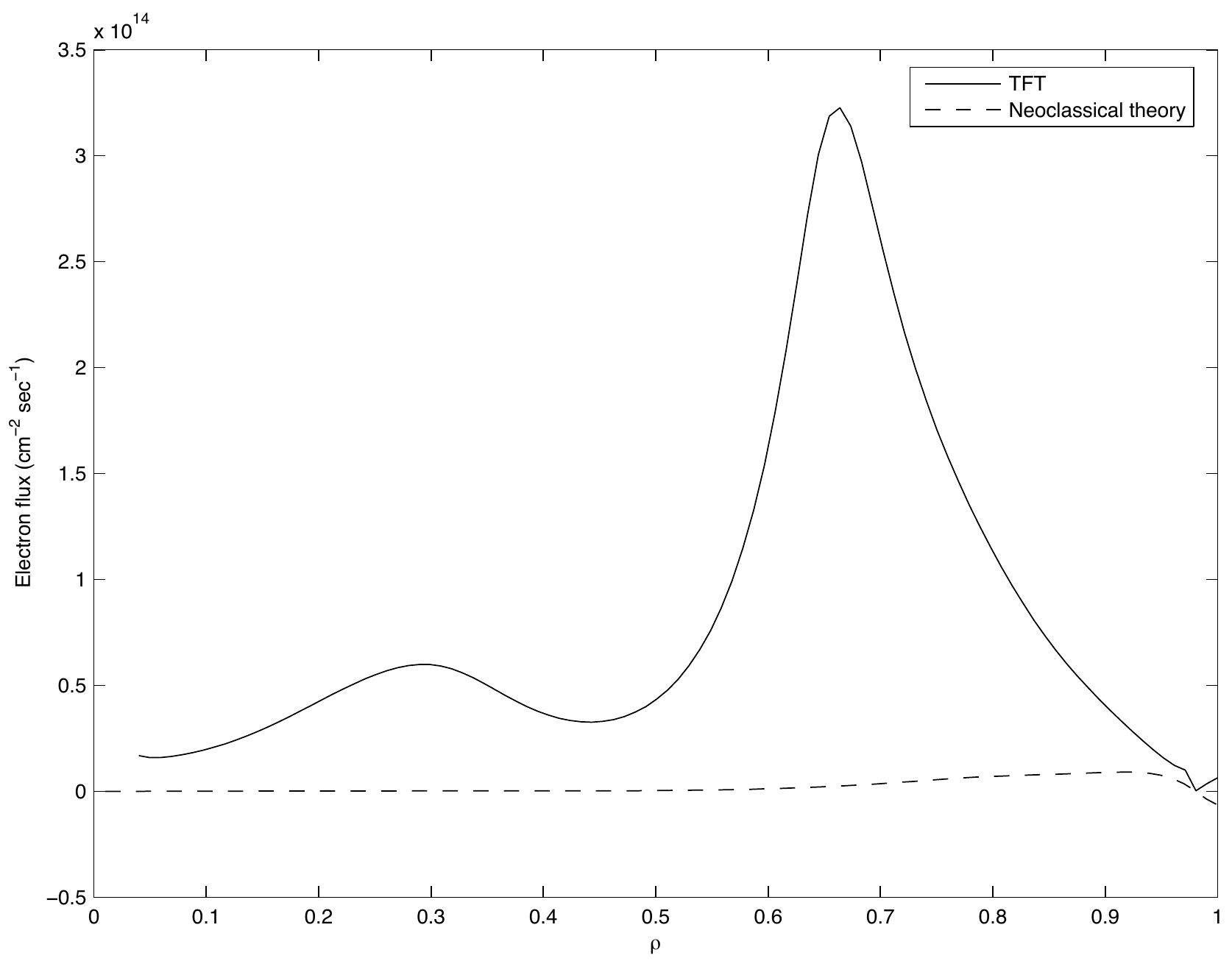}
\caption{ \label{Q1_PF}  L-mode JET plasmas, with $Z=1$, in the Pfirsch-Schl{\"{u}}ter regime. Predictions of the radial electron fluxes profiles according to the Thermodynamic Field Theory (TFT) and the neoclassical theory.}
\end{figure*}
%%%%%%%%%%%%%%%%%%%%%%%%%%%%
%%%%%%%%%%%%%%%%%%%%%%%%%%%%%%%%%%%%%%%%%%%%%%%%%%%%%%
\begin{figure*}[htb] 
\hspace{2cm}\includegraphics[width=7cm,height=6cm]{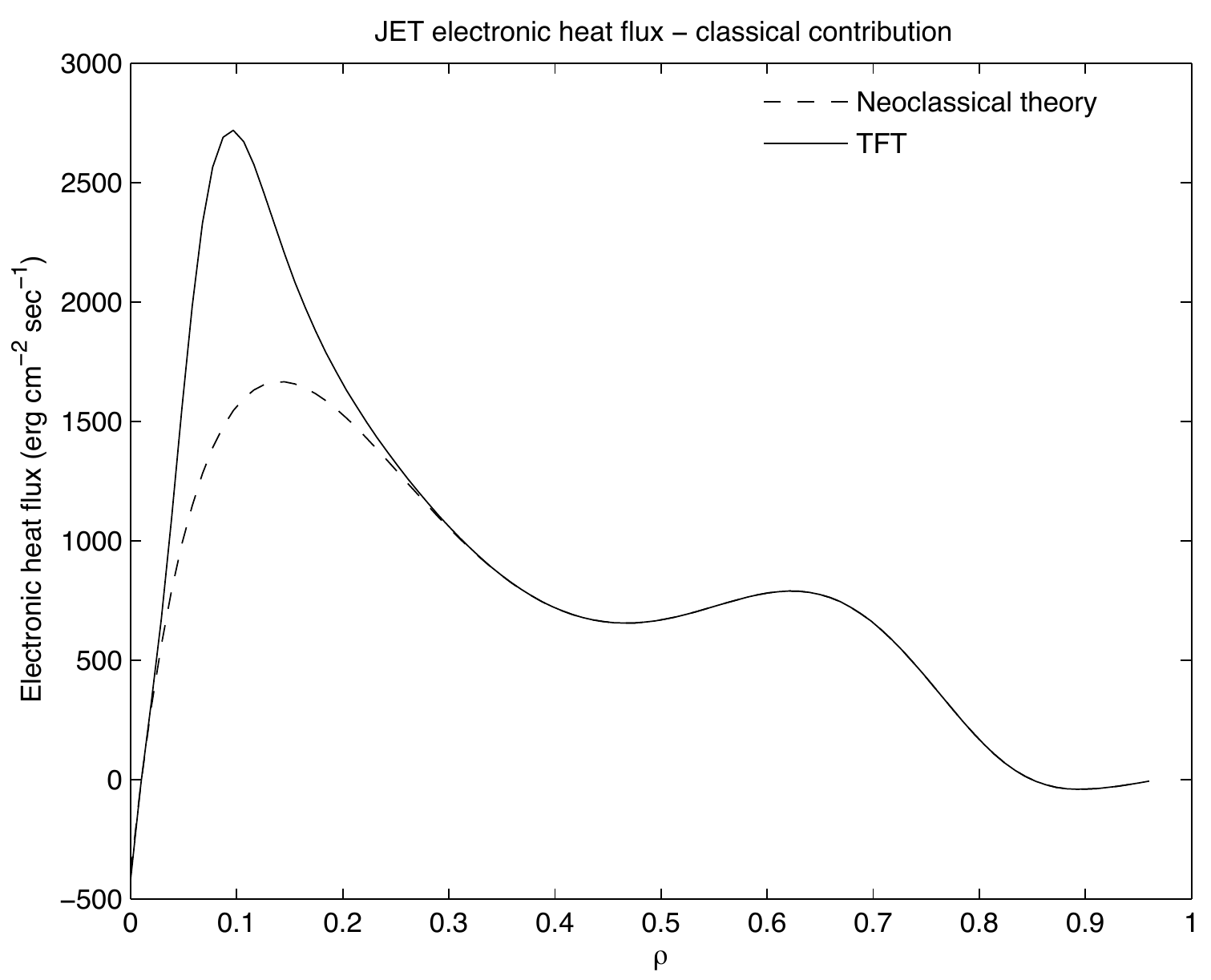}
\caption{ \label{Q3_CL} L-mode JET plasmas, with $Z=1$, in the classical regime. Predictions of the radial electron heat fluxes profiles according to the TFT and the neoclassical theory.}
\end{figure*}
%%%%%%%%%%%%%%%%%%%%%%%%%%%%%%%%%%%%%%%%%%%%%%%%%%%
%%%%%%%%%%%%%%%%%%%%%%%%%%%%%%%%%%%%%%%%%%%%%%%%%%%
\begin{figure*}[htb] 
\hspace{2cm}\includegraphics[width=7cm]{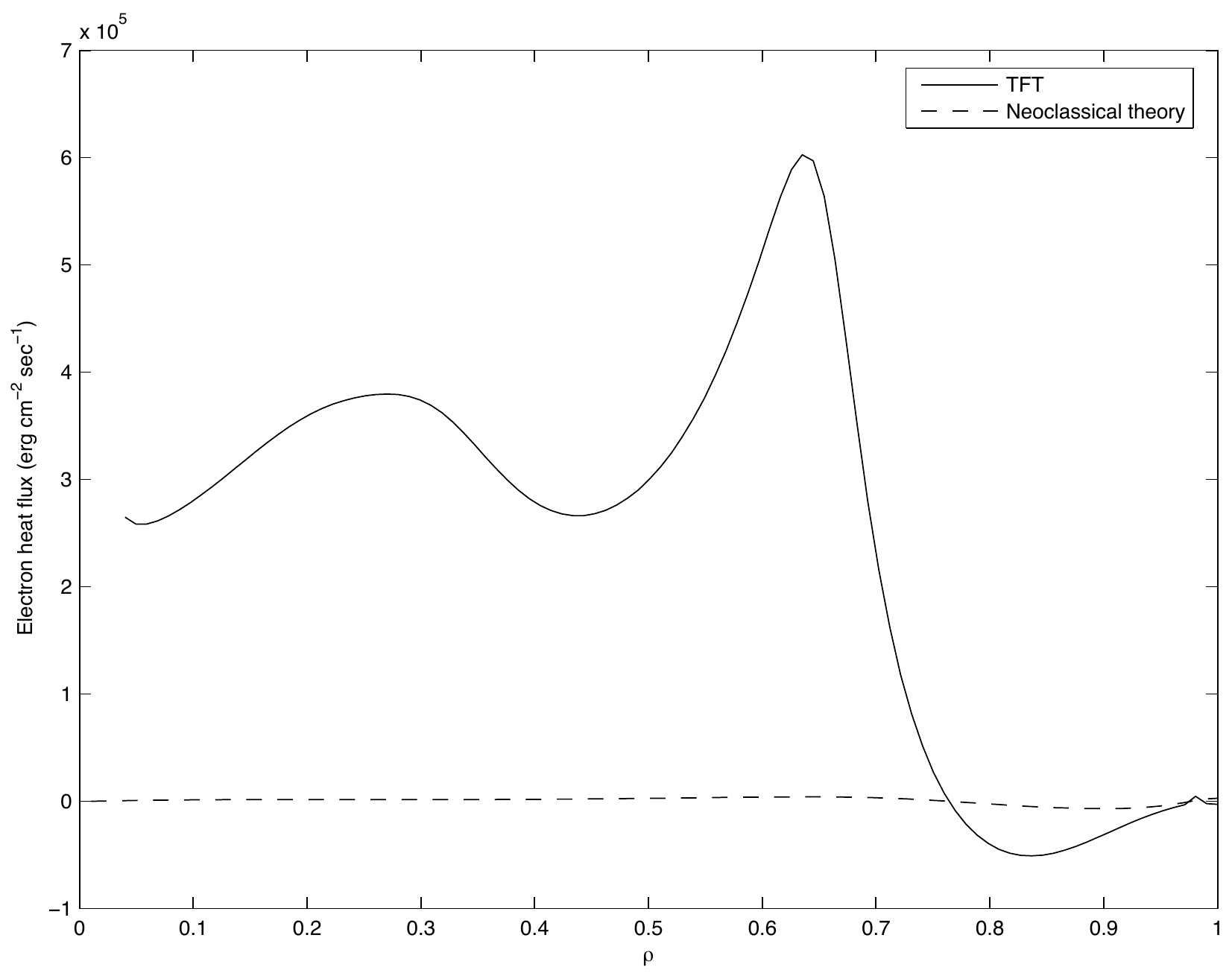}
\caption{ \label{Q3_PF} L-mode JET plasmas, with $Z=1$, in the Pfirsch-Schl{\"{u}}ter regime. Predictions of the radial electron heat fluxes profiles according to the TFT and the neoclassical theory.}
\end{figure*}
%%%%%%%%%%%%%%%%%%%%%%%%%%%%%%%%%%%%%%%%%%%%%%%%%%%
%%%%%%%%%%%%%%%%%%%%%%%%%%%%%%%%%%%%%%%%%%%%%%%%%%%
\begin{figure*}[htb] 
\hspace{2cm}\includegraphics[width=7cm]{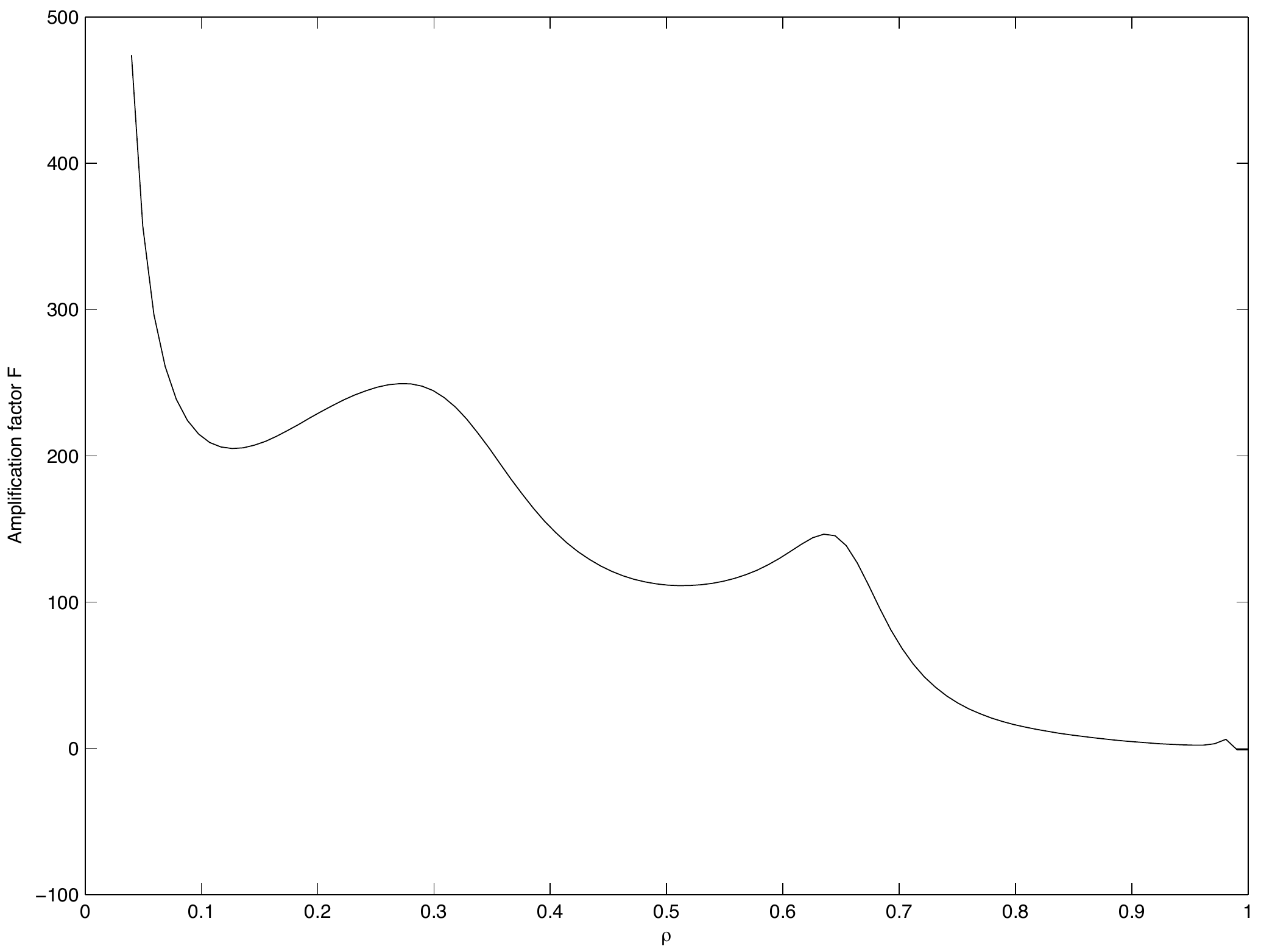}
\caption{ \label{F_M1} The amplification factor $\mathcal{F}_{{ps}}$ versus the minor radius $\rho$ for a L-mode JET plasmas, with $Z=1$. As we can see, this factor may reach values of order $10^2$.}
\end{figure*}
%%%%%%%%%%%%%%%%%%%%%%%%%%%%%%%%%%%%%%%%%%%%%%%%%%%
%%%%%%%%%%%%%%%%%%%%%%%%%%%%%%%%%%%%%%%%%%%%%%%%%%%%%%%%%%%%%%%%%%%%%%%%%%%%%%%%%%%%%%%%%%%%%%%%%%%%%%
\begin{figure*}[htb]
\hspace{0.0cm}\includegraphics[width=6cm]{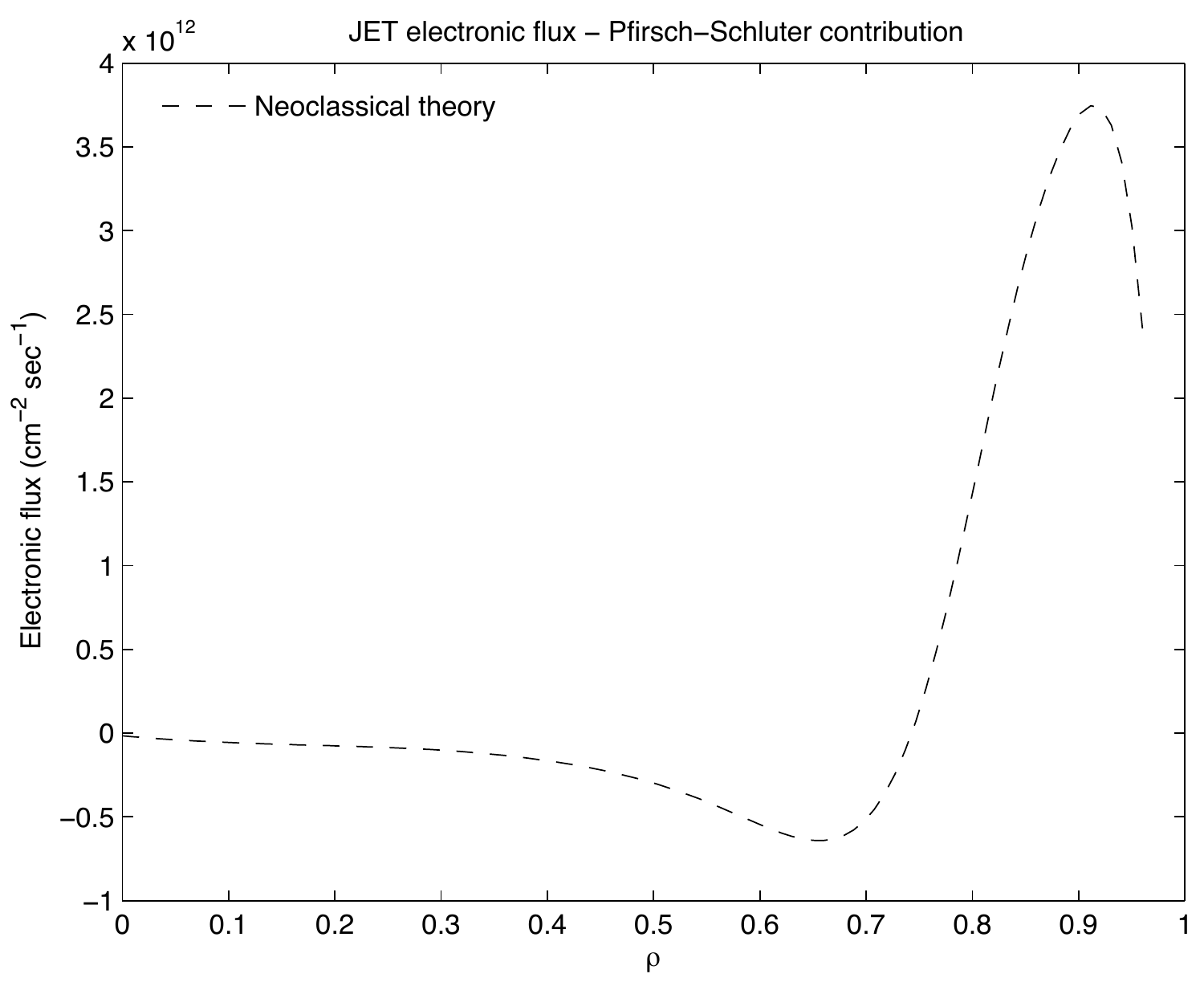}\includegraphics[width=6.2cm]{Q1DIM_M1.pdf}
\caption{ \label{comp_e_flux} The nonlinear Pfirsch-Schl{\"{u}}ter effect: comparison between the radial electron fluxes profiles according to the neoclassical theory (left picture) and the Thermodynamic Field Theory (right picture). Although displaying similar shapes, they strongly differ in magnitude.}
\end{figure*}
%%%%%%%%%%%%%%%%%%%%%%%%%%%%%%%%%%%%%%%%%%%%%%%%%%%
%%%%%%%%%%%%%%%%%%%%%%%%%%%%%%%%%%%%%%%%%%%%%%%%%%
%%%%%%%%%%%%%%%%%%%%%%%%%%%%%%%%%%%%%%%%%%%%%%%%%%%
\begin{figure*}[htb]
\hspace{0.0cm}\includegraphics[width=6cm]{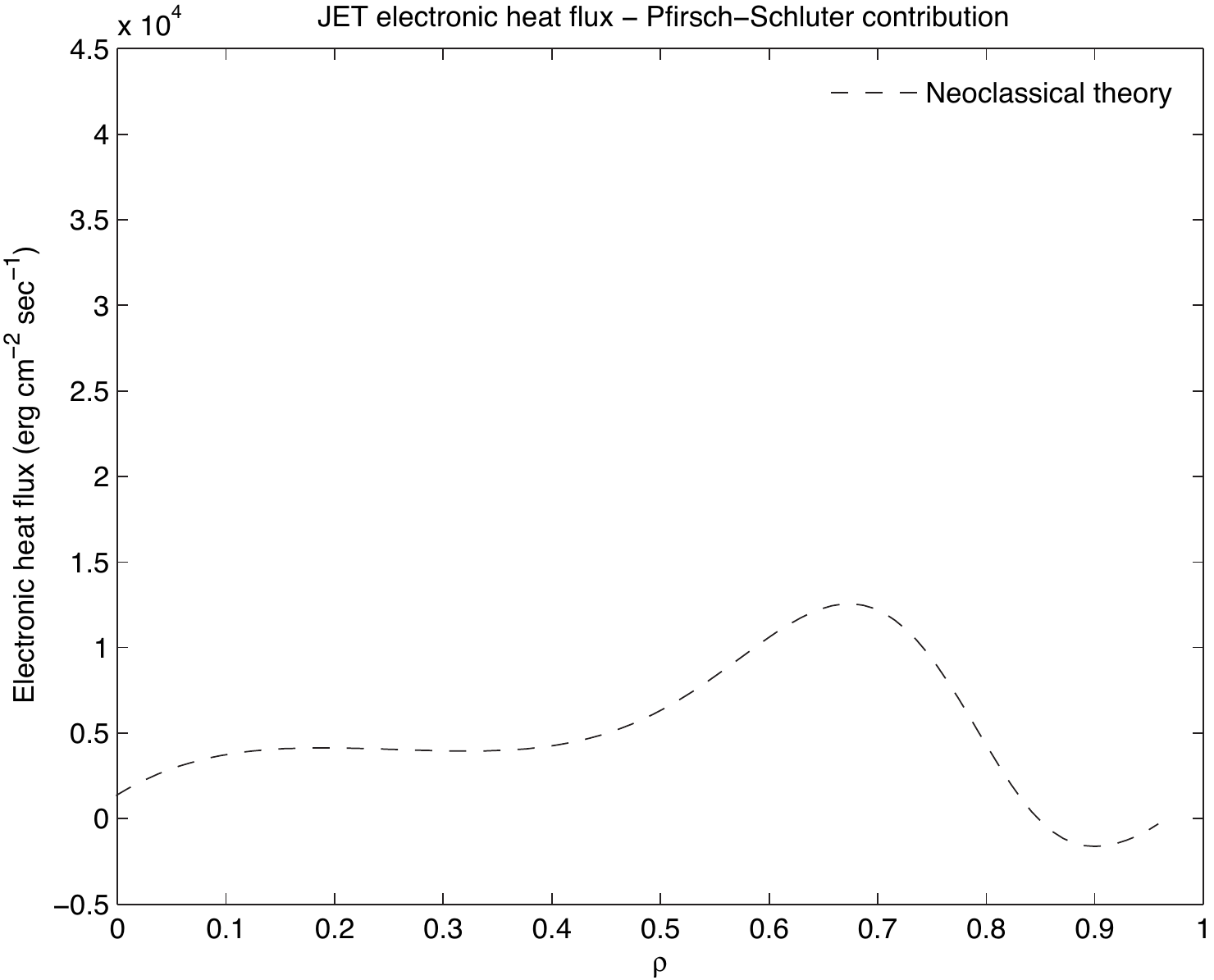}\includegraphics[width=6.2cm]{Q3DIM_M1.pdf}
\caption{ \label{comp_heat_flux} The nonlinear Pfirsch-Schl{\"{u}}ter effect: comparison between the (radial) electron heat  fluxes profiles computed by the neoclassical theory (left picture) and the Thermodynamic Field Theory (right picture). Also in this case, the profiles are qualitatively similar but, in magnitude, they differer by a factor, which may be of order $10^2$.}
\end{figure*}
%%%%%%%%%%%%%%%%%%%%%%%%%%%%%%%%%%%%%%%%%%%%%%%%%%%
%%%%%%%%%%%%%%%%%%%%%%%%%%%%%%%%%%%%%%%%%%%%%%%%%
\section{Conclusions}\label{conclusions}

We apply the thermodynamic field to study the classical and Pfirsch-Schl{\"u}ter (P-S) transport regimes in a magnetically confined plasma. A new set of nonlinear transport equations i.e., the flux-force relations, have been derived. These equations provide nonlinear corrections to the linear ("Onsager") transport coefficients. A quite encouraging result is the fact that a dissymmetry appears between the P-S  ion and electron transport coefficients: the latter submits to a nonlinear correction, which makes the radial electron coefficients much larger than the former. Such a correction is absent for ions. This is in qualitative agreement with experiments. In particular, we have shown that when a plasma is out of the linear region, the 
classical and the Pfirsch-Schl{\"u}ter transport coefficients are corrected by a universal function times the Onsager matrix. For the classical regime, the amplification function is (approximatively) expressed as the arc-tangent function of the thermodynamic forces while in case of Pfirsch-Schl{\"u}ter regime it corresponds to the amplification factor $\mathcal F_{ps}$. The theoretical predictions of TFT can be interpreted in terms of collisional mechanisms through heuristic kinetic model \cite{sonnino10}. Through such a model, a crude estimation of the expressions for the transport coefficients can be derived. We showed that these coefficients are approximatively given by the linear (Onsager) transport coefficients times an {\it amplification factor} expressed as
\begin{equation}\label{kinetic1}
\mathcal F_{ps}=\frac{\sigma^2_{eff.0}}{\sigma^2_{eff.}}\frac{\tau_{e0}}{\tau_{eps}}
\end{equation}
\noindent where $\tau_{eps}$ and $\sigma_{eff.}$ are, respectively, the electron collision time and the {\it effective electron cross section} in the P-S regime and $\tau_{e0}$ and $\sigma_{eff.0}$ are the electron collision time and the effective electron cross section estimated by the linear theory in this regime. In other words, the extra term found by the TFT is linked to an estimation of the {\it time-step} of the random walk of collisional particles. Combining our results with the kinetic model we find
\begin{equation}\label{kinetic2}
\tau_{eps}\simeq\frac{g-1}{{\tilde g}_{ps}-1}\Bigl(\frac{\sigma_{eff.0}}{\sigma_{eff.}}\Bigr)^2\tau_{e0}
\end{equation}
\noindent The electronic collision time decreases as the thermodynamic forces increase in magnitude. This implies that the electrical conductivity {\it decreases} as the intensity of the electric field increases, while the values of the electrical thermal and diffusion coefficients {\it increase} as the intensity of the thermodynamic forces increases \cite{sonnino10}. These results are in line with numerical simulations of a Lorentz gas subject to a strong electric field \cite{klages}. 

\noindent Concrete calculations for L-mode JET plasmas have also been reported. We found that, in the core of the plasma, the nonlinear classical transport coefficients may exceed the linear classical ones by a factor of order $2$. In this region of the plasma, for $Z=1$, the electron nonlinear Pfirsch-Schl{\"{u}}ter coefficients may exceed the linear ones by a factor of order $10^2$. 

\noindent This analysis is extended to the banana and plateau regimes in part II of this article.
 
\section{Appendix: The Gauge Invariance of the Field Equations}\label{gauge}
In the weak-field approximation, the small corrections $h_{\mu\nu}$ to the Onsager matrix, i.e., $g_{\mu\nu}(X)=L_{\mu\nu}+h_{\mu\nu}(X)$, satisfy the equations:
\begin{equation}\label{gauge1}
L^{\lambda\kappa}\frac{{\partial}^2h_{\mu\nu}}{\partial X^{\lambda} X^{\kappa}}+L^{\lambda\kappa}\frac{{\partial}^2h_{\lambda\kappa}}{\partial X^{\mu} X^{\nu}}-L^{\lambda\kappa}\frac{{\partial}^2h_{\lambda\nu}}{\partial X^{\kappa} X^{\mu}}-L^{\lambda\kappa}\frac{{\partial}^2h_{\lambda\mu}}{\partial X^{\kappa} X^{\nu}}=0
\end{equation}
\noindent These equations should be solved with the appropriate boundary conditions. These will be explicitly derived in appendix (\ref{Solutions}). In this section we clarify two important aspects: we provide the definition of gauge invariance of the field equations related to our problem and we prove that our choice of boundary conditions respects the principle of covariance.

\noindent The most general coordinate transformation that leaves the field weak is of the form
\begin{equation}\label{gauge2}
X^{\mu}\rightarrow X^{\prime\mu}=X^{\mu}+{\epsilon}^{\mu}(X)
\end{equation}
\noindent where $\frac{\partial {{\epsilon}^{\mu}(X)
}}{\partial X^{\nu}}$ is at most of the same order of magnitude as $h_{\mu\nu}(X)$. From Eq.(\ref{gauge2}), we have
\begin{eqnarray}\label{gauge3}
\frac{\partial {X^{\prime\mu}}}{\partial X^{\nu}}&=& {\delta}_{\nu}^{\mu}+\frac{\partial {{\epsilon}^{\mu}(X)}}{\partial X^{\nu}}\nonumber\\
\frac{\partial X^{\nu}}{\partial X^{\prime\mu}}&=& {\delta}_{\mu}^{\nu}-\frac{\partial {{\epsilon}^{\nu}(X)}}{\partial X^{\mu}}+O({\epsilon}^2)
\end{eqnarray}
\noindent we also find
\begin{equation}\label{gauge4}
g_{\mu\nu}^{\prime}(X)=g_{\mu\nu}^{\prime}(X^{\prime})+\frac{\partial {g_{\mu\nu}(X)}}{\partial X^{\lambda}}{\epsilon}^{\lambda}(X)+O({\epsilon}^2)
\end{equation}
\noindent and from the relation
\begin{equation}\label{gauge5}
g_{\mu\nu}^{\prime}(X^{\prime})=\frac{\partial {X^{\lambda}}}{\partial X^{\prime\mu}}\frac{\partial {X^{\kappa}}}{\partial X^{\prime\nu}}g_{\lambda\kappa}(X)
\end{equation}
\noindent we obtain
\begin{equation}\label{gauge6}
g_{\mu\nu}^{\prime}(X^{\prime})=g_{\mu\nu}(X)-g_{\lambda\nu}(X)
\frac{\partial {{\epsilon}^{\lambda}(X)}}{\partial X^{\mu}}-g_{\lambda\mu}(X)
\frac{\partial {{\epsilon}^{\lambda}(X)}}{\partial X^{\nu}}+O({\epsilon}^2)
\end{equation}
\noindent Taking into account Eq.(\ref{gauge5}), Eq.(\ref{gauge4}) becomes
\begin{equation}\label{gauge7}
g_{\mu\nu}^{\prime}(X)=g_{\mu\nu}(X)-g_{\lambda\nu}(X)
\frac{\partial {{\epsilon}^{\lambda}(X)}}{\partial X^{\mu}}-g_{\lambda\mu}(X)
\frac{\partial {{\epsilon}^{\lambda}(X)}}{\partial X^{\nu}}+
\frac{\partial {g_{\mu\nu}(X)}}{\partial X^{\lambda}}{\epsilon}^{\lambda}(X)
\end{equation}
\noindent From Eq.(\ref{gauge7}) we find
\begin{equation}\label{gauge9}
h_{\mu\nu}^{\prime}(X)=h_{\mu\nu}(X)-\frac{\partial {{\epsilon}_{\mu}(X)}}{\partial X^{\nu}}-
\frac{\partial {{\epsilon}_{\nu}(X)}}{\partial X^{\mu}}
\end{equation}
\noindent The meaning of Eq.~(\ref{gauge9}) is that the effect of an infinitesimal coordinate transformation on the tensor $h_{\mu\nu}(X)$ is that the new tensor $h_{\mu\nu}^{\prime}(X)$ equals the tensor $h_{\mu\nu}(X)-\frac{\partial {{\epsilon}_{\mu}(X)}}{\partial X^{\nu}}-
\frac{\partial {{\epsilon}_{\nu}(X)}}{\partial X^{\mu}}$ {\it at the same coordinate point}. 

Let us now consider the boundary conditions. As already stressed, the amplitudes of the thermodynamic forces are, in reality, bounded (see for example, figs~\ref{g1P}$\div$\ref{gi3} for a JET plasma). Therefore Eq.(\ref{gauge1}) should be solved with the appropriated boundary conditions. The choice of the boundary conditions cannot be arbitrary but should be done in order to respect the principle of covariance (other than the symmetry of the problem). In this paper we shall consider the case where, in coordinate $X^{\mu}$, the boundary $\partial$ is an hyper-sphere $B(R_{0})$ of radius $R_0$, centered in the origin of the axes. $h_{\mu\nu}(X)$ equals a constant, say with value $c_{\mu\nu}$, on this hyper-sphere. This case corresponds to the situation in which in the thermodynamic space one direction is not privileged with respect to another one. In order to re-obtain the Onsager laws when the system approaches equilibrium, we have also to require that the tensor $h_{\mu\nu}$ vanishes at the origin of the axes i.e., $h_{\mu\nu}({0})={0}$. In appendix (\ref{Solutions}), we show that these constrains, together with two other conditions, specify a well-posed Dirichlet's problem. 

\noindent We observe that, if $h_{\mu\nu}(X)$ satisfies Eq.(\ref{gauge1}) and the boundary conditions, so will also be $h'_{\mu\nu}(X)=h_{\mu\nu}(X)-\frac{\partial {{\epsilon}_{\mu}(X)}}{\partial X^{\nu}}-
\frac{\partial {{\epsilon}_{\nu}(X)}}{\partial X^{\mu}}$, obtained by performing a general coordinate transformation, with the tensor $\frac{\partial {{\epsilon}_{\mu}(X)}}{\partial X^{\nu}}+
\frac{\partial {{\epsilon}_{\nu}(X)}}{\partial X^{\mu}}$ vanishing on the hyper-sphere and at the origin of the axes. 

\noindent In conclusion, {\it The coordinate transformation 
\begin{equation}\label{gauge10}
X^{\mu}\rightarrow X^{\prime\mu}=X^{\mu}+{\epsilon}^{\mu}(X)
\end{equation}
\noindent with 
\begin{eqnarray}\label{gauge11}
&&\frac{\partial {{\epsilon}_{\mu}(X)}}{\partial X^{\nu}}+
\frac{\partial {{\epsilon}_{\nu}(X)}}{\partial X^{\mu}}{\Bigg\arrowvert}_ {B(R_{0})}=0\nonumber\\
&&\frac{\partial {{\epsilon}_{\mu}(X)}}{\partial X^{\nu}}+
\frac{\partial {{\epsilon}_{\nu}(X)}}{\partial X^{\mu}}{\Bigg\arrowvert}_ {X=0}=0
\end{eqnarray}
\noindent is the most general coordinate transformation, under the weak-field approximation. Moreover, Eq.(\ref{gauge9}) ensures that if $h_{\mu\nu}(X)$ is a solution of Eq.(\ref{gauge1}), so will also be $h'_{\mu\nu}(X)$ given by Eq.(\ref{gauge9}). $h_{\mu\nu}(X)$ and the transformed field $h'_{\mu\nu}(X)$ vanish as the system approaches equilibrium and, at the boundary, $h_{\mu\nu}(X)\mid_{B(R_{0})}=h'_{\mu\nu}(X)\mid_{B(R_{0})}=c_{\mu\nu}=const.$} This property is called the {\it gauge invariance} of the field equations. Notice that in the generalized polar coordinates $({\hat r},{\bold{\theta_i}},\phi)$ with $i=1\cdots n-2$, $r={\hat r} R_0$ and 
\begin{eqnarray}\label{gauge13}
X^1 &=& r\cos{\theta}_1\nonumber\\
X^2 &=& r\sin{\theta}_1\cos{\theta}_2\nonumber\\
X^3 &=& r\sin{\theta}_1\sin{\theta}_2\cos{\theta}_3\nonumber\\
\cdots & &\cdots\\
X^{n-1} &=& r\sin{\theta}_1\cdots r\sin{\theta}_{n-2}\cos\phi\nonumber\\
X^{n} &=& r\sin{\theta}_1\cdots r\sin{\theta}_{n-2}\sin\phi\nonumber
\end{eqnarray}
\noindent Eqs~(\ref{gauge11}) take the form
\begin{eqnarray}\label{gauge11a}
&&{\tilde\varepsilon}_{\mu\nu}(0,{\bold{\theta}}_i,\phi)={\tilde\varepsilon}_{\mu\nu}(1,{\bold{\theta}}_i,\phi)=0\qquad {\rm where}\nonumber\\
&&{\tilde\varepsilon}_{\mu\nu}(X)\equiv\frac{\partial {{\epsilon}_{\mu}(X)}}{\partial X^{\nu}}+
\frac{\partial {{\epsilon}_{\nu}(X)}}{\partial X^{\mu}}
\end{eqnarray}
\noindent Fig.(\ref{gaugeinv}) illustrates the gauge invariance property.
%%%%%%%%%%%%%%%%%%%%%%%%%%%%%%%%%%%%%%%%%%%%%%%%%%%
\begin{figure*}[htb] 
\hspace{1.5cm}\includegraphics[width=8cm]{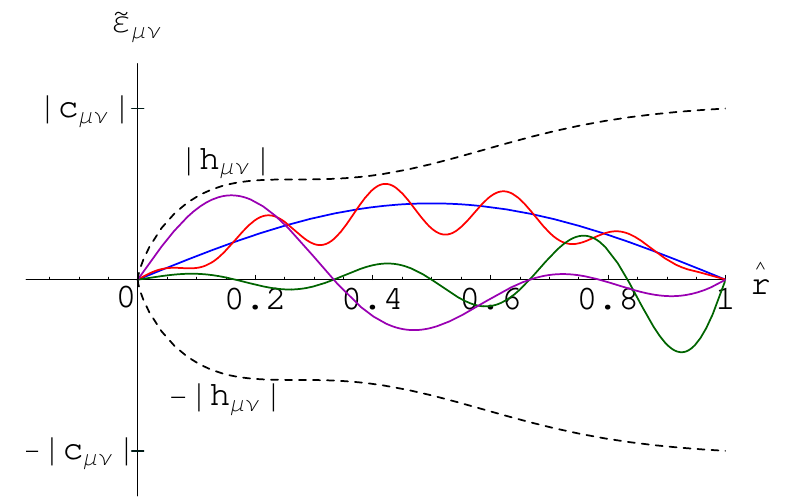}
\caption{ \label{gaugeinv} The gauge invariance of the field equations:  if $h_{\mu\nu}(X)$ is solution of the field equations (satisfying the boundary conditions and $h_{\mu\nu}(0)=0$), so will also be tensor $h'_{\mu\nu}(X)=h_{\mu\nu}(X)+{\tilde\varepsilon}_{\mu\nu}(X)$, with ${\tilde\varepsilon}_{\mu\nu}(X)$ any tensor of the type plotted above (continuous lines). ${\hat r}$ indicates the normalized radial polar coordinate (${\hat r}=r/R_0$).}
\end{figure*}
%%%%%%%%%%%%%%%%%%%%%%%%%%%%%%%%%%%%%%%%%%%%%%%%%%%

\noindent Now, to perform explicitly our calculations, we have to choose a particular gauge i.e., to decide in which particular coordinate system we want to work. It is often convenient to work in the {\it harmonic coordinate system} \cite{weinberg}, defined as\footnote{In literature, the harmonic coordinate system applies to the {\it hyperbolic} P.D.E., written in a covariant form. Here, we indicate the harmonic coordinate system as the one satisfying the {\it elliptic} P.D.E., written in a covariant form.\label{footgauge}}
\begin{equation}\label{gauge18}
{\Delta}^{2} X^{\mu} =0
\end{equation}
\noindent Since 
\begin{equation}\label{gauge19}
{\Delta}^{2} X^{\mu} =(g^{\lambda\kappa}X_{;\lambda}^{\mu})_{;\kappa}=-{\Gamma}_{\lambda\kappa}^{\mu}g^{\lambda\kappa}=-{\Gamma}^{\mu}
\end{equation}
\noindent considering $g^{\lambda\kappa}\frac{{\partial}^2 X^{\mu}}{\partial X^{\lambda} X^{\kappa}}=0$, Eq.(\ref{gauge19}) is satisfied if ${\Gamma}^{\mu}=0$, which, in the weak-field approximation, reduces to:
\begin{equation}\label{gauge20}
L^{\mu\lambda}\frac{{\partial}h_{\lambda\nu}}{\partial X^{\mu}}=
\frac{1}{2}L^{\mu\lambda}\frac{{\partial}h_{\mu\lambda}}{\partial X^{\nu}}
\end{equation}
\noindent Notice that this choice is always possible by performing the following coordinate transformations 
\begin{equation}\label{gauge21}
X^{\mu}\rightarrow X^{\prime\mu}=X^{\mu}+\hat{\epsilon}^{\mu}(X)
\end{equation}
\noindent with $\hat{{\epsilon}}^{\mu}(X)$ satisfying the differential equation
\begin{equation}\label{gauge22}
L^{\lambda\kappa}\frac{{\partial}^2{\hat{{\epsilon}}_{\nu}}}{\partial X^{\lambda} X^{\kappa}}=
L^{\lambda\kappa}\frac{{\partial}h_{\kappa\nu}}{\partial X^{\lambda}}-
\frac{1}{2}L^{\lambda\kappa}\frac{{\partial}h_{\lambda\kappa}}{\partial X^{\nu}}
\end{equation}
\noindent with boundary conditions
\begin{equation}\label{gauge23}
\frac{\partial {{\hat\epsilon}_{\mu}(X)}}{\partial X^{\nu}}+
\frac{\partial {{\hat\epsilon}_{\nu}(X)}}{\partial X^{\mu}}{\Bigg\arrowvert}_ {B(R_{0})}=0
\end{equation}
\noindent 
\begin{equation}\label{gauge23a}
\frac{\partial {{\hat\epsilon}_{\mu}(X)}}{\partial X^{\nu}}+
\frac{\partial {{\hat\epsilon}_{\nu}(X)}}{\partial X^{\mu}}{\Bigg\arrowvert}_ {X=0}=0
\end{equation}
\noindent If $h_{\mu\nu}$ does not satisfy Eq.(\ref{gauge18}), we can always find an appropriate $h'_{\mu\nu}(X)$ that does, by performing the coordinate transformation (\ref{gauge21}) with $\hat{{\epsilon}}^{\mu}(X)$ satisfying Eqs (\ref{gauge22})-(\ref{gauge23a}). Indeed, if $h_{\mu\nu}(X)$ satisfies Eq.(\ref{gauge1}), with boundary conditions $h_{\mu\nu}(X){\mid}_{B(R_{0})}=c_{\mu\nu}$ and $h_{\mu\nu}(0)=0$, from Eqs~(\ref{gauge22})-(\ref{gauge23a}), is deduced that
\begin{eqnarray}\label{gauge24}
&&L^{\lambda\kappa}\frac{{\partial}^2{h'_{\mu\nu}(X)}}{\partial X^{\lambda} X^{\kappa}}=0\nonumber\\
&&h'_{\mu\nu}(X){\mid}_{B(R_{0})}=c_{\mu\nu}=\mathrm{const.}\\
&&h'_{\mu\nu}(0)=0\nonumber
\end{eqnarray}
\noindent and
\begin{equation}\label{gauge25}
L^{\lambda\kappa}\frac{{\partial}h'_{\kappa\nu}(X)}{\partial X^{\lambda}}=
\frac{1}{2}L^{\lambda\kappa}\frac{{\partial}h'_{\lambda\kappa}(X)}{\partial X^{\nu}}
\end{equation}
\noindent We shall now prove that, in the limit of validity of the weak-field approximation, i.e., up to first order in $\epsilon$, the boundary conditions respect the principle of covariance. Let us first note that putting $g_{\mu\nu}(X)=L_{\mu\nu}+h_{\mu\nu}(X)$, from Eq.(\ref{gauge4}) we find
\begin{equation}\label{gauge8}
h_{\mu\nu}^{\prime}(X^{\prime})=h'_{\mu\nu}(X)+O({\epsilon}^2)
\end{equation}
\noindent In particular, at the leading order, in $X'=0$, we easily obtain $h'_{\mu\nu}(X')\mid_{X'=0}=0$. Indeed, from Eq.~(\ref{gauge2}) we have
\begin{equation}\label{gauge8a1}
h'_{\mu\nu}(X')\mid_{X'=0}=h'_{\mu\nu}(X_0)+O({\epsilon}^2)\quad {\rm where}\quad X_0^\mu+\epsilon^\mu(X_0)=0
\end{equation}
\noindent Therefore, at the leading order in $\epsilon$, we find
\begin{equation}\label{gauge8a2}
h'_{\mu\nu}(X')\mid_{X'=0}=h'_{\mu\nu}(-\epsilon(X_0))+O({\epsilon}^2)=h'_{\mu\nu}(X)\mid_{X=0}+O({\epsilon}^2)=0+O({\epsilon}^2)
\end{equation}
\noindent In the new coordinate system $X'^{\mu}$, $h'_{\mu\nu}(X')$ satisfies Eq.(\ref{gauge1}) and, up to the first order in $\epsilon$, remains constant, with constant equal to $c_{\mu\nu}$, on the hyper-sphere of radius $R_0$, centered at the origin of the axes $X'=0$. Indeed, indicating with $\partial$ and ${\partial}'$ the two boundaries in the $X^{\mu}$ and the $X'^{\mu}$ coordinate systems respectively, we can write
\begin{eqnarray}\label{gauge12}
\partial &:& \delta_{\mu\nu}X^{\mu}X^{\nu}=R_{0}^2\nonumber\\
{\partial}' &:& \delta_{\mu\nu}X'^{\mu}X'^{\nu}=R_{0}^2+
\delta_{\mu\nu}X'^{\mu}{\epsilon}^{\nu}+\delta_{\mu\nu}X'^{\nu}{\epsilon}^{\mu}+O({\epsilon}^{2})
\end{eqnarray}
\noindent In polar coordinates, boundaries $\partial$ and ${\partial}'$ are defined as
\begin{eqnarray}\label{gauge14}
\partial &:& r=R_{0}\nonumber\\
{\partial}' &:&r'=R_{0}+O({\epsilon}^{1/2})
\end{eqnarray}
\noindent From Eqs~(\ref{gauge9}) and (\ref{gauge8}) we find
\begin{equation}\label{gauge15}
h'_{\mu\nu}(X'){\Bigg\arrowvert}_ {{\partial}'}=h_{\mu\nu}(X){\Bigg\arrowvert}_ {\partial}-{\tilde\varepsilon}_{\mu\nu}(X){\Bigg\arrowvert}_ {\partial}=c_{\mu\nu}=\mathrm{const.}
\end{equation}
\noindent On the other hand, we also have:
\begin{eqnarray}\label{gauge16}
h'_{\mu\nu}(X'){\Bigg\arrowvert}_ {{\partial}'} &=&h'_{\mu\nu}(r', {\theta}'_1, {\theta}'_2,\cdots ,{\theta}'_{n-2}, {\phi}'){\Bigg\arrowvert}_ {{\partial}'}\nonumber\\
&=&h'_{\mu\nu}(R_{0}, {\theta}'_1, {\theta}'_2,\cdots ,{\theta}'_{n-2}, {\phi}'){\Bigg\arrowvert}_ {{\partial}'}+O({\epsilon}^{3/2})\\
&=&h'_{\mu\nu}(r', {\theta}'_1, {\theta}'_2,\cdots ,{\theta}'_{n-2}, {\phi}'){\Bigg\arrowvert}_ {r'=R_0}+O({\epsilon}^{3/2})\nonumber
\end{eqnarray}
\noindent where we have taken into account that tensors $h'_{\mu\nu}$ and $h_{\mu\nu}$ are quantities of the first order in $\varepsilon$. However, in polar coordinates, Eq.(\ref{gauge15}) reads:
\begin{eqnarray}\label{gauge17}
&&h'_{\mu\nu}(r', {\theta}'_1, {\theta}'_2,\cdots ,{\theta}'_{n-2}, {\phi}'){\Bigg\arrowvert}_ {r'=R_0}=
h_{\mu\nu}(r, {\theta}_1, {\theta}_2,\cdots ,{\theta}_{n-2}, {\phi}){\Bigg\arrowvert}_ {r'=R_0}-
\nonumber\\
&&{\tilde\varepsilon}_{\mu\nu}(r', {\theta}'_1, {\theta}'_2,\cdots ,{\theta}'_{n-2}, {\phi}'){\Bigg\arrowvert}_ {r'=R_0}\!\!\!\!\!\!\!\!\!\!\!\!\!\!\!+O({\epsilon}^{3/2})=c_{\mu\nu}+O({\epsilon}^{3/2}))
\end{eqnarray}
\noindent Summarizing, {\it If we have to solve Eq.(\ref{gauge1}) in a particular coordinate system $X^{\mu}$ where $h_{\mu\nu}(X)$ is constant, say $c_{\mu\nu}$, on the hyper-sphere of radius $R_0$ and $h_{\mu\nu}(0)=0$ then, after a coordinate transformation (\ref{gauge10})-(\ref{gauge11}), also the new tensor $h'_{\mu\nu}(X')$ is solution of Eq.(\ref{gauge1}) and, in the limit of validity of the weak-field approximation (i.e., up to the first order order in $\epsilon$), it also assumes the same constant value $c_{\mu\nu}$, on the hyper-sphere having the same radius $R_0$, in the space $X'^{\mu}$. Moreover, at the origin of the axes $X'^\mu=0$, we also have $h'_{\mu\nu}(0)=0$}. In other words, up to the first order, the covariance principle is respected. Fig.~(\ref{covariance}) illustrates the covariance principle.

%%%%%%%%%%%%%%%%%%%%%%%%%%%%%%%%%%%%%%%%%%%%%%%%%%%
\begin{figure*}[htb] 
\hspace{1.5cm}\includegraphics[width=10cm]{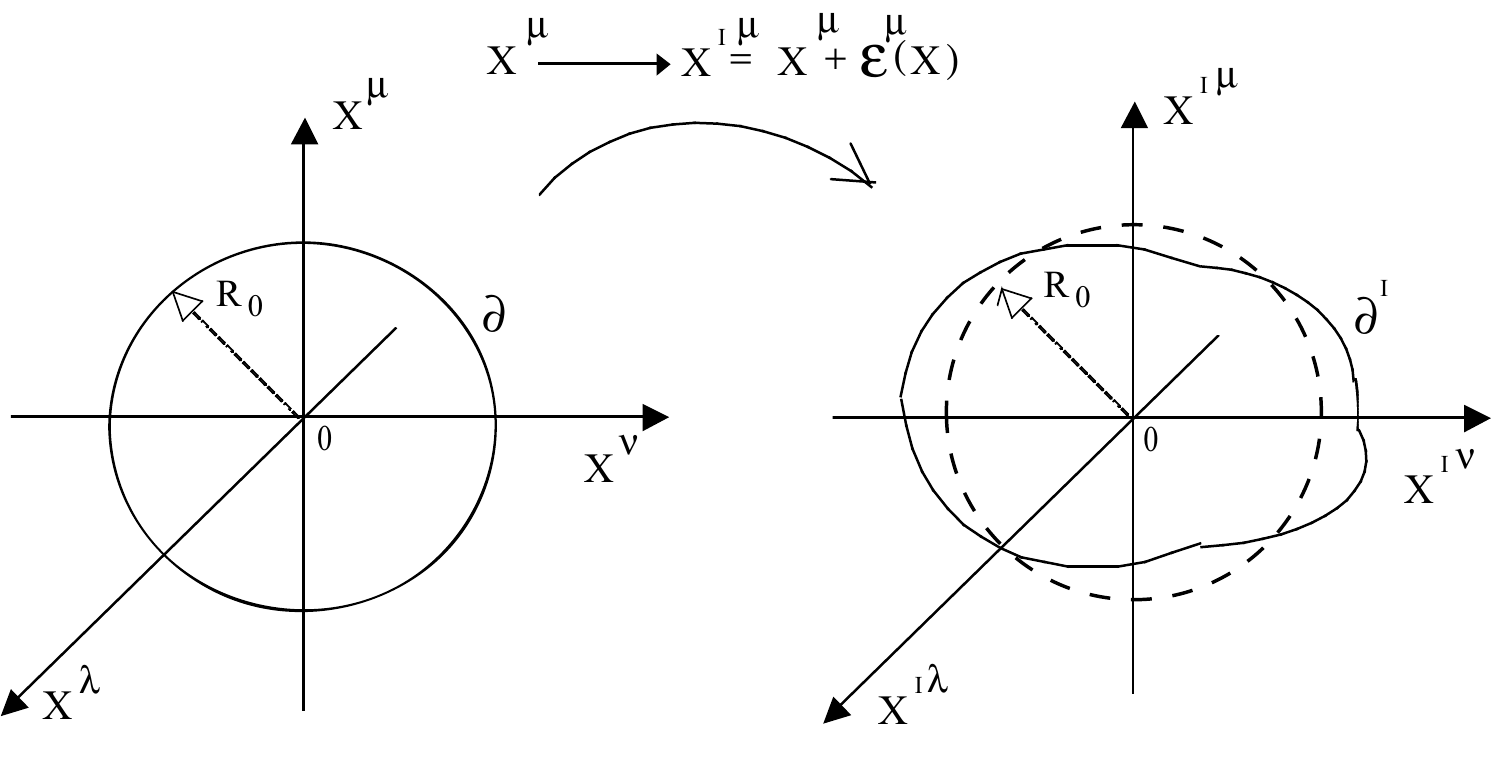}
\caption{ \label{covariance} The covariance principle: If $h_{\mu\nu}(X)$ is solution of Eq.~(\ref{gauge1}) satisfying the boundary conditions $h_{\mu\nu}(X)\mid_{B(R_{0})}=c_{\mu\nu}=const.$ on a hyper-sphere of radius $R_0$, centered at the origin of the axes $X=0$, and $h_{\mu\nu}(0)=0$, then, after an infinitesimal coordinate transformation, also $h'_{\mu\nu}(X')$ is solution of Eq.~(\ref{gauge1}) and, up to the first order in $\epsilon$, it assumes the same constant value $c_{\mu\nu}$ on the hyper-sphere of radius $R_0$ centered at the origin of the axes $X'=0$. Moreover, $h'_{\mu\nu}(0)=0$.}
\end{figure*}
%%%%%%%%%%%%%%%%%%%%%%%%%%%%%%%%%%%%%%%%%%%%%%%%%%%

\noindent Notice that, with the gauge choice (\ref{gauge21})-(\ref{gauge23a}), from Eqs~(\ref{gauge3}), (\ref{gauge8}) and (\ref{gauge8a2}), we derive
\begin{eqnarray}\label{gauge26}
&&L^{\lambda\kappa}\frac{{\partial}^2{h'_{\mu\nu}(X')}}{\partial X'^{\lambda} X'^{\kappa}}=0+O({\hat{\epsilon}}^2)\nonumber\\
&&L^{\lambda\kappa}\frac{{\partial}h'_{\kappa\nu}(X')}{\partial X'^{\lambda}}=
\frac{1}{2}L^{\lambda\kappa}\frac{{\partial}h'_{\lambda\kappa}(X')}{\partial X'^{\nu}}+O({\hat{\epsilon}}^2)\\
&&h'_{\mu\nu}(X'){\mid}_{B(R_{0})}=c_{\mu\nu}+O({\hat{\epsilon}}^{3/2})=\mathrm{const.}+O({\hat{\epsilon}}^{3/2})\nonumber\\
&&h'_{\mu\nu}(0)=0+O({\hat{\epsilon}}^{2})\nonumber
\end{eqnarray}
\noindent 

\section{Appendix: Solutions of the Field Equations}\label{Solutions}
As seen in the last section, in a convenient choice of the coordinate system (the gauge choice) and after having performed an orthogonal coordinate transformation, our problem reduces to finding the solutions of Laplace equation. In this section we shall derive the appropriate Dirichlet boundary conditions. The explicit solution for the two-dimensional case, can be found in appendix~(\ref{2-solution})\footnote{The solution in a 3-dimensional thermodynamic space, is presented in ref.\cite{sonnino9}.\label{solutions}}. 

\noindent The Laplace equation, in two or three dimensions, reads:
\begin{equation}\label{Laplace}
{\Delta}^{2} h_{\mu\nu}(\mathbf{X}) =0
\end{equation}
\noindent Solution of Eq.(\ref{Laplace}) should however satisfy the following conditions:
\begin{description}
\item[a)] The solution should vanish at the origin of the axes: 
\begin{equation}\label{a}
h_{\mu\nu}(\mathbf{0}) ={\bf 0}
\end{equation}
\item[b)] The solution should be constant, say with value ${c}_{\mu\nu}$, on the spherical surface (possibly, except in a set of measure zero where it can assume values different from ${c}_{\mu\nu}$):
\begin{equation}\label{b}
h_{\mu\nu}(\mathbf{X}){\mid}_{B(R_{0})}=c_{\mu\nu}\neq 0
\end{equation}
\noindent {where} $B(R_{0})$ indicates a ball of radius $R_{0}$ centered at the origin of the axes;
\item[c)] The solution should be invariant with respect to the permutation of the axes $x$, $y$ and $z$;
\item[d)] Inside the sphere, the solution should be of class $\mathit{C^{2}}$ (possibly, except in a set of measure zero where it should be at least of class $\mathit{C^{0}}$).
\end{description}
\noindent Now, our problem consists in re-writing these four conditions in order to have a well-posed Dirichlet's problem. First, we proceed by analyzing the simple case of Laplace's equation where the solution is of class $\mathit{C^{2}}$ in all points inside the sphere and constant on the spherical boundary, i.e.,:
\begin{equation}\label{sphere}
{\Delta}^{2} h_{\mu\nu}(\mathbf{X}) =0\,\,\,\,\,\,\,\,\,\,\, \mathrm{where}\,\,\,\,\,
h_{\mu\nu}(\mathbf{X}){\mid}_{B(R_{0})}=c_{\mu\nu}\neq 0
\end{equation}
\noindent As known, from the {\it mean value theorem} \cite{zachmanoglou}, the solution is
\begin{equation}\label{sphere1}
h_{\mu\nu}(\mathbf{X}) = c_{\mu\nu} = \mathrm{const.}
\end{equation}
\noindent in all points inside the sphere. Solution (\ref{sphere1}) does not satisfy the condition a) and for this it should be discarded. Thus the solution can be found by performing domain decomposition and then apply the appropriate boundary conditions in such a way as to obtain a harmonic function satisfying the conditions a), b), c) and d). However, these cuts cannot be  arbitrary but they have:
\begin{description}
\item[i)] to respect the symmetry of the problem;
\item[ii)] to be performed in such a way that, after having obtained the solution in a portion of the sphere, we shall then be able to reconstruct the entire solution, valid for the whole sphere, making use of Schwartz's principle (see ref \cite{courant}).
\end{description}
\noindent Our solution, corresponding to the function which satisfies Eq.(\ref{Laplace}) together with conditions a), b), c) and d), will be obtained by performing the {\it minimum} number of cuts satisfying conditions i) and ii). 
\noindent In this purpose, let us cut the sphere with a plane passing through the origin and solve the P.D.E. (\ref{Laplace}) with the following boundary conditions:
\begin{equation}\label{sphere2}
h_{\mu\nu}(R_{0},\theta , \phi) = \left\{ \begin{array}{ll}
c_{\mu,\nu} = \mathrm{const.}& \mbox{if $ 0\leq\theta < \frac{\pi}{2}$}\\
v_{\mu\nu}(\theta,\phi) & \mbox{if $\frac{\pi}{2} < \theta \leq\pi$}
\end{array}
\right.
\end{equation}
\noindent where $v_{\mu\nu}(\theta,\phi)$ are functions of the variables $\theta$ and $\phi$. Note that cutting the sphere with a plane, passing through the origin of the sphere, and not with a generic surface, is the only possibility that we have if we want to respect the condition i). Of course without loss of generality, by performing a rotation of the axes, it is always possible to reach the plane $z=0$. Now, denoting with $\tilde{h}_{\mu\nu}(r,\theta,\phi)$ the solution of Eq. (\ref{Laplace}) with boundary conditions (\ref{sphere2}) in the upper hemisphere $0\leq\theta < \frac{\pi}{2}$, we are able to write the entire solution valid in all points of the sphere thanks to Schwartz's principle:
\begin{equation}\label{sphere3}
h_{\mu\nu}(r,\theta, \phi)= \left\{ \begin{array}{ll}
\tilde{h}_{\mu\nu}(r,\theta,\phi) & \mbox{if $0\leq\theta < \frac{\pi}{2}$}\\
- \tilde{h}_{\mu\nu}(r, \theta,\phi) & \mbox{if $\frac{\pi}{2} < \theta \leq\pi$}\\
\end{array}
\right.
\end{equation}
\noindent Solution (\ref{sphere3}) is of class $\mathit{C^{0}}$ on the plane $z=0$ if we have
\begin{equation}\label{sphere4}
\tilde{h}_{\mu\nu}(r,\frac{\pi}{2},\phi) = - \tilde{h}_{\mu\nu}(r,\frac{\pi}{2},\phi)\quad\quad\quad\quad\mathrm{i.e.,}\quad\quad\quad 
\tilde{h}_{\mu\nu}(r,\frac{\pi}{2},\phi) = 0
\end{equation}
\noindent In conclusion, the solution (\ref{sphere3}), obtained by applying Schwartz's principle, is of class $\mathit{C^{0}}$ inside the sphere if it {\it vanishes} on the plane $z=0$. This constraint respects the condition a). On the other hand, to find a harmonic function, which vanishes on the plane $z=0$ and is $constant$ on the upper spherical surface, corresponds to the solution of a well-posed Dirichlet's problem. In this case, the solution exists, it is unique and can be easily found by solving, for example, the P.D.E. (\ref{Laplace}) with the following boundary conditions:
\begin{equation}\label{sphere4a}
h_{\mu\nu}(R_{0},\theta , \phi) = \left\{ \begin{array}{ll}
c_{\mu,\nu} & \mbox{if $ 0\leq\theta < \frac{\pi}{2}$}\\
-c_{\mu,\nu} & \mbox{if $\frac{\pi}{2} < \theta \leq\pi$}
\end{array}
\right.
\end{equation}
\noindent Using Schwartz's principle, the harmonic solution, having the same value constant on the entire spherical surface (except in a set of measure zero), reads \cite{morse}:
\begin{eqnarray}\label{sphere5-1}
&&\!\!\!\!\!\!\!\!\!\!\!\!\!\!\!\!\!h_{\mu\nu}(r,\theta,\phi) = \\
& &\!\!\!\!\!\!\!\!\!\!\!\frac{c_{\mu\nu}}{2}\sum_{n=1} ^{\infty} [ P_{n-1}(0)-P_{n+1}(0)+P_{n+1}(-1)-P_{n-1}(-1)]\!P_{n}(\cos\theta)\!\biggl (\frac{r}{R_{0}}\biggr )^{n}\nonumber 
\end{eqnarray}
\noindent for $0\leq\theta < \frac{\pi}{2}$, and 
\begin{eqnarray}\label{sphere5-2}
&&\!\!\!\!\!\!\!\!\!\!\!\!\!\!\!\!\!h_{\mu\nu}(r,\theta,\phi) = \\
& &\!\!\!\!\!\!\!\!\!\!\!\!\!\!-\!\frac{c_{\mu\nu}}{2}\sum_{n=1} ^{\infty} [ P_{n-1}(0)-P_{n+1}(0)+P_{n+1}(-1)-P_{n-1}(-1)]\!P_{n}(\cos\theta)\!\biggl (\frac{r}{R_{0}}\biggr )^{n}\nonumber 
 \end{eqnarray}
 \noindent for $\frac{\pi}{2} < \theta \leq\pi$. In the last expressions, $P_{n}(\cos(\theta))$ indicate {\it Legendre's polynomials}. As we can see, expressions (\ref{sphere5-1}) and (\ref{sphere5-2}) do not represent the solution we are looking for because they satisfy the conditions a), b) and d) but not condition c). However, a solution, which is also invariant with respect to the permutation of the three axes, can be evaluated by cutting the sphere with three perpendicular planes, passing through the origin, and requiring that the solution vanishes also on the planes $x=0$ and $y=0$ (in addition to the plane $z=0$). For this we have to solve the well-posed Dirichlet's problem in the first 3-dimensional quadrant with boundary conditions $zero$ on the three planes $x=0$, $y=0$ and $z=0$ and, $constant$ on the spherical boundary. The entire solution, valid for all quadrants, will be reconstructed using Schwartz's principle.  Therefore we seek a harmonic solution, which is constant on each spherical surface of the 3-dimensional quadrants, with value alternatively equal to $+c_{\mu\nu}$ and to $-c_{\mu\nu}$ (going counter clock-wise starting from the first 3-dimensional quadrant) i.e.: 
\begin{equation}\label{sphere6}
h_{\mu\nu}(R_{0},\theta ,\phi) = \left\{ \begin{array}{ll}
c_{\mu,\nu} & \mbox{if $ 0 < \phi<\frac{\pi}{2}$}\\
-c_{\mu,\nu} & \mbox{if $\frac{\pi}{2} < \phi < \pi$}\\
c_{\mu,\nu} & \mbox{if $ - \pi < \phi< - \frac{\pi}{2}$}\\
- c_{\mu,\nu} & \mbox{if $- \frac{\pi}{2} < \phi< 0$} 
\end{array}
\right.
\end{equation}
\noindent for $ 0\leq \theta<\frac{\pi}{2}$, and 
\begin{equation}\label{sphere7}
h_{\mu\nu}(R_{0},\theta ,\phi) = \left\{ \begin{array}{ll}
- c_{\mu,\nu} & \mbox{if $ 0 < \phi<\frac{\pi}{2}$}\\
c_{\mu,\nu} & \mbox{if $\frac{\pi}{2} < \phi < \pi$}\\
- c_{\mu,\nu} & \mbox{if $ - \pi < \phi< - \frac{\pi}{2}$}\\
c_{\mu,\nu} & \mbox{if $- \frac{\pi}{2} < \phi< 0$} 
\end{array}
\right.
\end{equation}
\noindent for $\frac{\pi}{2}< \theta\leq\pi$. 
The solution of our problem, satisfying the conditions a), b), c) and d), can be then obtained by applying Schwartz's principle to the expression valid for the first 3-dimensional quadrant. The validity of the above method is not limited only to the three dimensional case. Indeed, it can be easily generalized to find the solution in a space having $n$-dimensions. In particular, for the 2-dimensional case, we have to solve Eq.(\ref{Laplace}) submitted to the following boundary conditions:
\begin{equation}\label{sphere8}
h_{\mu\nu}(R_{0},\theta) = \left\{ \begin{array}{ll}
c_{\mu,\nu} & \mbox{if $ 0 < \theta<\frac{\pi}{2}$}\\
-c_{\mu,\nu} & \mbox{if $\frac{\pi}{2} < \theta < \pi$}\\
c_{\mu,\nu} & \mbox{if $ - \pi < \theta< - \frac{\pi}{2}$}\\
- c_{\mu,\nu} & \mbox{if $- \frac{\pi}{2} < \theta< 0$} 
\end{array}
\right.
\end{equation}
\noindent and then apply Schwartz's principle. This procedure is shown in the following subsection. The 3-dimensional solution can be found in the appendix of the second part of this work \cite{sonnino9}.

\subsection{The Two-Dimensional Problem}\label{2-solution}
In this section we shall obtain the harmonic solution, in two-dimensions, satisfying the conditions a), b), c) and d) mentioned in the previous section. The solution of this problem can be found in ref. \cite {sonnino}. Here, we shall re-obtain the same solution using the method illustrated in the last section. Let us then start solving Eq.(\ref{Laplace}) with the following boundary conditions (see Fig.\ref{2D}):
%%%%%%%%%%%%%%%%%%%%%%%%%%%%%%%%%%%%%%%%%%%%%%%%%%%%%%%%%%%%%%%%%%%%%%%%%%%%%

\begin{figure*}[htb] 
\hspace{3cm}\includegraphics[width=12.0cm,height=5.5cm]{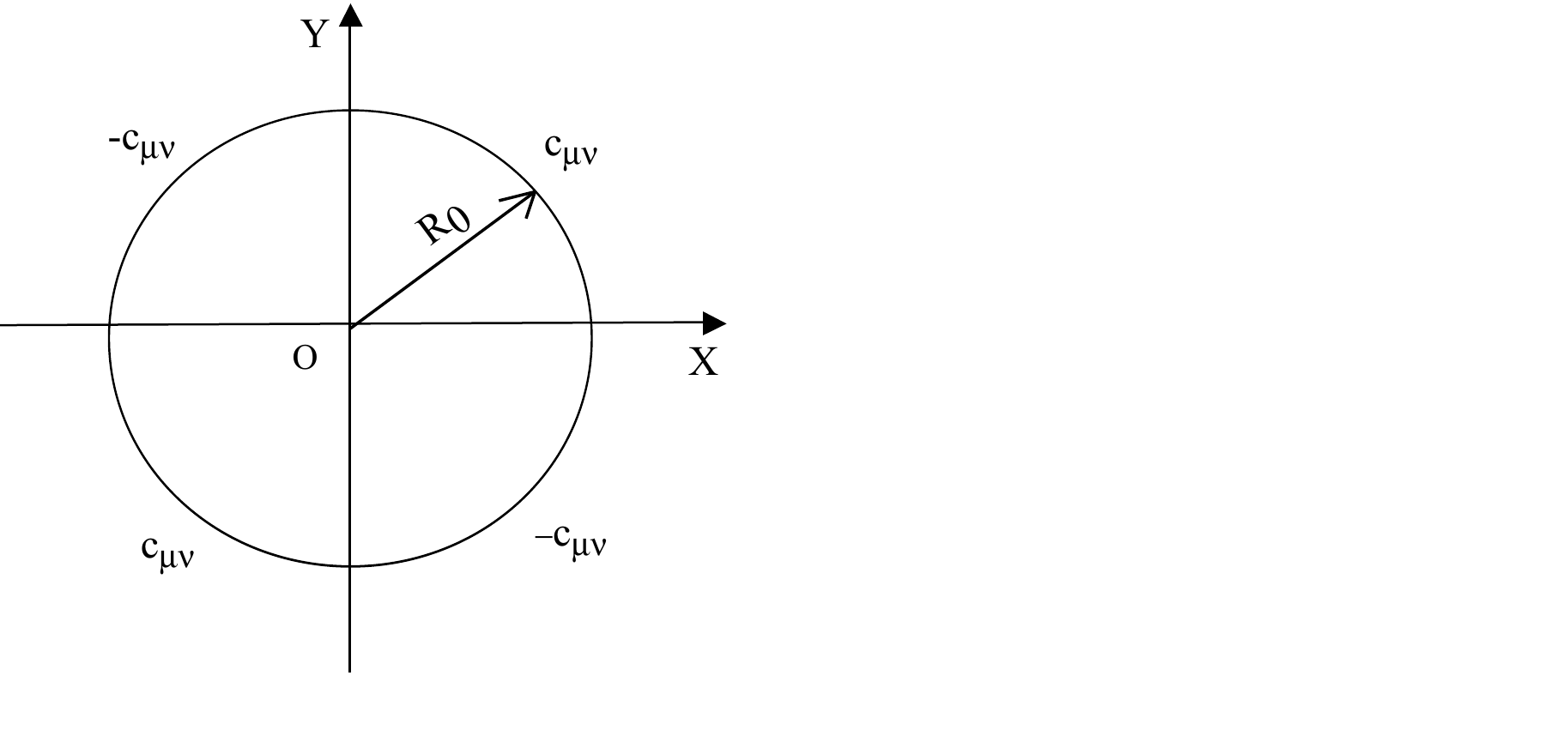}
\caption{ \label{2D} The 2-dimensional boundary conditions. }
\end{figure*}

%%%%%%%%%%%%%%%%%%%%%%%%%%%%%%%%%%%%%%%%%%%%%%%%%%%%%%%%%%%%%%%%%%%%%%%%%%%%%%%

\begin{equation}\label{circle}
h_{\mu\nu}(R_{0},\theta) = \left\{ \begin{array}{ll}
c_{\mu,\nu} & \mbox{if $ 0 < \theta<\frac{\pi}{2}$}\\
-c_{\mu,\nu} & \mbox{if $\frac{\pi}{2} < \theta < \pi$}\\
c_{\mu,\nu} & \mbox{if $ - \pi < \theta< - \frac{\pi}{2}$}\\
- c_{\mu,\nu} & \mbox{if $- \frac{\pi}{2} < \theta< 0$} 
\end{array}
\right.
\end{equation}
\noindent The solution of Eq. (\ref{Laplace}) with boundary conditions (\ref{circle}) can be written in general as  \cite{zachmanoglou}:
\begin{equation}\label{circle1}
h_{\mu\nu}(r,\theta) = \frac{a_{0}}{2}+\sum_{n=1}^{\infty}\ (a_{n}\cos n\theta  +b_{n}\sin n\theta )\ \biggl (
\frac{r}{R_{0}}\biggr )^{n}
\end{equation}
\noindent with
\begin{eqnarray}\label{circle2}
a_{0} & = & 2\int_{-\pi}^{\pi}u_{\mu\nu}(R_{0},\theta)\ d\theta\nonumber \\
a_{n} & = & \frac{1}{\pi}\int_{-\pi}^{\pi}u_{\mu\nu}(R_{0},\theta)\cos (n\theta )\ d\theta\qquad
\qquad\mathrm{for}\qquad\mathrm{n}\geq 1 \\
b_{n} & = & \frac{1}{\pi}\int_{-\pi}^{\pi}u_{\mu\nu}(R_{0},\theta)\sin (n\theta )\ d\theta \qquad
\qquad\mathrm{for}\qquad\mathrm{n}\geq 1 \nonumber
\end{eqnarray}
We can easily compute the integrals in Eq.(\ref{circle2}) and obtain
\begin{eqnarray}\label{circle3}
a_{n} & = & 0\qquad\qquad\qquad\qquad\qquad\qquad\quad\ \ \mathrm{(n=0,1,2}\cdots )\\
b_{n} & = & -c_{\mu\nu}\frac{8\cos (\frac{n\pi}{2})\sin (\frac{n\pi}{4})^2}{n\pi}\qquad\qquad\mathrm{(n=1,2}\cdots )\nonumber
\end{eqnarray}
\noindent Therefore solution $u_{\mu\nu}(r,\theta)$ can be written as
\begin{equation}\label{circle4}
h_{\mu\nu}(r,\theta) = -\frac{8 c_{\mu\nu}}{\pi}\sum_{n=1}^{\infty}\frac{\cos (\frac{n\pi}{2})\sin (\frac{n\pi}{4})^2}{n}\ \sin (n\theta )\biggl (\frac{r}{R_{0}}\biggr )^{n}
\end{equation}

\noindent which can be brought into the form
\begin{equation}\label{circle5}
h_{\mu\nu}(r,\theta) = \frac{4 c_{\mu\nu}}{\pi}\sum_{n=1}^{\infty}\frac{\sin (2(n-1)\theta)}{2n-1}\biggl (\frac{r}{R_{0}}\biggr )^{n}
\end{equation}
\noindent However, solution (\ref{circle5}) can be compacted obtaining \cite{gradshteyn}
\begin{equation}\label{circle6}
h_{\mu\nu}(r,\theta) = \frac {2  c_{\mu\nu}}{\pi}\arctan \biggl [\frac{2{\hat r}^2\sin (2\theta )}{1-{\hat r}^4}\biggr ]\quad\qquad\mathrm{where}\quad{{\hat r}\equiv\frac{r}{R_{0}}}
\end{equation}
\noindent or, in coordinate $x$ and $y$:
\begin{equation}\label{circle7}
h_{\mu\nu}(x, y) = \frac {2  c_{\mu\nu}}{\pi}\arctan \biggl [\frac{4R_{0}^2 x y}{R_{0}^4-(x^{2}+y^{2})^2}\biggr ]
\end{equation}
\noindent Solution (\ref{circle7}) is valid only in the quadrants $xy > 0$. The general solution, valid in all quadrants, can be obtained using Schwartz's principle:
\begin{equation}\label{circle8}
h_{\mu\nu}(x, y) = \frac {2  c_{\mu\nu}}{\pi}\arctan \biggl [\frac{4R_{0}^2 \mid x y\mid}{R_{0}^4-(x^{2}+y^{2})^2}\biggr ]
\end{equation}
\noindent As seen in appendix (\ref{gauge}), the coefficients $c_{\mu\nu}$ are determined imposing that solution (\ref{circle8}) satisfies the gauge invariance. In section (\ref{ps}) we obtained $c_{\mu\nu} = \chi L_{\mu\nu}$ where $\chi$ is a parameter independent of the thermodynamic forces and $L_{\mu\nu}$ indicates the Onsager matrix. In the P-S regime, parameter $\chi_{ps}$ is determined by imposing the supplementary condition e) reported in section (\ref{NLps}): {\it the losses profiles are continuous functions}. This condition, imposes that $\chi_{ps}=-1$. In the classical regime, this parameter can be determined by a simple kinetic model illustrated in ref.~\cite{sonnino10}. We obtain $\chi_{cl}=1$.

\vskip 0.5truecm

\section{Acknowledgments}
One of us (GS) would like to pay tribute to the memory of Prof. I Prigogine, a pioneer researcher in this area, and to Prof. R. Balescu, who gave him the opportunity to exchange most interesting views in different areas  of plasma physics. 

\noindent GS would like to thank his hierarchy at the European Commission and the members of the EURATOM Fusion Association, Belgian State at U.L.B.

\noindent GS is also very grateful to Dr U. Finzi, of the European Commission, for his continuing encouragement, for reading some sections and making helpful suggestions. 

%%%%%%%%%%%%%%%%%

%%%%%%%%%%%%%%%%%%%%%%%%%%%%%%%%%%%%%%%%%%%%%%%%%%%%%%%%%%%%%%%%%%%%%%%%%%%%%%%%%%%%%%%%%%%

\end{document}